\documentclass{aa}  

\usepackage{graphicx}
\usepackage{txfonts}
\usepackage[colorlinks=true,allcolors=blue]{hyperref}
\usepackage{color}
\usepackage{colortbl}
\begin{document}

   \title{Give to Ursa Minor what is Ursa Minor's:\\ an updated census of the RR Lyrae population in the Ursa Minor dwarf galaxy 
   based on {\it Gaia} DR3}

   \subtitle{}

   \author{A. Garofalo\inst{1},
          G. Clementini\inst{1},
          F. Cusano\inst{1},
          T. Muraveva\inst{1} and
          L. Monti\inst{1}}

   \institute{INAF,
Osservatorio di Astrofisica e Scienza dello Spazio di Bologna, via Piero Gobetti 93/3,
40129 Bologna, Italy\\
              \email{alessia.garofalo@inaf.it}
             }

   \date{Received xxx; accepted xxx}

 
  \abstract
   {} 
   {We use RR Lyrae stars identified by the {\it Gaia} third data release (DR3) to explore the outskirts of the  Ursa Minor (UMi) dwarf spheroidal galaxy (dSph) and update the census of its variable star population.}
   {We adopted different tools based on the {\it Gaia} DR3 astrometric and photometric data (proper motions, Period-Wesenheit-Metallicity relations, spatial distribution, colour-magnitude diagram and stellar isochrone fitting) to discriminate between different types of variable stars,  and to identify UMi members
   .}
   {We find a total of 129 RR Lyrae stars and Anomalous Cepheids (ACs) that belong to UMi. 
   We report 47 new RR Lyrae stars (46 bona fide and 1 candidate) and 5 new ACs (4 bona fide and 1 candidate), including new possible members in the extreme periphery of the galaxy at a distance of $\sim$ 12 half-light radii.
    We reclassified 13 RR Lyrae stars identified by the  {\it Gaia} DR3 Specific Object Study pipeline for Cepheids and RR Lyrae stars (SOS Cep\&RRL), using data from the literature and {\it Gaia} astrometry and photometry. Specifically, we assigned these 13 DR3 RR Lyrae stars to 10 Anomalous Cepheids and 3 double-mode RR Lyrae 
    (RRd)
    , respectively.
   From the average luminosity of the RR Lyrae stars we derive for UMi a distance modulus of $(m-M)_{0}$ 19.23 $\pm$ 0.09 mag in excellent agreement with the literature. 
   Finally, we investigated whether some of UMi's variable stars 
   might be members of 
   the ultra-faint stellar cluster Mu{\~n}oz~1, that lies at a projected distance of  45$\arcmin$ from UMi's centre. 
    Based on the properties of the variable stars (distances, colours and metallicities), we find unlikely that they 
   belong to the cluster.}
   {}

\titlerunning{Updated census of RR Lyrae in Ursa Minor galaxy}
   \maketitle
%

\section{Introduction}

 RR Lyrae stars are the most frequent pulsating variable stars among Local Group (LG) galaxies. 
Since they are old (typical age $>$10 Gyr) and have predominately sub-solar metallicity, although spanning a quite large total range in iron abundance of $-$2.5$<[\rm{Fe/H}]<$0.2 dex, 
they are widely used as tracers of old and metal-poor stellar populations in the Milky Way (MW) and in the LG (\citealt{Smith-1995,Catelan-2015} and reference therein).
They 
provide an important {\it asset} in the study of resolved stellar populations, because 
from their properties 
we can infer 
indirect information such as distance, age, metallicity and reddening on the stellar population of the host galaxies (\citealt{Sarajedini-2011} and reference therein).\\
The {\it Gaia} 
third data release (DR3; \citealt{Vallenari-2023}), 
published positions and distances for nearly two billion sources observed 
during the initial 34 months of science operations, and multi-epoch photometry allowing the identification of more than 10 million variable sources down to a limiting magnitude G $\sim$ 21 mag.  
A final catalogue of 270,891 RR Lyrae stars 
processed through the Specific Object Study pipeline for Cepheids and RR Lyrae stars (SOS Cep\&RRL, \citealt{Clementini-2023} and reference therein) has been published in {\it Gaia} DR3. They are distributed all sky tracing the MW outer halo, its stellar streams and the old component of its satellites as far as 100 kpc from us.
So far this is 
the largest, most homogeneous, and parameter-richest catalogue of all-sky RR Lyrae stars published in the $G$ magnitude range from $\sim$ 7 to $\sim$ 21 mag.  
For almost all of them {\it Gaia} DR3 provided astrometry (individual parallaxes and proper motions), photometry (epoch and mean $G$, $G_{BP}$ and $G_{RP}$ magnitudes), pulsation characteristics (period, amplitude, Fourier parameters etc.) and  based on the latter:  
individual photometric metallicity and $G$ absorption for nearly half of the sample (for details see \citealt{Clementini-et-2019,Clementini-2023}). 
{\it Gaia}, and {\it Gaia} DR3 in particular, 
has had a huge impact on the role of RR Lyrae stars as standard candles and stellar population tracers.
Based on the {\it Gaia} DR3 parallaxes the luminosity-metallicity (LZ) relation followed by RR Lyrae stars in the visual, and the period-luminosity-(metallicity),  and period-Wesenheit-(metallicity) (PW(Z)) relations followed in the infrared 
were recalibrated 
(\citealt{Prudil-et-al-2024} in the optical bands, 
\citealt{Garofalo-et-al-2022,Li-et-al-2023,Prudil-et-al-2024} in the {\it Gaia} bands, 
and 
\citealt{Zgirski-et-al-2023,Prudil-et-al-2024} in the infrared). 
In addition, using the pulsation properties of the {\it Gaia} DR2 and DR3 RR Lyrae stars, empirical relations were re-determined in the {\it Gaia} bands to estimate individual photometric metallicities for these varaible stars \citep{Iorio-et-al-2021,Li-et-al-2023,Muraveva-et-al-2024}. Several studies have also been 
undertaken on the spatial distribution of RR Lyrae stars in MW dwarf spheroidal (dSph) and ultra-faint dwarf satellites, and to trace the stellar population of a possible extended stellar halo (\citealt{Tau-et-al-2024} and reference therein) based on {\it Gaia} DR3 data.
\\RR Lyrae stars are very frequently identified  in dSph galaxies,  
being these systems typically dominated by old (population II) stars. The dSphs are 
the most common type of galaxies in the LG and 
they are often found to surround 
larger galaxies such as the MW and M31.
The $\Lambda$ Cold Dark Matter ($\Lambda$CDM)  model \citep{SearleZinn-1978} of hierarchical formation of structures predicts that massive galaxies grow from accretion of smaller systems. 
The dSph satellites of the MW 
may have been involved in the formation and evolution of our Galactic halo and be the witnesses and leftovers of the MW assembling process. Studying their resolved stellar populations allow us to improve our knowledge of their properties and shed light on whether 
they contribute to the Galactic halo formation and evolution.
With {\it Gaia} 
providing astrometric data like positions and proper motions and, at the same time,  photometric data in 3 optical bands for sources distributed all sky, it has become possible, as never before, to explore the extreme outskirts of the MW satellites (dSphs and ultra-faint dSphs) and find member stars up to 9 or 10 half-light radii (r$_{h}$) away from the centre of the galaxies as for instance in Tucana~II \citep{Chiti-2021}, in Sculptor and Ursa Minor \citep{Jensen-et-al-2024}.
\\The Ursa Minor (UMi) dSph galaxy was discovered by \citet{Wilson-1955} as one of the smallest and faintest galaxies (M${_V}=-$8.8 mag; \citealt{McConnachie-2012} and reference therein) in the LG. UMi and other 5 dSphs  (Sculptor, Draco, Fornax, Leo~I and Leo~II) were known as Sculptor systems after Sculptor, the first galaxy of this type to be discovered. Since then, gigantic progress has been made in 
the study of structural and dynamical parameters of dSph galaxies in and around the LG, thanks in particular to wide-field surveys and high-resolution capabilities 
 become available at  
ground-based and space telescopes, and the increased computational capabilities to manage  the resulting data flow.
UMi, being one of the first MW satellites to be discovered, has been extensively studied over the decades with photometric and spectroscopic observations.\\
It is widely known that UMi appears to be dominated by a very old (t $>$ 10 Gyr) and metal-poor ([Fe/H] $\sim-$2 dex) stellar population (see \citealt{Mateo-1998,Mighell-1999,Dolphin-2002, Bellazzini-et-al-2002}). However, already \cite{Bellazzini-et-al-2002} study based on photometric data collected at the Italian  Telescopio Nazionale Galileo (TNG) showed 
that UMi has a sizeable spread in metal content ($\sigma_{\rm[Fe/H]}$=0.10 dex) due to a   significant self-enrichment occurred 
in this system in quite a short timescale, a few Gyr, at very early epochs.
In addition they noticed the presence of a massive and extended CDM halo that 
might have inhibited the formation of 
significant substructures in UMi's stellar component. This study also showed 
that the galaxy innermost regions are strongly structured and asymmetrical.
\\\citet{Kirby-et-al-2010} published 
multi-element abundances from medium-resolution spectroscopy 
with DEIMOS at the Keck Observatory for almost 3 thousand individual stars
in 8 MW dSphs. 
In particular, they provide metal abundances  for 212 red giant branch (RGB) stars of UMi spanning the from  $-$3.81 to $-$0.66 dex in [Fe/H]. In a following paper focused on the metallicity evolution of individual dSphs
\citet{Kirby-et-al-2011} 
infer for UMi a mean metallicity  [Fe/H]=$-2.13 \pm 0.01$ dex with spread ($\sigma_{\rm[Fe/H]}$) of 0.34 dex. 
 They also looked for 
 a metallicity gradient, 
 but found a 
 negligible slope of [Fe/H] with radius,   
 in their sample which, however, only covers  an  inner region within 3 half radii of the galaxy. Furthermore, they noticed that unlike 
 other dSphs that typically show at least a hint of a metal-poor tail in their metallicity distribution, UMi has a metal-rich tail, which could perhaps be contaminants, i.e. metal-rich Galactic stars.
 More recently, \cite{Pace-et-al-2020}, published the largest spectroscopic dataset of UMi's stars more than doubling the number of known spectroscopic members. They confirmed the membership for 892 stars by analysing more than 1400 objects observed with DEIMOS at the Keck Observatory. This wide dataset allowed them to find two distinct populations with different chemical, kinematic, and spatial distributions.
The first population is kinematically cold, with a velocity dispersion $\sigma{_v}$= 4.9$^{+0.8}
_{-1.0}$ kms$^{-1}$, less metal-poor ([Fe/H] = $-$2.05 $\pm$ 0.03 dex), and centrally concentrated, whereas the second population is kinematically hot, $\sigma{_v}$= 11.5$^{+0.9}_{-0.8}$ kms$^{-1}$, more metal-poor ([Fe/H] = $-$2.29$^{+0.05}_{-0.06}$ dex), and spatially more extended. In particular, \cite{Pace-et-al-2020} found that the spatial distribution of members in the outskirts of UMi, where the metal-poor stars dominate, seems more spherical than the ellipticity  $\epsilon$= 0.55 reported in \cite{McConnachie&Venn-2020} for UMi’s global stellar component (this value is also higher than typical values found for other dSphs, $\epsilon<$ 0.45, \citealt{Munoz-et-al-2018}). They inferred two separate values: $\epsilon$= 0.75 for the metal-rich and $\epsilon$= 0.33 for the metal-poor stars, respectively. \citet{Sestito-et-al-2023} using
high
resolution spectra (R$\sim$ 40000) obtained with GRACES at the Gemini-Northern telescope, measured radial velocities and metallicities for 5 member stars located from $\sim$5 to $\sim$12 times the half-light radius 
(r$_{h}$) 
of UMi, concluding that this dSph has a more extended structure than previously thought. 
In \citet{Pace-et-al-2020} the most distant member of UMi had been found near $\sim 5.5
\times$r$_{h}$, however, \citet{Sestito-et-al-2023} results show that UMi extends out to a projected elliptical distance of $\sim$ 12$\times$r$_{h}$, or $\sim$ 4.5 kpc from the centre. This distance is 
close to the tidal radius of UMi derived by \citet{Pace-et-al-2020}, 5-6 kpc.\\
To conclude the overview on UMi and 
its outer halo, \cite{Munoz-et-al-2012}, serendipitously discovered an ultra-faint stellar system, likely a Galactic 
extremely faint globular cluster, Mu{\~n}oz~1, 
with the wide-field MegaCam imager on the
Canada-France-Hawaii Telescope (CFHT). 
Mu{\~n}oz~1 is located 45$\arcmin$ ($\sim$ 30 kpc) from the center of UMi at a heliocentric distance of 45$\pm 0.5$ kpc from us and despite the very close angular separation, the system is not associated with the dSph.
\\From the variability point of view, the first studies of UMi variable star population was led by \citet{vanAgt-1967} which 
identified 93 variable stars in the galaxy publishing photometry for 39 of them and periods for 29 of them.  Later, \citet{Kholopov-1971} used van Agt’s photometry to derive periods for an additional 10 stars.
The most recent and complete study of UMi variable sources has been published by \citet{nemec88}. They identified 95 variable stars determining accurate pulsation properties for 89 of them: 82 were classified as RR Lyrae stars and 7 as Anomalous Cepheids (ACs). 
All the variable stars identified in UMi so far are located within the inner regions of the galaxy. With this work, we aim to reclassify and derive more accurate properties for the already known variable stars in UMi. Furthermore, using {\it Gaia} DR3 we explore the outer regions of this dSph to update the census of UMi's variable star members. In particular, by better defining the distribution and extension of UMi's RR Lyrae stars we get hints on the structure and geometry of UMi's ancient stellar population and its potential links with the MW stellar halo. \\
The paper is organized as follows: the starting sample of variables potentially members of UMi selected using {\it Gaia} DR3 data is described in Section~\ref{sec:data}. Section~\ref{sec:gs} introduces the Gold sample of RR Lyrae stars that we used to determine the distance to UMi. In Sections~\ref{sec:ac} and~\ref{sec:new} we describe methods and tools adopted to confirm UMi's known and new ACs and RR Lyrae stars. Their updated census 
is presented in Section~\ref{sec:final} where we also discuss the possibility that some variables might belong to Mu{\~n}oz~1. Finally, Section~\ref{sec:concl} summarises the paper's main results and conclusions.

\section{Data from Gaia DR3}\label{sec:data}
\begin{table}
	\centering
    \caption{Centre coordinates and structural parameters of UMi adopted in this work reported with the respective references}
	\label{tab:galaxy_prop}
	\begin{tabular}{lll} 
		\hline
  Property&Value&Ref\\
  \hline
  RA$_0$&15:09:11.34  &\citet{Irwin-1995}\\
  DEC$_0$&+67:12:51.7&\citet{Irwin-1995}\\
  r$_h$ (arcmin)&17.32 $\pm$ 11 & \citet{Sestito-et-al-2023}\\
    r$_h$ (pc)&382 $\pm$ 53  & \citet{Sestito-et-al-2023}\\
  $\epsilon$&0.55 $\pm$ 0.01 &\citet{Sestito-et-al-2023}\\
  $\phi$ (deg)&50 $\pm$ 1&\citet{Sestito-et-al-2023}\\
		\hline
	\end{tabular}
\end{table}
\begin{figure*}
\centering
    \includegraphics[width=\textwidth]{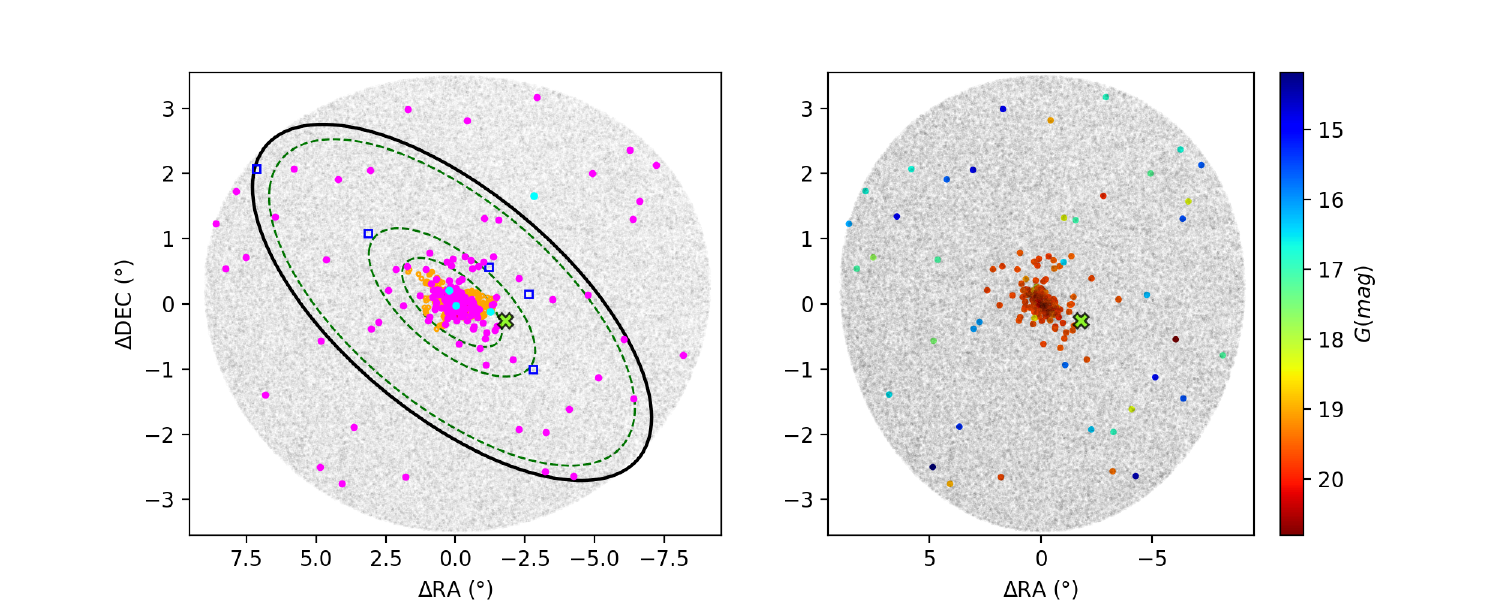}
    \caption{{\it Left panel:} Map of sources within a circle of 3.5 degrees in radius from UMi's centre coordinates, that we retrieved from the {\it Gaia} DR3 archive.  
    Four ellipses 
    show 3 (inner ellipse), 5, 11 and 12 times the galaxy r$_h$ (outer ellipse, black solid line) adopting the position angle $\phi$,
r$_h$, and ellipticity from \citet{Sestito-et-al-2023}, and the centre coordinates of \citet{Irwin-1995} (see Table~\ref{tab:galaxy_prop}). A lime cross indicates the position of the cluster Mu\~noz~1. 
Magenta filled circles mark sources classified as RR Lyrae stars in the {\it Gaia} DR3  \texttt{vari\_rrlyrae} table that are located inside and outside 12 times the galaxy r$_h$. Four cyan circles mark further candidate RR Lyrae stars listed in the {\it Gaia} DR3 \texttt{vari\_classifier\_result} table. Red and orange open circles mark spectroscopically confirmed members of UMi respectively from \citet{Kirby-et-al-2010} and \citet{Pace-et-al-2020}, blue empty squares mark the furthest RGB members of UMi from \citet{Sestito-et-al-2023}. {\it Right panel:} same as in the left panel but showing with filled circles only sources classified as RR Lyrae stars in the  \texttt{vari\_classifier\_result} and \texttt{vari\_rrlyrae} tables, colour-coded according to their mean $G$ magnitude. 
}
    \label{fig:map1}
\end{figure*}
\begin{figure}
\centering
  \includegraphics[width=\hsize]{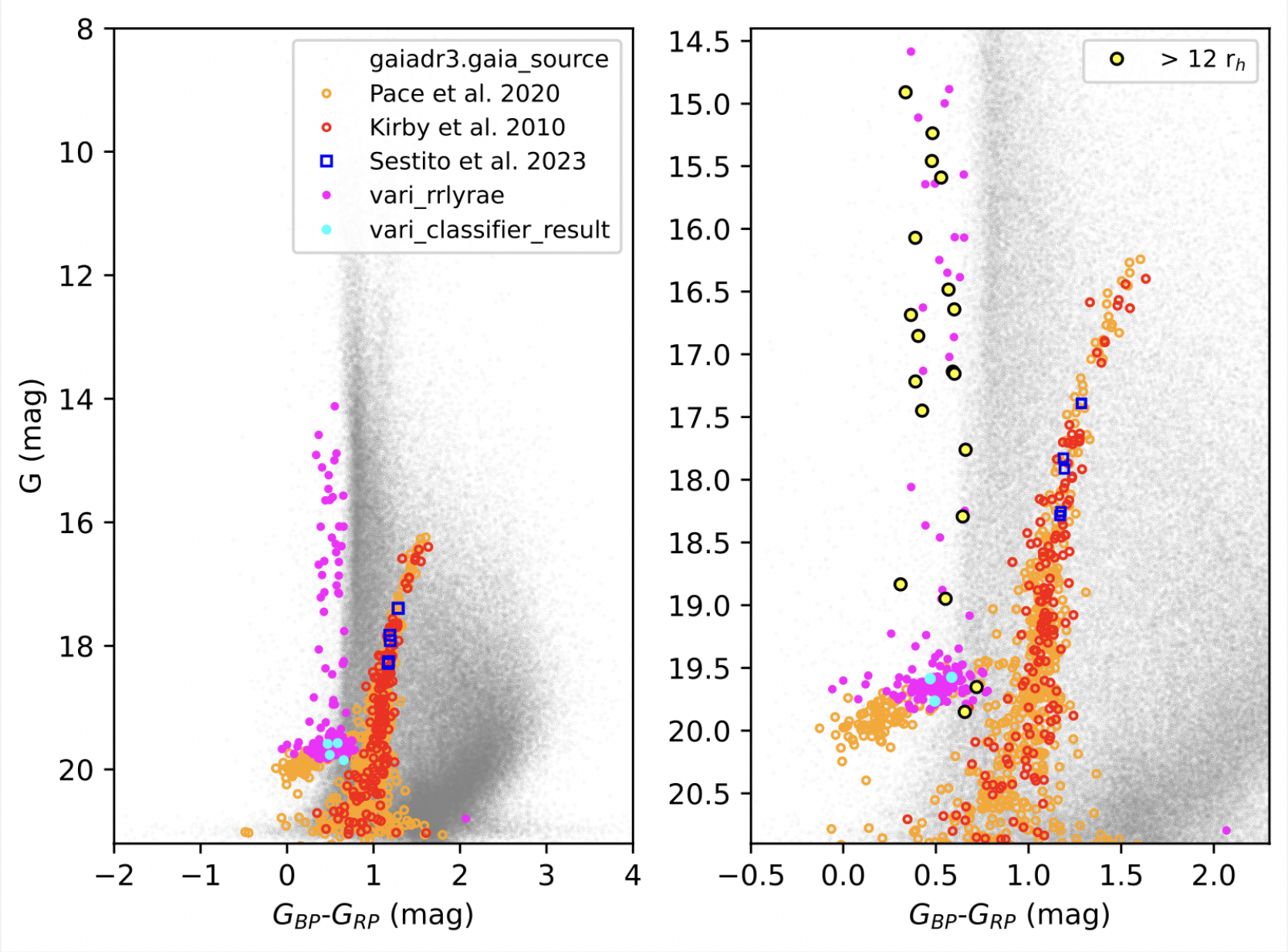}
    \caption{{\it Left:} $G$, G$_{BP}$-G$_{RP}$ CMD of all sources in our catalog of $\sim$ 160 thousand objects within the area delimited by our selection around the centre of UMi. Magenta- and cyan-filled circles mark the RR Lyrae stars classified by {\it Gaia} DR3. Red, orange and blue symbols mark UMi's members from \citet{Kirby-et-al-2010}, \citet{Pace-et-al-2020} and \citet{Sestito-et-al-2023}, respectively.
    {\it Right:} Zoom in of the left panel. The yellow-filled black circles highlight 20 
    sources classified as RR Lyrae by {\it Gaia} DR3, located beyond 12 times the elliptical half-light radius of the galaxy.}
    \label{fig:cmdall}
\end{figure}
We retrieved from the {\it Gaia} DR3 catalogues all sources contained in a circle of 3.5 degrees in radius from UMi centre coordinates (Table\ref{tab:galaxy_prop}). We chose 
this area because it coincides with 12 half-light radii from UMi's centre.
This is approximately 
the most external radius where UMi member stars have been found by \citet[$\sim11.7\times$ r$_{h}$]{Sestito-et-al-2023}. 
The query\footnote{SELECT source$\_$id, ra,  ra$\_$error, dec, dec$\_$error, parallax, parallax$\_$error, phot$\_$g$\_$mean$\_$mag, phot$\_$bp$\_$mean$\_$mag, phot$\_$rp$\_$mean$\_$mag, pmra, pmra$\_$error, pmdec, pmdec$\_$error
FROM gaiadr3.gaia$\_$source
WHERE  1 = CONTAINS(  POINT(227.24200, +67.22210), CIRCLE(ra, dec, 3.5))\\} to the {\it Gaia} archive returned a total of 160,689 sources. 
Figure~\ref{fig:map1} shows the spatial distribution of these sources.
The inner green dashed ellipse represents the area enclosed within 3 times UMi's r$_{h}$ 
according to  \citet{Sestito-et-al-2023}. 
The outer ellipses track areas which cover 5, 11 and 12 times UMi's r$_{h}$, respectively. 
In the map a lime cross indicates  
the centroid of the faint cluster Mu\~noz~1 with respect to UMi's centre.
To more safely identify stars belonging to UMi 
we cross-matched the 160,689 sources extracted from the {\it Gaia} DR3 catalogue against the list of 212 RGB members of UMi spectroscopically confirmed by 
\citet{Kirby-et-al-2010}. We recovered 166 of them ($\sim$ 78\%). Their 
metallicity ranges from $-$1.43 to $-$3.81 dex.
These spectroscopically confirmed members all lie in the inner area defined by UMi's r$_{h}$.  
They are marked by red open circles in Fig.~\ref{fig:map1}, but 
are totally covered by the variable stars (magenta-filled circles in Fig.~\ref{fig:map1}). 
We also counter-identified member stars of UMi published by \citet{Pace-et-al-2020} (Red Giant Branch $-$ RGB, and Horizontal Branch $-$ HB stars; orange-filled circles in the left panel of Fig.~\ref{fig:map1} and following figures) and \citet{Sestito-et-al-2023} (RGB stars; blue empty squares) recovering 705 sources from \citet{Pace-et-al-2020} with probability $>0.95$ to be UMi's members. Among them, 157 objects are in common with the RGB stars that we cross-matched from \citet{Kirby-et-al-2010}. 
The magenta filled circles in the left panel of Fig.~\ref{fig:map1} mark 164 
RR Lyrae stars validated and characterized by the {\it Gaia} SOS Cep \& RR Lyrae pipeline (\texttt{vari\_rrlyrae}; \citealt{Clementini-2023}), whereas 
cyan filled circles mark 4 RR Lyrae candidates identified by the general classifier of the {\it Gaia} variability pipeline, which relies  on supervised machine learning techniques  (\texttt{vari\_classifier\_result}; \citealt{Rimoldini-et-al-2023}) 
(see Section~\ref{sec:gs} for details). 
Adopting for UMi the structural parameters provided in  Table~\ref{tab:galaxy_prop}, we find that the
majority of RR Lyrae stars (126) are placed within the area covering 5 times the elliptical half-light radius of the galaxy, 22 are in the most peripheral areas between 5 and 12 times the r${_h}$), and 
20 RR variables are beyond the extreme outskirts of the galaxy, traced by the outer ellipse corresponding to 12 elliptical half$-$light radii.
In the right panel of the Fig.~\ref{fig:map1} we highlight the spatial distribution of the 168 sources in this area classified as RR Lyrae stars by  {\it Gaia} DR3 colour-coding them according to their mean $G$ magnitude.  
The majority of the RR Lyrae stars beyond 12 half$-$light radii very likely are not UMi's members and rather belong 
to the MW field 
(see discussion in Section~\ref{sec:ac}). \\
The colour–magnitude diagram (CMD) of the 160 thousand sources around UMi's centre we selected from the {\it Gaia} DR3 archive is 
shown in the left panel of Fig.~\ref{fig:cmdall}.
As in Fig.~\ref{fig:map1} magenta and cyan filled circles mark RR Lyrae classified in the {\it Gaia} DR3 \texttt{vari\_rrlyrae} and \texttt{vari\_classifier\_result} tables, respectively. The right panel of the figure provides a zoom-in view of the CMD. 
The 20 
most external variable stars in Fig.~\ref{fig:map1} 
are now highlighted by yellow filled circles. The luminosity of many of them
is incompatible with the mean magnitude of UMi's HB ($<G_{HB}>\sim$ 19.6 mag) confirming that it's unlikely that they are RR Lyrae stars belonging to this galaxy. However, some of them may still be variable stars of UMi of a different type, they could indeed be Classical Cepheids or, more likely, Anomalous Cepheids. 
Finally, the variable star with {\it Gaia} \texttt{source\_id}=1669496020369252864 (named V169, later in the text) was dropped from the sample because its mean magnitude ($G \sim$ 21 mag) and ($G_{BP} - G_{RP}$) colour are not compatible with the distance and CMD of UMi making it unlikely that it could be 
an RR Lyrae star belonging to this dSph\footnote{ 
The star 
is also placed well beyond 12 r${_h}$ from the centre of UMi ($\Delta$RA=$-$6.05 $\deg$, $\Delta$DEC=$-$0.54 $\deg$).}. 
\section{RR Lyrae stars in UMi: the Gold sample}\label{sec:gs} 
\begin{figure}
\includegraphics[width=\hsize]{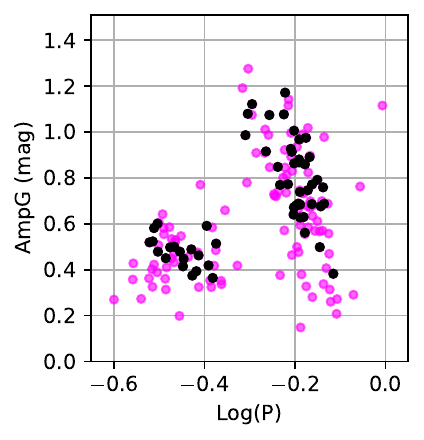}
    \caption{Period-amplitude (Bailey) diagram in the $G$ band for 164 
    variable stars in the area that we selected around UMi's centre  which are 
    included in \texttt{vari\_rrlyrae} table. Black dots mark the 57 RR Lyrae stars of the Gold sample.}
    \label{fig:bailey}
\end{figure}

\begin{figure*}
\includegraphics[width=9cm]{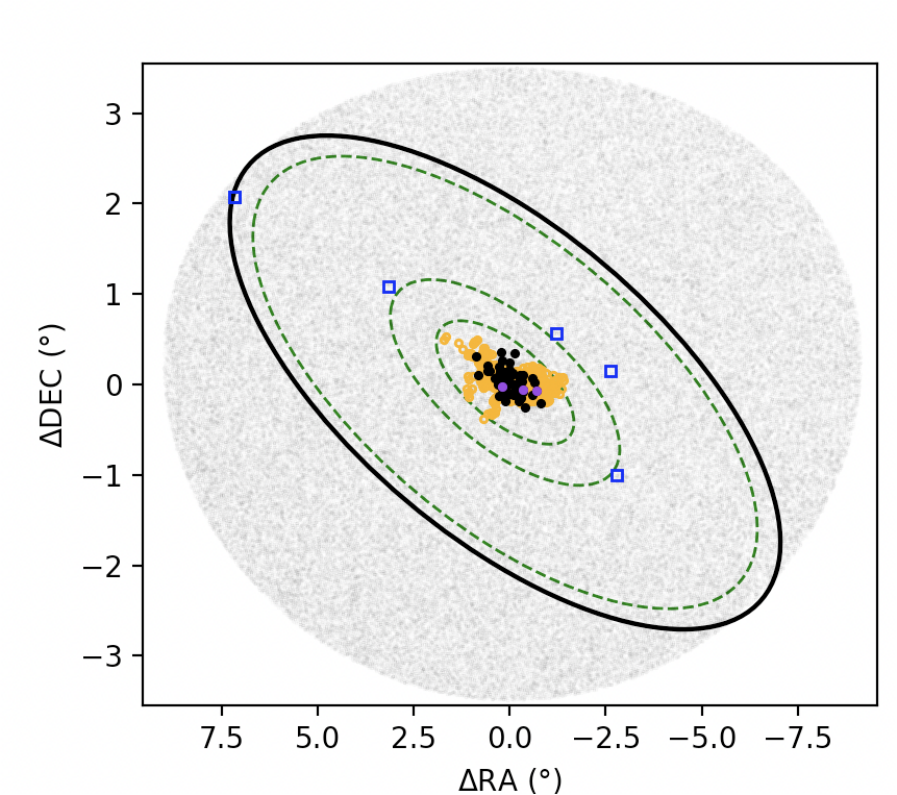}\includegraphics[width=5.8cm]{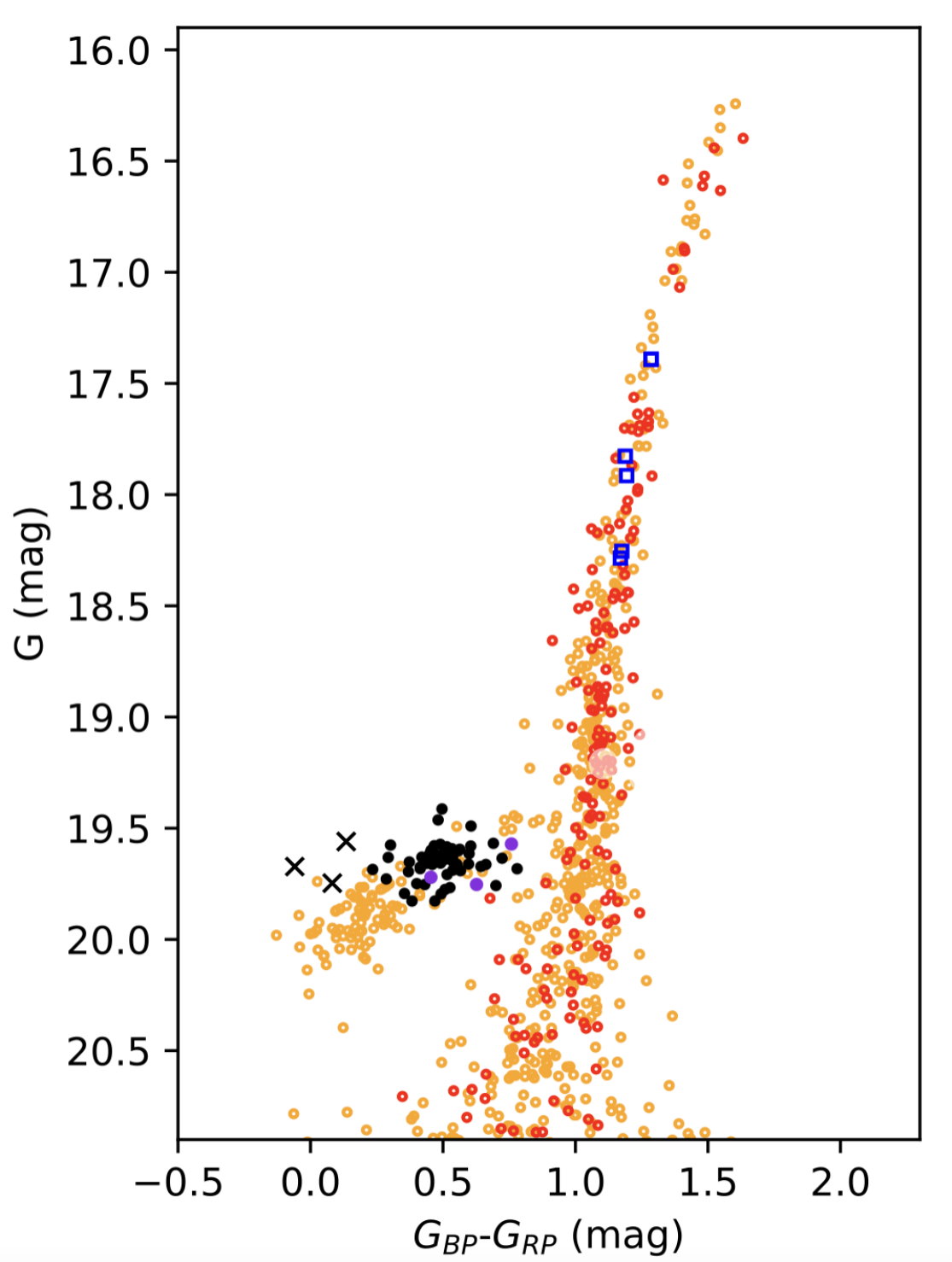}
    \caption{{\it Left panel}: Same as in the left panel of Fig.~\ref{fig:map1} but showing only the 57 RR Lyrae stars in the Gold sample plotted as black solid circle and using violet filled circles for V43, V60 and V63. We computed revised mean magnitudes for these 3 stars (see text for details). Black crosses mark V43, V60 and V63 position on the CMD using the unrevised magnitudes.  {\it Right panel}: $G$ vs $G_{BP}-G_{RP}$ CMD of UMi's spectroscopically confirmed members and the 57 RR Lyrae stars in the Gold sample. Symbols are as in the left panel. 
    }
    \label{fig:map_cmd_gold}
\end{figure*}

\begin{figure*}
\center
\includegraphics[width=6.4cm]{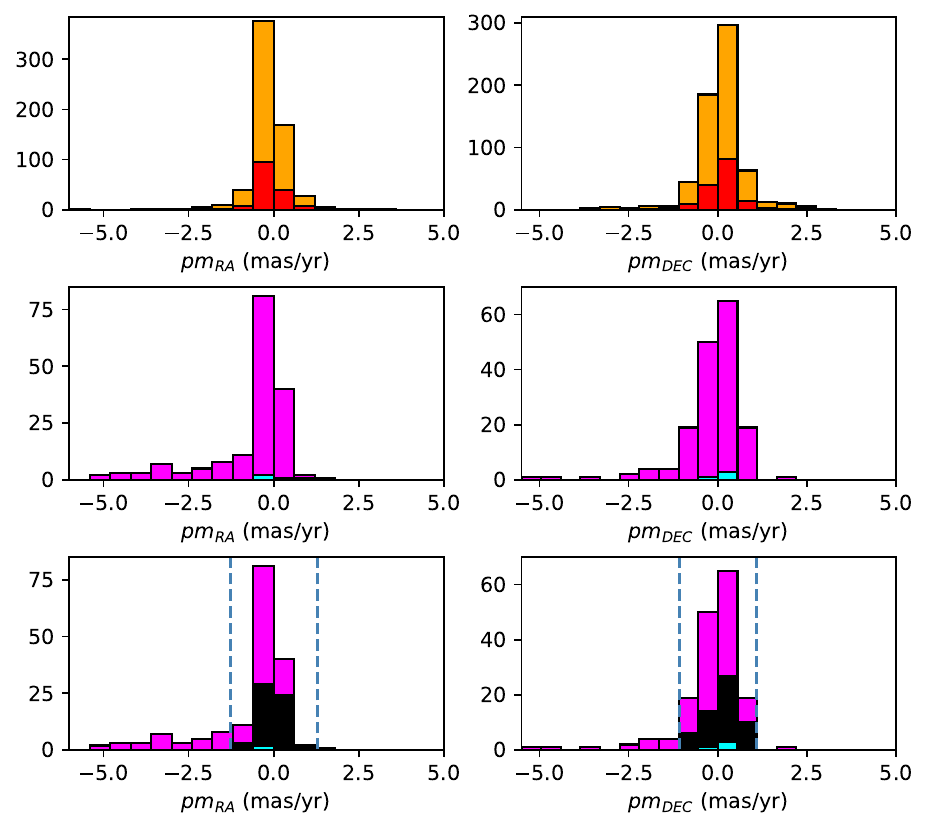}
~\includegraphics[width=11.3cm]{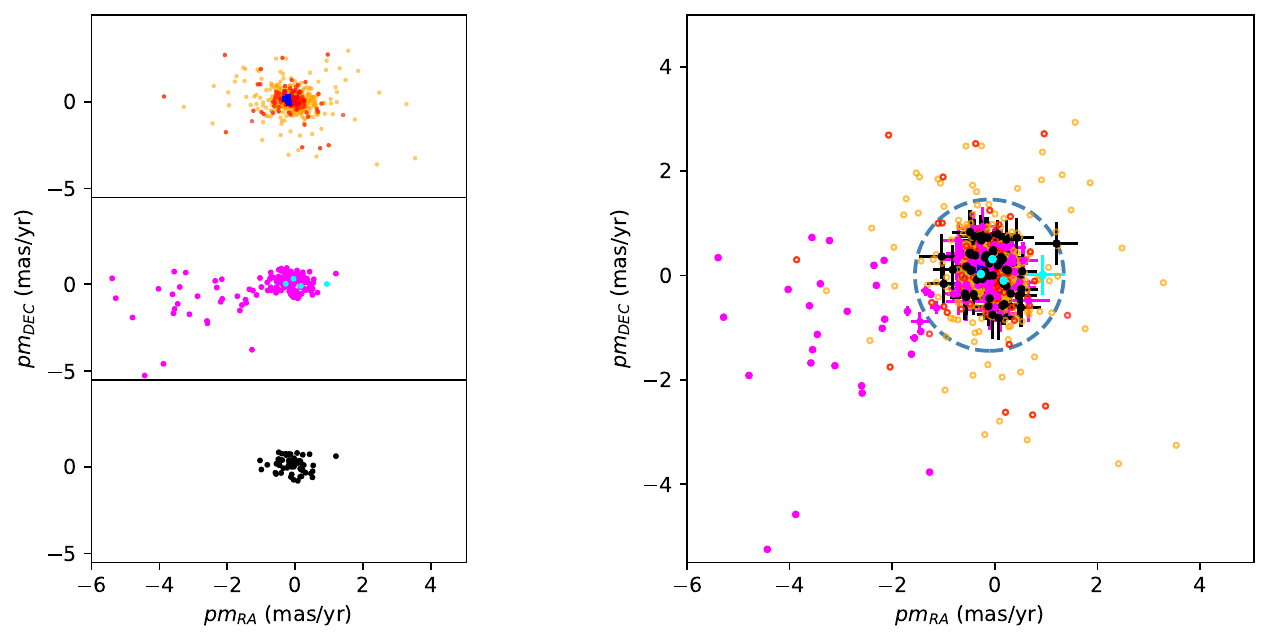}
    \caption{{\it Left panels}: Distribution of {\it Gaia} DR3 proper motion components for UMi RGB member stars (top panels) from 
    \citealt{Kirby-et-al-2010} and \citealt{Pace-et-al-2020}), 
    for the whole sample of variable stars classified as RR Lyrae by {\it Gaia} (middle panels), and compared with UMi's known RR Lyrae stars in the Gold sample (bottom panels). {\it Central panels}: {\it Gaia} DR3 proper motions maps for the three datasets shown in the left panels. The five RGB stars from \citet{Sestito-et-al-2023} are marked by blue open circles in the top-central panel and in the right panel of the figure. {\it Right panel}: Proper motion map for all three datasets together. Proper motions of RR Lyrae stars included in the Gold sample are plotted with their uncertainties. The blue dashed lines in the bottom-left panels and the blue dashed circle in the right panel mark the proper motion distribution range for RR Lyrae in the Gold sample. 
    }
    \label{fig:pmgold_histoemap}
\end{figure*}

\begin{figure*}
\center
\includegraphics[width=\textwidth]{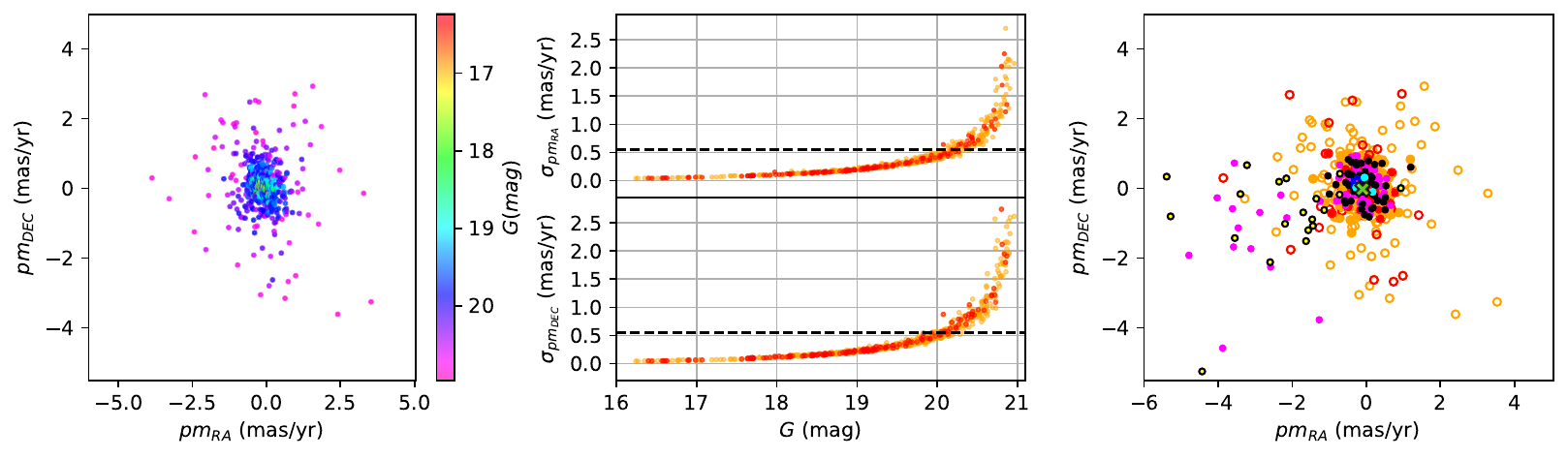}
    \caption{{\it Left panel:} Map of the {\it Gaia} DR3 proper motion components for member stars of UMi spectroscopically confirmed by \citet{Kirby-et-al-2010}, \citet{Pace-et-al-2020} and \citet{Sestito-et-al-2023}. Sources are colour-coded according to their $G$ magnitudes in the  {\it Gaia} DR3 catalogue. {\it Central panels:} Errors of the {\it Gaia} DR3 proper motion components plotted versus $G$ magnitude for the sources in the left panel. Dashed lines show the 0.55 mas/yr threshold that we adopted 
    to distinguish between reliable and less reliable proper motions. {\it Right panel:} {\it Gaia} DR3 proper motions of sources around UMi's centre. Sources are colour-coded as in Figs.~\ref{fig:cmdall} and ~\ref{fig:pmgold_histoemap}. In particular, filled orange and red circles are RGB member stars of UMi with proper motion uncertainties smaller than 0.55 mas/yr. A lime cross marks the mean proper motions of the faint cluster Mu\~noz~1 
    according to 
    \citet{Vasiliev-Baumgardt-2021}.}
    \label{fig:pm_members}
\end{figure*}

 To distinguish RR Lyrae stars (or variables of other types) belonging to UMi from field variables, we started by identifying UMi's known RR Lyrae stars in our initial dataset of 168 objects, based on previously published studies. 
 We adopted, as a reference, the catalogue published by \citet[hereafter N88]{nemec88} that is, so far, the most recent and complete inventory of UMi's variable stars.
 Nemec and collaborators have identified in UMi 82 RR Lyrae stars (47 fundamental mode $-$ RRab, and 35 first overtone $-$ RRc, pulsators), 7 ACs and 5 variables without an accurate period determination (V14, V30, V65, V88 and V94), whereas two additional variables, V74 and V76,  are outside their photographic plates. 
They reclassify and confirm nearly all the 
candidate variables previously identified in UMi by \citet{vanAgt-1967}. However, five of them were not confirmed to vary: V46, V85, V87, V89, and V91. These stars are only mentioned without a discussion and no identifying information (location on charts or coordinates) is provided.
We have followed the nomenclature 
and numbering 
by N88 and counter identified 82 variable stars from N88's catalogue among the 168 
RR Lyrae stars classified in {\it Gaia} DR3,  
using a spatial cross-match radius of 2.5$\arcsec$. 
Then we defined a clean sample of 57 confirmed RR Lyrae stars which have $|P_{Gaia}-P_{N88}|<$0.003 days\footnote{This is the period criterion adopted by \citet{Clementini-2023} to select known RR Lyrae stars and define a reference sample to validate new RR Lyrae stars in {\it Gaia} DR3.} and the same classification in type (RRab or RRc) in both catalogues. These 57 sources form our Gold sample of RR Lyrae stars in UMi.\\
Figure~\ref{fig:bailey} shows the period-amplitude diagram of the 57 RR Lyrae stars in the Gold sample (black filled circles) with respect to the sample of 164 variable stars (magenta filled circles - \texttt{vari\_rrlyrae} table, for which periods and amplitudes are available) in the area that we selected around UMi's centre. The position, hence pulsation properties, of the Gold sample stars correctly reflects their classifications in type: sources with shorter periods and smaller amplitudes are classified as RRc while sources classified as RRab stars are characterized by longer periods and larger amplitudes. 
Periods and classifications published by N88 are based on an average of 20-22 phase points spanning a temporal baseline of more than 20 years.  
{\it Gaia} DR3 data, on the other hand, covers a time range of only 34 months. 
The number of phase points in the $G$ light curves  used by the SOS Cep\&RRL pipeline to determine the period and classification of the variable stars 
included in  the area that we selected around UMi's centre ranges from 16 for V169 to 62 for V178 (also named I later on).  
 Hence, differences between the {\it Gaia} DR3 and the N88 periods of UMi's 
 known variables (see Sections~\ref{sec:ac} and ~\ref{sec:new}), are more 
 likely due to the different number of phase points than to incorrect cross-identification of the sources, although N88 had a larger time baseline they had on average fewer phase points than  {\it Gaia} DR3 to infer the pulsation period.\\
The average number of phase points for the Gold sample RR Lyrae stars is 37, with minimum of 31 and maximum of 43 observations. As described in \citet{Clementini-2023}, 
the search for a secondary periodicity by the SOS Cep\&RRL pipeline was activated only for sources having 40 or more measures in the DR3 $G$-band  time series data and, with residuals from the 
 best-fitting model of the $G$ light curve folded with the primary periodicity larger than 
 0.05 mag.
It is therefore possible that some of the 
RR Lyrae stars classified as c-type may have a second periodicity not searched for 
and actually be double-mode RR Lyrae (RRd stars). The Gold sample contains 19 RRc stars, we searched for a secondary periodicity
6 RRc stars in the Gold sample having \texttt{num\_clean\_epochs\_g} $<$ 40 and period $>0.36$ d, the first overtone period range where RRd stars are more often found. We conclude that V49 and V81 could be candidate RRd stars.\\
Identification and properties of the 57 RR Lyrae stars in the Gold sample are provided in Table~\ref{tab:rrnemec}. 
 The Gold sample includes 38 fundamental mode, 17 first overtone and 2 candidate double-mode RR Lyrae stars.  The average period of the RRab stars is $<{\rm P_{ab}}>$=0.637 $\pm 0.062$ days (average on 38 stars), the average period of the RRc stars is ${\rm <P_{c}>=0.3514 \pm 0.037}$ days (average on 17 stars), or ${\rm<P_{c+d}>=0.3560 \pm 0.038}$ days (average on 19 stars) including the candidate RRd stars.
\\
We have used the Gold sample as reference 
to reclassify the known variables, 
to identify new RR Lyrae members of UMi (see Section~\ref{sec:new}), and to clean 
the galaxy CMD 
(see Section~\ref{sec:final}) along with 
the {\it Gaia} DR3 proper motions and sources belonging to UMi from the literature.
In the Gold sample we have included V30. This star is classified by N88 as an RR Lyrae with uncertain period between two  most plausible values: 0.33 and 0.25 days. We adopt for V30 the longer period that is in perfect agreement with {\it Gaia} DR3 estimation. Among the remaining 6 variable stars without an accurate period determination 
from N88, source V76 is one of the 20 
counter-identified RR Lyrae stars that do not meet the Gold sample criterion. These 20 sources are discussed in Section~\ref{sec:new}. Stars V14, V74, V88 and V94 are not in the \texttt{vari\_rrlyrae} table, hence we do not have pulsation properties from {\it Gaia} but we have  photometric information from the {\it Gaia} general catalogue (see Section~\ref{sec:new} for discussion). Finally, V65 is not in the {\it Gaia}  general catalog either, but the star is included among the Type II Cepheids in the General Catalogue of Variable Stars \citep{Samus-2017}.\\
 The spatial distribution of the 57 RR Lyrae stars in the Gold sample is shown by black dots in the left panel of Figure~\ref{fig:map_cmd_gold}.  
 All the Gold sample RR Lyrae are contained in the area enclosing 3 r{$_h$} of the galaxy and have mean $G$ magnitudes and $G_{BP}-G_{RP}$ colours consistent with them being RR Lyrae members of UMi tracing the galaxy's HB, 
 as shown 
 in the right panel of Fig~\ref{fig:map_cmd_gold}. Three sources, stars V60, V63 and V43 have bluer colours ($G_{BP}-G_{RP}\sim0$ mag) and  $<G_{BP}>$ magnitudes about half magnitude brighter than the other RR Lyrae stars.
 We have re-analysed their 
 time series data confirming the pulsation periods published in DR3 but obtaining  
 new $<G>, <G_{BP}>$ and $<G_{RP}>$ mean magnitudes (see the new pulsation characteristics 
 and light curves for these stars in Appendix~\ref{sec:appendix}) that correctly place 
 the three stars within the instability strip on the HB, as shown in the right panel of Fig~\ref{fig:map_cmd_gold} where the 3 sources are plotted with violet filled circles on the CMD when using their revised magnitudes and colours 
 (see Section~\ref{sec:dist}).\\ 
Proper motions published in {\it Gaia} DR3 can help to identify UMi's members.
The right panel of Fig.~\ref{fig:pmgold_histoemap} shows the map of {\it Gaia} DR3 proper motion components for UMi's 
RGB stars cross-identified from the \citet{Sestito-et-al-2023}, \citet{Pace-et-al-2020}, and \citet{Kirby-et-al-2010} catalogues (blue, orange and red symbols, respectively) along with the proper motions of the whole sample of 168 variables (magenta filled circles) and the proper motions of the RR Lyrae stars in the Gold sample highlighted by black filled circles.
The central panels of the figure show, from top to bottom, the proper motion maps of the three datasets separately.  
 Finally, the left panels of Fig.~\ref{fig:pmgold_histoemap} show the distribution of the two components of proper motions for the corresponding datasets. 
 The proper motion distribution of the RR Lyrae stars in the Gold sample, highlighted by blue dashed lines in the bottom-left plots and a blue dashed circle in the map on the right, seems to track perfectly the proper motions of UMi. In contrast, the distribution is rather broad for the whole sample of variable stars indicating that some of them are probably not members of the dSph. The distribution of RGB stars on the proper motion map as well as on the histograms also extend beyond the RR Lyrae Gold sample. We have tried to better understand the nature of this distribution in order to check whether the proper motions of RGB stars with values further away from the main concentration are reliable 
 or not. In the left panel of Figure~\ref{fig:pm_members} the DR3 proper motions of UMi's RGB stars are colour-coded according to their $G$ magnitude showing that 
 sources with $G> 20$ mag are also the furthest with respect to the main concentration of sources that is centered on the average values of UMi's proper motions ($\mu_{\alpha}=-0.124^{\pm+0.004}_{-0.004}$ $\mu_{\delta}=0.071^{+0.005}_{-0.0005}$ \citealt{Battaglia-et-al-2022}). As shown in the central panels of the same figure, these faint sources also have larger uncertainties on the proper motion components. In particular, uncertainties increase with magnitude, becoming larger at $G \sim 20$ mag, which is approximately 0.5 mag fainter than the average $<G>$ magnitude of UMi's RR Lyrae stars, where the errors are around 0.55 mas/yr for both components. 
 If we select RGB stars with most reliable proper motions, those 
 with errors smaller than 0.55 mas/yr (filled circles), they are all concentrated and their distribution is totally coincident with the distribution of the Gold sample RR Lyrae stars (see right panel of Fig.~\ref{fig:pmgold_histoemap}) 
 that, according to {\it Gaia} DR3 proper motions, are quite well positioned in 
 the region 
 where UMi's member stars 
 are located.\\
In the following, we use the RR Lyrae stars  from the Gold sample 
along with RGB stars that quite solidly define the area of UMi member stars, as a proxy for identifying additional UMi members in the entire sample of 168 variable stars. 


\subsection{Distance determination}\label{sec:dist}
\begin{figure}
	\includegraphics[width=\hsize]{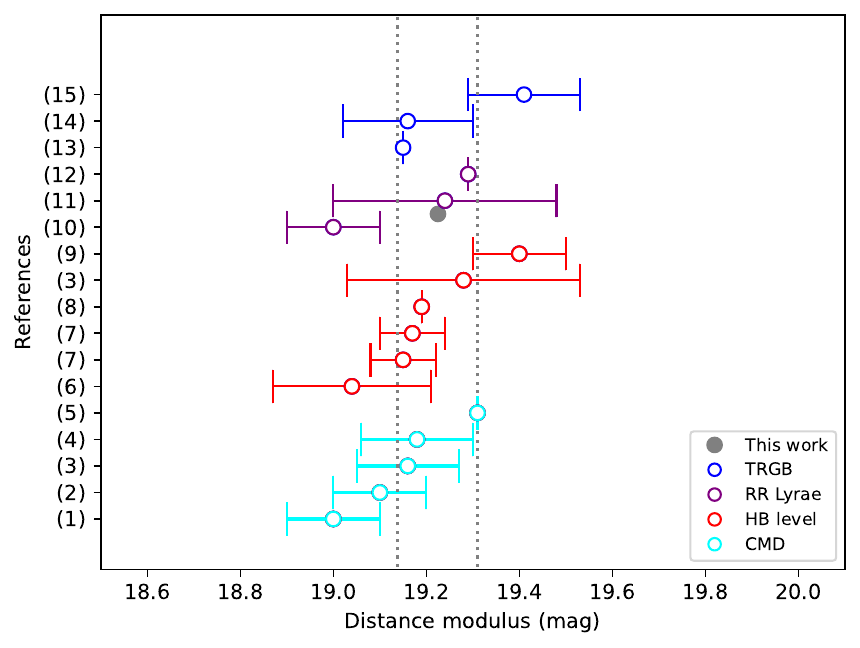}
    \caption{Comparison of UMi distance modulus 
    (grey filled circle), obtained using the Gold sample RR Lyrae stars and the $M_{G}-$[Fe/H] relation,  with literature measurements obtained from different methods and distance indicators: CMD (cyan), HB level (red), RR Lyrae stars (purple) and RGB Tip (blue). The dotted lines show the uncertainty ($\pm$0.09 mag) associated with our distance modulus estimate (see text for details).
    {\bf References:} (1) \citet{Olszewski-1985}; (2) \citet{Kalirai-et-al-2013}; (3) \citet{Dolphin-2002}; (4) \citet{Mighell-1999}; (5) \citet{Weisz-et-al-2014}; (6) \citet{Irwin-1995}; (7) \citet{Ruhland-et-al-2011}; (8) \citet{Webbink1985}; (9) \citet{Carrera-et-al-2002}; (10) \citet{Cudworth-et-al1986}; (11) N88; (12) \citet{Tammann2008}; (13) \citet{Frayn-2003}; (14) \citet{Tully-et-al-2013}; (15) \citet{Bellazzini-et-al-2002}.
   }
    \label{fig:dist}
\end{figure}

\begin{figure}
    \includegraphics[width=\hsize]{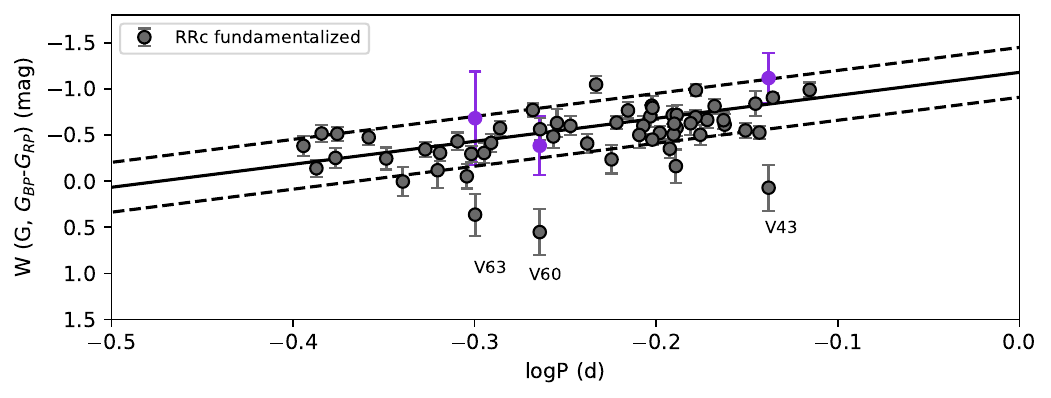}
    \caption{Comparison of the Wesenheit magnitudes of UMi Gold sample RR Lyrae stars 
    and the PW$_{G,G_{BP},G_{RP}}$Z relation of \citet{Garofalo-et-al-2022} (solid line) $\pm3\sigma$ (dashed lines),  fixed to the distance modulus, 19.23 $\pm$ 0.09 
    mag that we derived 
    using the $M_{G}-\rm{[Fe/H]}$ relation. 
Violet-filled circles mark the RR Lyrae stars V43, V60 and V63 plotted with revised mean magnitudes after our analysis of their light curves (see text for details).  }
    \label{fig:pwGOLD}
\end{figure}

As a first step, we measured the distance to UMi 
from the Gold sample RR Lyrae stars adopting as fiducial the luminosity-metallicity relation in the $G$ band, $M_{G}-$[Fe/H], of \citet[eq.19]{Garofalo-et-al-2022} that is based on 291 Galactic RR Lyrae stars with parallaxes published in {\it Gaia} Early DR3 (EDR3).  We adopted as interstellar extinction the value E(B$-$V)=0.03 mag as traditionally used for UMi by many authors \citep{Mateo-1998,Mighell-1999,Bellazzini-et-al-2002,McConnachie-2012}. 
This assumption is a reasonable compromise between 0.028$\pm$0.001 mag and 0.032$\pm$0.001 mag that we derived using \citet{Schlafly-and-Finkbeiner-2011} and \citet{Schlegel-98} maps covering an area of about 3 degrees around UMi centre. 
The observed mean $G$ magnitudes were then dereddened using $A_{G}$=0.078 $\pm 0.001$ mag as derived from the relation $A_{G}/A_{V}$= 0.840 of \citet{Bono-at-al-2019}.\\
Several estimates of UMi average metallicity exist in the literature. We have adopted the recent estimate by 
\citet{Pace-et-al-2020}, who derived a mean metallicity of [Fe/H]=$-$2.13 $\pm$ 0.02 dex (with  $\sigma_{\rm [Fe/H]}$ = 0.35 $\pm 0.01$ dex dispersion) using the largest sample of stars in UMi, 892 members. 
 This value is in excellent agreement with  \citet{Kirby-et-al-2011} mean metallicity of  $<{\rm [Fe/H]}>= -2.13 \pm 0.01$ dex (with 
$\sigma_{\rm[Fe/H]}=0.34$ dispersion) based on medium-resolution spectra of stars in the inner part of the galaxy. Adopting [Fe/H]=$-2.13 \pm 0.02$ dex, we measure for UMi a distance modulus of $19.23 \pm 0.09$ mag (corresponding to d = 70 $\pm$ 3 kpc).
The uncertainty on the distance modulus (hence distance), was obtained from error propagation considering the photometric uncertainties of the $G$ magnitudes, the intrinsic scatter of the adopted $M_{G}-$[Fe/H] relation, errors of 10$^{-4}$ days in the period, the uncertainty of the $A_{G}$ estimation,  and the uncertainty on the metallicity value.
Our distance modulus for UMi is in excellent agreement with the value derived by N88 [$(m-M)_{0}$ = 19.24 $\pm$ 0.24 mag, corresponding to d = 70 $\pm$ 9 kpc]. 
In Figure~\ref{fig:dist} we have compared our distance modulus of UMi with literature 
estimates based on independent distance indicators collected in the NASA/IPAC Extragalactic Database (NED).  
 We find that our result is in good agreement with the majority of the literature estimates. Our distance modulus is $\sim$0.26 mag 
 longer than \citet{Cudworth-et-al1986} and \citet{Olszewski-1985} results. In their paper \citet{Olszewski-1985} noticed that their modulus is shorter than earlier values. On the other hand, \citet{Cudworth-et-al1986} estimate is based on the mean magnitude of a smaller number of RR Lyrae stars 
 ($\sim$ 40) observed in a region of UMi inner than in N88 work. 
The study of N88 has so far been the most complete in terms of identification and characterization of the RR Lyrae population in UMi, therefore we consider the excellent agreement between our distance modulus and the one derived by N88 a solid proof of the validity of our results.\\
As a sanity check, we have plotted 
UMi's Gold sample RR Lyrae stars on the Period-Wesenheit-Metallicity (PWZ) relation in the {\it Gaia} bands (PW$_{G,G_{BP},G_{RP}}$Z), adopting the mean metallicity $<\rm[Fe/H]>=-2.13$ dex, and the coefficient of the PWZ relation in table 9 of \citet{Garofalo-et-al-2022}\footnote{\citet{Garofalo-et-al-2022} PWZ relation is calibrated on the {\it Gaia} DR3 parallaxes of 169 Galactic field RR Lyrae stars distributed all sky, spanning the metallicity range $-$2.84 $<\rm{[Fe/H]}<$0.07 dex and luminosities from 9 to 17 mag in the $G$ band.}. Apparent Wesenheit magnitudes in the {\it Gaia} three bands were obtained from \citet{Ripepi-at-al-2019} relation:
\begin{equation}
	W\left(G,G_{\rm{BP}},G_{\rm{RP}}\right)=G-\lambda\left(G_{\rm{BP}}-G_{\rm{RP}}\right)\,,
	\label{eq:Wesenheit-mag-def}
\end{equation}
where the $G$, $G_{BP}$ and $G_{RP}$ mean magnitudes were taken from the \texttt{vari\_rrlyrae} table and 
the appropriate $\lambda$ value for RR Lyrae stars is $\lambda=$ 1.922 \citep{Garofalo-et-al-2022}. 
We then used 
the distance modulus of UMi 
derived in Sect.~\ref{sec:dist} (19.23 $\pm$ 0.09 mag) to scale 
the apparent Wesenheit magnitudes to absolute magnitudes
\footnote{We did not directly use the {\it Gaia} DR3  parallaxes of UMi's Gold sample RR Lyrae stars to  
obtain the absolute Wesenheit magnitudes because at the 
distance (and faint magnitude) of UMi's RR Lyrae they could be less solid.}.
Figure~\ref{fig:pwGOLD} shows the position of the Gold sample RR Lyrae on the $PWZ$ plane, 
after transforming the 
first overtone pulsators into fundamental mode pulsators
 by adding +0.127 to the logarithm of the periods \citep{Iben-1974}. 
The solid line shows the \citet{Garofalo-et-al-2022} PWZ relation with dashed lines indicating the  $\pm 3\sigma$  deviations. Three RR Lyrae stars, V43, V60 and V63, lie significantly below the PWZ relation. As noticed in Sec.~\ref{sec:gs}, these stars have brighter $<G_{BP}>$ magnitudes and, in turn, fainter Wesenheit magnitudes than the other RR Lyrae in the Gold sample. We re-determined the intensity-averaged magnitudes of these sources from analysis of their {\it Gaia} light curves with  Graphical Analyzer of Time Series (GRATIS), a private software developed at the Bologna Observatory by P. Montegriffo (see, e.g., \citealt{Clementini-et-al-2000}). 
We confirmed the periods and classification published in {\it Gaia} DR3 which 
are also consistent with N88 results (see 
Appendix~\ref{fig:lc_app_1}), found new $G$ magnitudes that do not differ significantly from those published in the \texttt{vari\_rrlyrae} table\footnote{The difference in the distance modulus of UMi using the recalculated $G$ magnitudes would be 0.001 mag. This difference is well within the uncertainty of our distance modulus.}, while obtaining  revised Wesenheit magnitudes that place the three stars within $\pm3\sigma$ from the PWZ relation 
(violet-filled symbols in Fig.~\ref{fig:pwGOLD}).

\begin{table*}
 \caption{
 Identification and properties of the 57 RR Lyrae stars in the Gold sample. 
 Sources are ordered by increasing ID number, adopting the nomenclature from N88.
 Column (1): {\it Gaia} DR3 sourceid,  column (2): N88 sourceid, 
 columns (3) and (4): {\it Gaia} DR3 coordinates$^{*}$ (ra and dec). 
 Pulsation period (column 5), pulsation mode (column 6), intensity-averaged $G$ magnitude (column 7), $G$-band peak-to-peak amplitude 
 (column 8) and number of epoch in the $G$-band (column 9) are from the {\it Gaia} DR3 \texttt{vari\_rrlyrae} table. 
 Photometric [Fe/H] values (column 10) are from \cite[see Section~\ref{sec:final}]{Muraveva-et-al-2024}. 
 }
 \label{tab:rrnemec}
 \small
 \begin{tabular}{clccccccccc}
  \hline
  {\it Gaia} DR3 source\_id & name & ra& dec& P&Type& $G$ &AmpG&
    $\# epochG$&$\rm[Fe/H]$\\
  &  & (deg) &(deg)& (days) &  &(mag)& (mag)&&(dex)\\
  \hline
1645480968432583168 & V2 & 228.1678 & 67.5216& 0.648 & ab&19.595 & 0.684&37&$-$2.22\\
1645474504506279040 & V4 & 227.8815 & 67.4263& 0.716 & ab&19.663 & 0.498&38&$-$1.75\\
1645486637789393024 & V5 & 227.4816 & 67.4752& 0.767 & ab&19.568 & 0.382&40&$-$\\ 
1645465983291163264 & V8 & 227.2937 & 67.4517 & 0.554 &ab&19.613 & 1.074&39&$-$2.07\\
1645485156025159552&V10& 227.5857 & 67.4051 & 0.618 &ab&19.620 & 0.927&37&$-$1.88\\
1645472619016137600&V15& 227.8719 & 67.3682 &0.645 & ab&19.599 & 0.968&41&$-$2.04\\
1645472584656396672&V17&227.8310&67.3619&0.719&ab&19.623 & 0.676&41&$-$2.11\\
1645469831581809664&V18&228.1098&67.3186&0.600&ab&19.584 & 1.172&36&$-$1.83\\
1645449872869325568&V19&227.5394&67.3577&0.342&c&19.685 & 0.500&39&$-$2.01\\
1645461550884867712&V20&227.4499&67.3582&0.628&ab&19.663 & 1.006&41&$-$1.71\\
1645461482165377408&V22&227.4583&67.3342&0.306&c&19.827 & 0.523&37&$-$1.73\\
1645449593695899264&V23&227.5805&67.3236&0.663&ab&19.463 & 0.561&38&$-$\\
1645449456256933376&V24&227.5419&67.3107&0.301&c&19.795 & 0.519&39&$-$\\
1645460623171913344&V25&227.2222&67.3389&0.647&ab&19.615 & 0.627&38&$-$\\
1645460558747945216&V26&227.2602&67.3217&0.628&ab&19.672 & 0.863&38&$-$1.61\\
1645448494184240384&V28&227.6738&67.2782&0.308&c&19.657 & 0.581&39&$-$2.13\\
1645448562903707776&V30&227.5958&67.2691&0.334&c&19.695 & 0.497&39&$-$1.78\\
1645449078299793920&V31&227.4973&67.2879&0.358&c&19.753 & 0.447&40&$-$\\
1645459523660256256&V32&227.3192&67.2939&0.627&ab&19.758 & 0.672&38&$-$1.69\\
1645463268871746816&V33&227.0436&67.3164&0.621&ab&19.624 & 0.914&33&$-$1.57\\
1645463410606222976&V34&226.9517&67.3148&0.642&ab&19.675 & 0.688&37&$-$\\
1645460180790815232&V35&227.0330&67.2835&0.659&ab&19.691 & 0.630&31&$-$1.83\\
1645454202196333568&V36&226.9859&67.2503&0.707&ab&19.579 & 0.792&34&$-$2.22\\
1645448150586830208&V37&227.5258&67.2482&0.679&ab&19.580 & 0.891&39&$-$2.06\\
1645459420581026304&V38&227.3385&67.2725&0.357&c&19.793 & 0.415&39&$-$\\
1645447429032321024&V39&227.4277&67.2474&0.578&ab&19.630 & 0.848&40&$-$1.78\\
1645447502047359488&V40&227.3263&67.2362&0.327&c&19.767 & 0.450&36&$-$\\
1645448081867341312&V41&227.4650&67.2319&0.490&ab&19.775 & 0.986&35&$-$\\
1645444925066984704&V42&227.5951&67.2095&0.647&ab&19.631 & 0.739&43&$-$\\
1645446226441431424&V43&227.4935&67.1874&0.728&ab&19.561 & 0.759&39&$-$2.11\\
1645446127657821312&V44&227.4849&67.1677&0.382&c&19.658 & 0.393&42&$-$\\
1645447188514116096&V45&227.3569&67.2038&0.507&ab&19.827 & 1.122&34&$-1.30$\\
1645458836465430272&V48&227.1821&67.2185&0.687&ab&19.612 & 0.686&36&$-$2.04\\
1645458772041537408&V49&227.1874&67.2122&0.415&c$^{**}$&19.661 & 0.364&34&$-$\\
1645453167108584448&V50&227.0648&67.2089&0.688&ab&19.646 & 0.771&31&$-$2.00\\
1645441970129471488&V51&227.5822&67.0873&0.634&ab&19.634 & 0.680&38&$-$2.36\\
1645445543542253952&V52&227.3502&67.1010&0.597&ab&19.576 & 1.077&36&$-$1.96\\
1645439732450810752&V53&227.2849&67.1086&0.731&ab&19.490 & 0.687&36&$-$2.44\\
1645439874185418368&V54&227.2062&67.1013&0.673&ab&19.592 & 0.746&35&$-$2.17\\
1645452823511172224&V55&227.1370&67.1719&0.663&ab&19.635 & 0.859&35&$-$2.08\\
1645440969401422080&V58&227.1144&67.1566&0.351&c&19.682& 0.479&36&$-$\\
1645452514273508992&V60&226.9606&67.1575&0.544&ab&19.670 & 0.915&31&$-$\\
1645454648872907008&V63&226.6047&67.1479&0.374&c&19.749 & 0.374&36&$-$\\
1645452411194275200&V64&226.9480&67.1312&0.644&ab& 19.661 & 0.880&35&$-$\\
1645435574922418688&V66&227.4034&67.0250&0.626&ab&19.602 & 0.640&41&$-$2.26\\
1645438804737835264&V68&227.0013&67.0612&0.422&c&19.572 & 0.514&37&$-$\\
1645437778242088832&V69&227.0621&67.0271&0.314&c&19.749 & 0.601&37&$-$\\
1645451346042353920&V70&226.7000&67.0959&0.585&ab&19.682 & 0.769&33&$-$2.21\\
1645437842665115648&V73&226.8859&66.9624&0.403&c&19.413 & 0.591&42&$-$\\
1645356856761796096&V75&226.4976&67.0122&0.497&ab&19.728 & 1.079&37&$-$1.56\\
1645457947407902464&V77&226.7213&67.2819&0.667&ab&19.621 & 0.975&35&$-$2.05\\
1645462444238074880&V78&227.2943&67.3813&0.372&c&19.652 & 0.489&36&$-$\\
1645442485525547136&V81&227.4783&67.1125&0.386&c$^{**}$&19.689 & 0.462&39&$-$\\
1645452312410710144&V83&227.0560&67.1518&0.406&c&19.640 & 0.419&34&$-$\\
1645456190766318336&V99&226.6550&67.2373&0.609&ab&19.614 & 0.773&34&$-$1.66\\
1693505918347240192&V100&227.1773&67.5580&0.314&c&19.708 & 0.478&35&$-$\\
1645490309985954304&V101&227.5333&67.5726&0.646&ab&19.637 & 0.867&41&$-$2.10\\
  \hline
 \end{tabular}
   \tablefoot{$^{*}$ Right ascension and declination are at J2016.0, the reference epoch (\texttt{ref\_epoch}) for both {\it Gaia} EDR3 and the full {\it Gaia} DR3.
   \\
   $^{**}$ RRd candidate (see text for details).}
\end{table*}
\section{Anomalous Cepheids in UMi}\label{sec:ac}
\begin{figure}
	\includegraphics[width=\hsize]{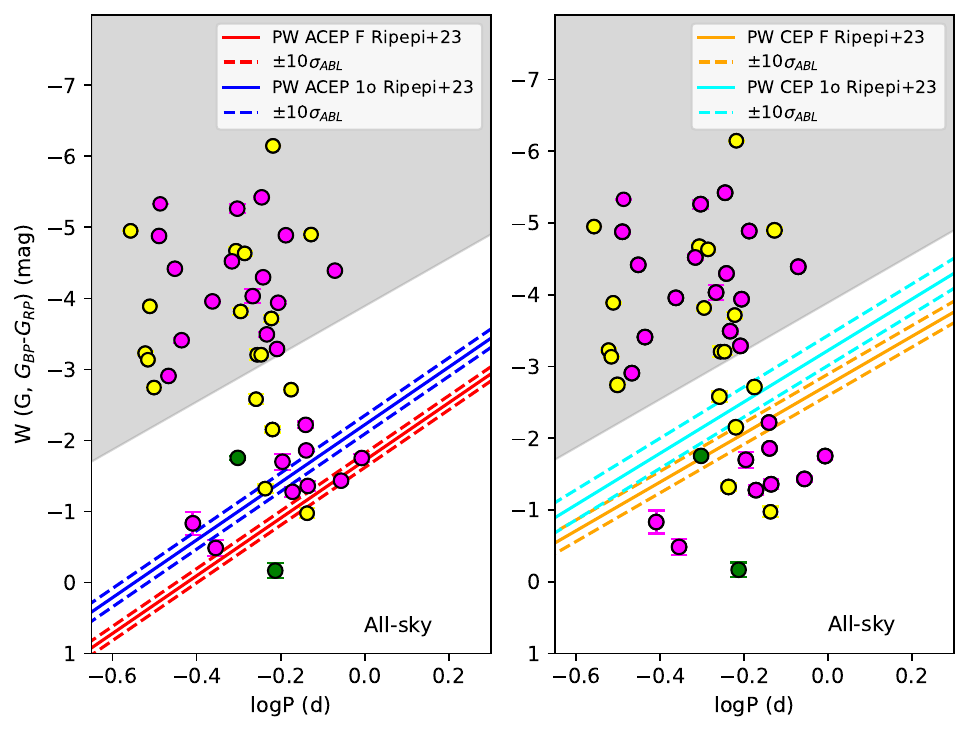}
    \includegraphics[width=\hsize]{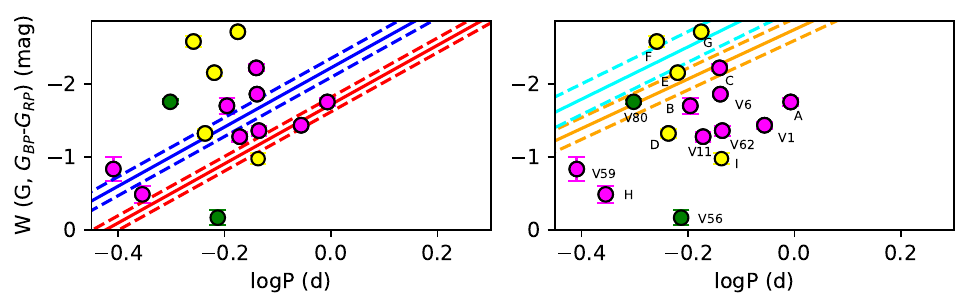}
    \caption{{\it Upper panels}: PW$_{G,G_{BP},G_{RP}}$  diagrams for 43 
    variable stars that are considerably brighter than UMi's HB ($G<$19.25 mag).
     Magenta and yellow filled circles mark sources within and beyond 12r$_{h}$ from the galaxy centre.
    Solid lines show the PW$_{G,G_{BP},G_{RP}}$ relations for MW ACs (ACEP; left panels) and CCs (CEP; right panels) from \citet{Ripepi-2023}, with their related uncertainties (dashed lines). 
    PW$_{G,G_{BP},G_{RP}}$ relations for 
    fundamental mode pulsators (lower lines) and 
    first-overtone pulsators (upper lines) are displayed in different colours. 
    {\it Lower panels}: Zoomed-in view (between 0.0 and $-$2.7 in W) of the PW$_{G,G_{BP},G_{RP}}$ diagrams in the upper panels. 
Green filled circles mark stars V56 and V80, classified as ACs by N88, but not confirmed as RR Lyrae stars or Cepheids in {\it Gaia} DR3.  
Pulsation periods and intensity-averaged magnitudes from {\it Gaia} data are not available for these two stars (see discussion in the text).}
    \label{fig:pw}
\end{figure}

\begin{figure}
    \centering
    \includegraphics[width=\columnwidth]{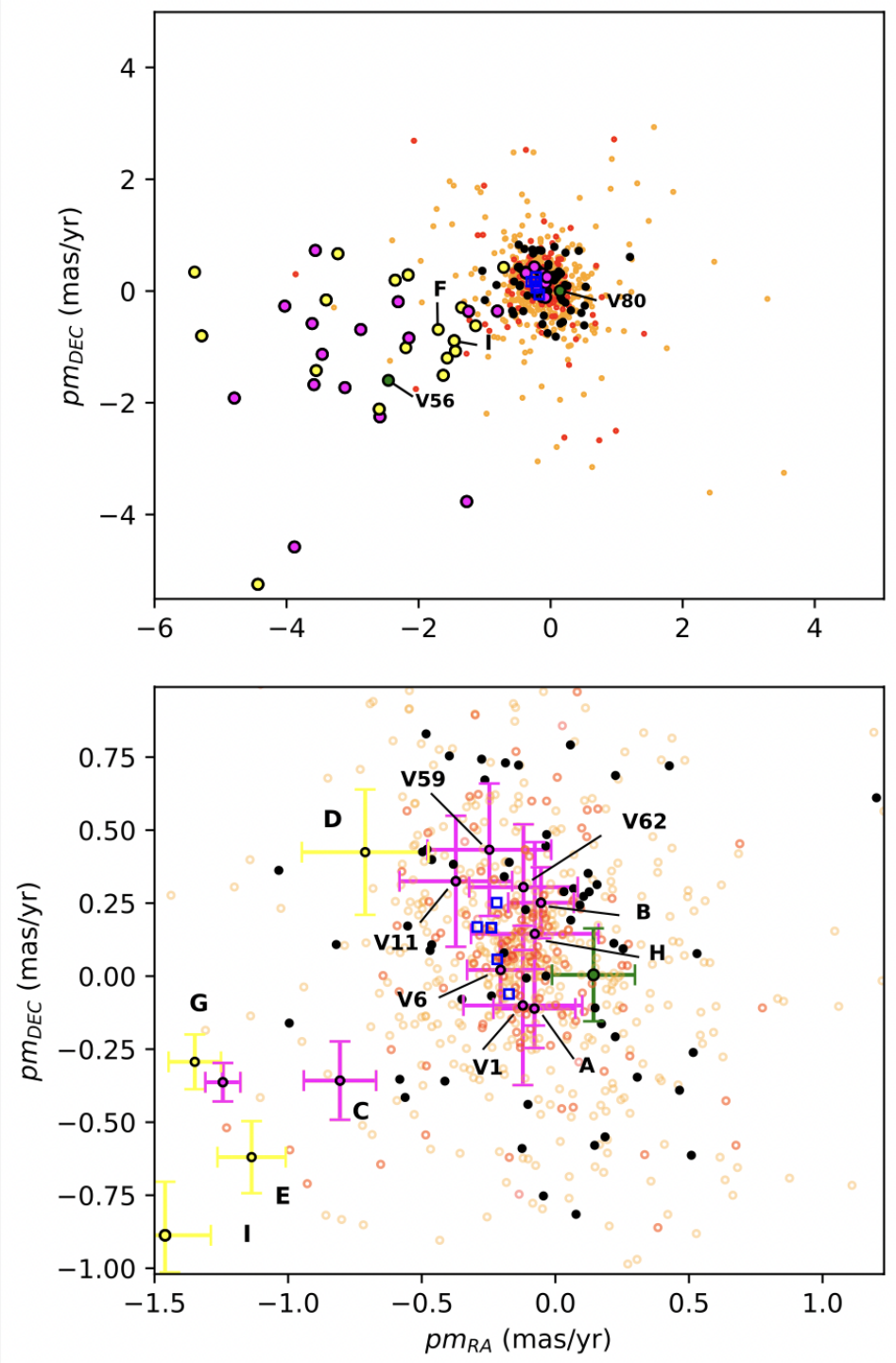}
    \caption{{\it Upper panel:} Proper motion diagram of UMi's Gold RR Lyrae stars (black filled circles) and 43 
    variable stars significantly brighter than UMi's HB, that are placed within (magenta filled circles) and beyond (yellow filled circles) 
    12 r$_{h}$ from the centre of the galaxy.
{\it Lower panel:} Zoom-in of the upper panel. The labeled sources are discussed in Section~\ref{sec:ac}. Green filled circles mark V56 and V80 for which we do not confirm N88's classification as ACs.}
    \label{fig:pm_acs}
\end{figure}

\begin{figure*}
    \centering
    \includegraphics[width=9.7cm]{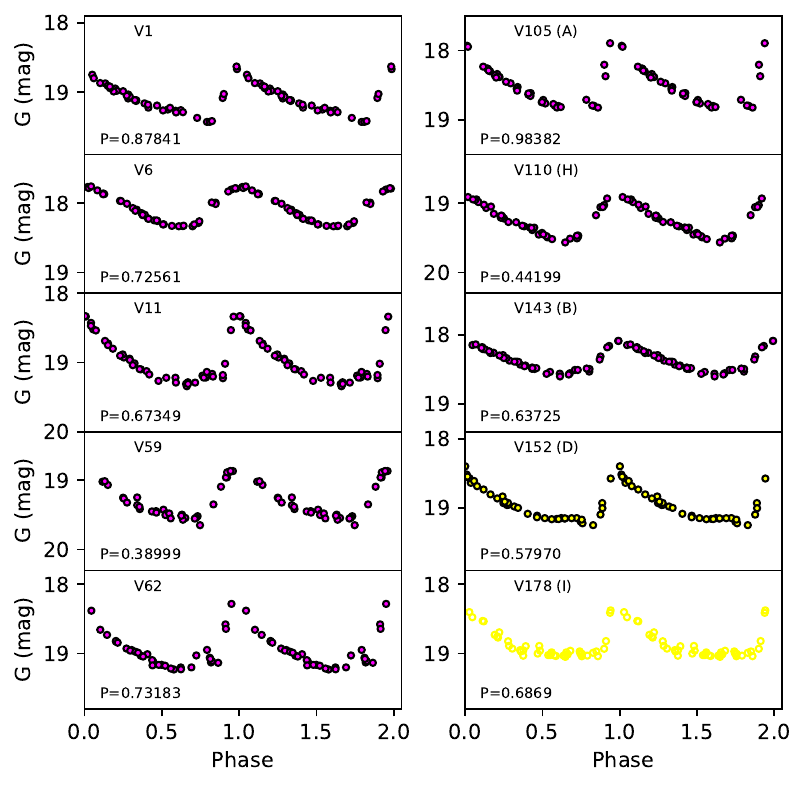}
    ~\includegraphics[width=5.1cm]{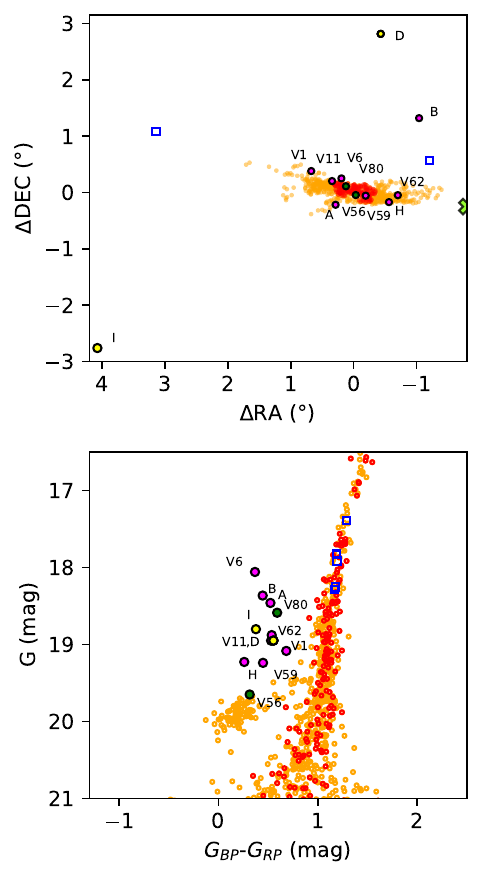}
    \caption{{\it Left and central panels:} {\it Gaia} DR3 $G$-band light curves of ACs in UMi. Five of them were already known from N88 that we have confirmed. Additional 4 new ACs plus 1 AC candidate have been identified in this work. 
    Variable stars are ordered by increasing number according to N88's and our identifications. Magenta- and yellow-filled phase points mark stars within and beyond 12 times the galaxy half-light radius, respectively. Pulsation periods adopted to fold the light curves are from {\it Gaia} DR3 \texttt{vari\_rrlyrae} table, except for (I) V178 whose light curve is folded with the period derived in our analysis of the time series data (see details in the text). {\it Right panels:} Spatial distribution (top) and position on the CMD (bottom) of the 10 ACs (9 confirmed plus 
    1 candidate) with respect to UMi's members. Green filled circles mark V80 and V56, for which 
    we do not confirm N88's classification as ACs.}
    \label{fig:lc_acs}
\end{figure*}

 In UMi, there are three ACs identified by 
 \citep{vanAgt-1967,Kholopov-1971} and later confirmed  by N88: V6, V56 and V59. N88 added 4 new ACs (V1, V11, V62, V80) bringing to 7 the number of ACs identified in UMi.
Three further variable stars identified by N88, V19, V73 and V79, have mean magnitudes 0.20-0.27 mag brighter than UMi RR Lyrae and were classified by these authors as probable RR Lyrae stars evolved off the Zero Age HB \citep[ZAHB; see for example][]{Lee-1990}. These three sources are classified by {\it Gaia} as RR Lyrae stars and included in the DR3 \texttt{vari\_rrlyrae} table. 
In particular, the $G$ mean magnitudes of V19 (see Table~\ref{tab:rrnemec}) and V79 (see Section~\ref{sec:new} and Figs.~\ref{fig:cmdeco_note} and~\ref{fig:pw_rr_note}) are perfectly consistent with the $<G>$ magnitude defined by UMi's Gold sample RR Lyrae stars. On the other hand,  V73 is the brightest RR Lyrae in the Gold sample with 
 $G$ mean magnitude 0.24 mag brighter than the other stars, supporting N88's hypothesis that the star is an evolved RR Lyrae. We note that the 
 distance modulus of UMi obtained from the Gold sample RR Lyrae stars is almost unchanged if V73 is removed from the sample. 
In the selected area covering 3.5 degrees around the center of UMi, out of a total sample of 168 variable stars classified as RR Lyrae stars in  {\it Gaia} DR3, at least 43 sources 
have $G$ mean magnitude $\sim 5\sigma$ brighter than the HB level of UMi ($<G>=19.653 \pm 0.081$ mag, according to the mean $G$ magnitude of the Gold sample RR Lyrae stars).
These 43 variable stars with $G \leq$ 19.25 mag could either be RR Lyrae belonging to the MW halo, or ACs belonging to UMi  
not correctly classified in {\it Gaia} DR3. We 
checked whether variable stars classified as Classical  Cepheids (CCs) or ACs and published in the {\it Gaia} DR3 \texttt{vari\_cepheid} table \citep{Ripepi-2023} are included in the area that we selected around the centre of the galaxy, but found none. 
We then tried to discriminate the nature of the 43 bright 
sources from the comparison with the $PW_{(G,G_{\rm BP},G_{\rm RP})}$ relations for ACs and CCs derived by \citet{Ripepi-2023} and 
adopting the distance modulus of UMi obtained from the Gold sample RR Lyrae stars (Section~\ref{sec:dist}).
	\label{eq:Wesenheit-mag-def}
In Fig.~\ref{fig:pw} we show the position of the 43 bright variable stars on the PW$_{G,G_{BP},G_{RP}}$ plane. Solid lines in the upper-left panels mark the relations from \citet{Ripepi-2023} for first-overtone and fundamental mode ACs (ACEP 1O and ACEP F, respectively) in the MW with their uncertainties (dashed lines), and in the upper-right panels the corresponding relations for Galactic CCs (CEP 1O and CEP F, respectively).  
The lower panels of Fig.~\ref{fig:pw} show the zoomed-in view (between 0.0 and $-$2.7 in W$_{G,G_{BP},G_{RP}}$) of the PW relations in the upper panels. Magenta- and yellow-filled circles mark sources 
within and beyond 12r$_{h}$, respectively.
A grey-shaded area in the upper panels highlights the 29 brightest sources among the 43 variable stars significantly above the HB. 
With $W(G, G_{BP}-G_{RP})$ values brighter than $-2.5$ mag, these variable stars in the grey-shaded area appear to be much more luminous than the Wesenheit index of ACs and CCs at the distance of UMi, hence it is unlikely that they 
belong to the galaxy.
 This conclusion is further 
 supported by their positions on the proper motion diagram in Fig.\ref{fig:pm_acs}: they are distributed all over the area going from the nearest star (\texttt{source\_id} 1643960889310752128 with \texttt{pm\_ra}$=-$1.460  and \texttt{pm\_dec}$=-$0.887 mas/yr) to the furthest star, outside the diagram (\texttt{source\_id} 1646045808171028352 with \texttt{pm\_ra}$=-$7.866 \texttt{pm\_dec}$=-$2.051 mas/yr).\\
 The 14 bright variable sources below the grey-shaded area in large majority lie close to, or,  
 in the best cases, well follow within the errors  the PW$_{G,G_{BP},G_{RP}}$ relations for ACs of \citet{Ripepi-2023},  
 as shown in the lower-left panel of Fig.~\ref{fig:pw}.
 We classify most of them as confirmed or candidate ACs belonging to UMi, a classification also confirmed from 
 their position on the proper motion diagram in Figure~\ref{fig:pm_acs}, and the location on the spatial distribution map and the CMD shown in the 
 right panels of Figure~\ref{fig:lc_acs}.
We comment on confirmed or candidate ACs individually here below. The first 5 sources are variable stars already studied by N88. We assigned to the remaining nine sources not in N88's list identifiers in the form of alphabetic capital letters and V-numbers continuing N88 nomenclature of increasing number with increasing the distance from UMi centre.
\begin{itemize}
    \item[$-$] {\bf V1}: N88 classified this star as AC with P=0.470 days. The source is listed as RRab star with period P=0.878 days in the {\it Gaia} DR3 \texttt{vari\_rrlyrae} table.
    This coincides with an alias period found by N88 for the star. The $G$ band light curve of V1 (see Figure~\ref{fig:lc_acs}), is well sampled and nicely folded by the period published in the \texttt{vari\_rrlyrae} table, that we believe is the correct period for the star.
    V1 is plotted in Figure~\ref{fig:pw} according to the period published in the \texttt{vari\_rrlyrae} table and found to fall on the PW$_{G,G_{BP},G_{RP}}$ for fundamental mode ACs. We therefore confirm 
    the classification of V1 as ACEP F. 
    \item[$-$]{\bf V6}: the source is classified as  RRab star with P=0.726 days in the {\it Gaia} DR3 
    \texttt{vari\_rrlyrae} table.
    N88 found the same period for V6, but
     classified the star AC because it is too bright to be an RR Lyrae belonging to UMi. The mean $G$ magnitude of V6 is almost 1.6 mag brighter than the mean luminosity of UMi RR Lyrae stars in the Gold sample. The star falls on the PW$_{G,G_{BP},G_{RP}}$ for first overtone ACs.
     Hence, also for V6 we confirm the classification as AC (ACEP 1O). 
    \item[$-$]{\bf V11}: N88 classified this star as AC, whereas the star is included in the \texttt{vari\_rrlyrae} table of {\it Gaia} DR3 with a pulsation period that differs by 0.02 days from that published in N88. Based on the position close to the PW$_{G,G_{BP},G_{RP}}$ for fundamental mode ACs we confirm the classification as AC of V11.
    \item[$-$]{\bf V59}: the source is classified RRab star in the DR3 \texttt{vari\_rrlyrae} table with  period in perfect agreement with that found by N88  who however classify the star AC.
    Based on the star location close to the PW$_{G,G_{BP},G_{RP}}$ for first overtone ACs we confirm N88 classification of V59 as AC of UMi. 
    \item[$-$]{\bf V62}: As for the 4 variable stars discussed above, V62 is included in the \texttt{vari\_rrlyrae} table of {\it Gaia} DR3 with period differing by only 0.03 days from N88's one, who, however, classified the source as AC. 
    V62 lies close to the PW$_{G,G_{BP},G_{RP}}$ for fundamental mode ACs hence we confirm its classification as AC  (ACEP F). 
    \item[$-$]{\bf (A) V105}: {\it Gaia} $G,G_{BP}$ and $G_{RP}$ light curves for the star (Fig.~\ref{fig:lc_acs}) are well sampled and nicely folded by the period published in the DR3 \texttt{vari\_rrlyrae} table. 
    The star position right on the PW$_{G,G_{BP},G_{RP}}$ for fundamental mode ACs (bottom-left panel of Fig.~\ref{fig:pw}), the proper motions (bottom panel of Fig.~\ref{fig:pm_acs}), as well as the position on the spatial distribution map and CMD (
    right panels of Figure~\ref{fig:lc_acs}), allow us to conclude that A could be a new AC belonging to UMi.
    \item[$-$]{\bf (H) V110}: As for A, all properties of H 
    lead us to classify the star as a new AC of UMi.
    \item[$-$]{\bf (B) V143}: this source is located within 11 r$_{h}$ from UMi centre ($\delta_{RA}=-1.04^{\circ}$, $\delta_{DEC}=1.31^{\circ}$). Similarly to A and H, based on the star position near the PW$_{G,G_{BP},G_{RP}}$ relation for first overtone ACs and the star location on the proper motion diagram, also B could be a new AC member of UMi.
    \item[$-$]{\bf (D) V152:} 
    this source is classified as RR Lyrae by {\it Gaia} DR3 and other surveys, (Pan-STARRS, PS1; \citealt{Sesar-et-al-2017} and  Zwicky Transient Facility, ZTF; \citealt{Bellm-et-al-2019}), that also find for the star the same period published in the {\it Gaia} DR3 \texttt{vari\_rrlyrae} table. Although the star is located well beyond 12 r$_{h}$ from UMi's centre ($\delta_{RA}=-0.43^{\circ}$, $\delta_{DEC}=2.81^{\circ}$), the proper motions and the position near the PW$_{G,G_{BP},G_{RP}}$ relation for fundamental mode ACs leave the possibility open 
    that this source may be an AC of UMi, one of the furthest we have identified.
    \item[$-$]{\bf (C) V162:} As D, star C is  classified RR Lyrae in {\it Gaia} DR3 as well as also in the PS1 and ZTF surveys. However, 
    the luminosity more than 1 mag brighter than UMi's  HB rules out that it could be an RR Lyrae belonging to the galaxy. 
    The star is also too bright to be an ACs being 
    $\sim$0.5 mag above 3$\sigma$ from the PW$_{G,G_{BP},G_{RP}}$ relation for first overtone ACs in UMi, 
    while it could be a fundamental mode CC in the galaxy, but this seems a rather unlikely possibility, given the very old and metal poor population of UMi.
    Star C 
    is placed within 12 elliptical half-light radii from the galaxy centre, but its 
    proper motions are barely consistent with those of the other RR Lyrae stars in the galaxy. We think  
    that classification as a field RR Lyrae is  the most plausible for this star.
    \item[$-$]{\bf (E) V174, (F) V166 and (G) V175:} These stars are among the 20 variable sources found beyond  12 times the UMi's r$_{h}$ 
    (Fig.~\ref{fig:map1}). They are significantly brighter (from 1.4 mag for E to over 2 mag for G) than the mean $G$ magnitude of the Gold sample RR Lyrae stars tracing the galaxy HB, as shown in Fig.~\ref{fig:cmdall}.
    Their positions on the PW$_{G,G_{BP},G_{RP}}$ relations exclude them from being ACs of UMi. Their proper motions also show poor consistency with the proper motions of UMi's Gold sample RR Lyrae and other member stars 
    (Fig.~\ref{fig:pm_acs}). E and F are also classified as field RR Lyrae stars, with the same pulsation period as published in the DR3 \texttt{vari\_rrlyrae} table, by ZTF, PS1 and the CATALINA \citep{Drake-et-al-2014} surveys. 
    We thus conclude that these three variable stars do not belong to UMi.
    \item[$-$]{\bf I V178:} Like E, F and G this source is located beyond 12 
    r$_{h}$ and its proper motions are poorly 
    consistent with those of the 
    Gold sample RR Lyrae stars. The source light curves, folded according to the period in the DR3  \texttt{vari\_rrlyrae} table, P=0.7292 days, are rather noisy in all bands. We re-analyzed the star time series data finding a slightly shorter period (P= 0.6869 days, logP=$-$1.631) and a revised Wesenheit magnitude (W$_{G,G_{BP},G_{RP}}$=$-$1.13 mag) that 
    place the source closer to the PW$_{G,G_{BP},G_{RP}}$ relations of UMi's ACs.
    Therefore, we do not rule out that star I could be an AC candidate in UMi.
    
    
\end{itemize}

Among variables classified by N88 as ACs, two stars, V56 and V80, are contained in the {\it Gaia} DR3 main catalogue but not in the tables of Cepheids or RR Lyrae stars. We 
attempted 
to clarify the nature of these variable sources adopting the periods in N88 (their table VII) 
and 
the mean magnitudes and astrometric information available in the DR3 main catalogue.
The upper-right panel of Fig.~\ref{fig:lc_acs} 
shows that V56 and V80 are quite close to the centre of UMi.  
V80 proper motion components (upper panel of Fig.~\ref{fig:pm_acs}) are also consistent with those of the Gold sample RR Lyrae stars and the bulk of UMi's members.
On the other hand, the proper motions of V56 have rather large errors ($\sigma_{pm_{RA}}$=0.946 mas/yr and $\sigma_{pm_{DEC}}$=1.044 mas/yr) and the high RUWE\footnote{RUWE for V56 is 3.48. if {\it Gaia} Renormalized Unit Weight Error (RUWE) significantly greater than 1.4 could indicate that the source is non-single or otherwise problematic for the astrometric solution. A description of this parameter can be found in section 14.1.2 of the {\it Gaia} Data Release 2 Documentation release 1.2 \url{https://gea.esac.esa.int/archive/documentation/GDR2/}} 
of the 
astrometric solution for this star 
signals that they are unreliable.  
Thus, we cannot use them to confirm or rule out 
 that V56 is a member of UMi.
 The location of V80 and V56  on the  PW$_{G,G_{BP},G_{RP}}$ relations in Fig.~\ref{fig:pw}, adopting the periods of N88, shows that they are not ACs belonging to UMi, 
 but $G$ magnitude and colour of V56 
 (see bottom-right panel of Fig.~\ref{fig:lc_acs})
 are consistent with the star being an RR Lyrae of the galaxy (see discussion in Sect.~\ref{sec:new}).
\citet{Kholopov-1971} claims V80 to be a binary star, but N88 ruled out this hypothesis based on  their data.
The star is included in the \texttt{vari\_classifier} table of {\it Gaia} DR3 and classified as variable AGN. 
We have analyzed the {\it Gaia} 
light curves of V80 finding 
luminosity variations smaller than 0.2 mag in all the three bands and 
a light variation in the $G$ band  compatible with a very long period (P$\gtrsim$ 900 days).
On the basis of the above evidences we have not included V56 and V80 in the list of UMi's ACs.
\\
To summarize, 
we finally classify as new ACs belonging to UMi
stars: A, H, B and D, while classify star I 
AC candidate. Adding these sources 
to the five already known ACs that we have confirmed (V1, V6, V11, V59, V62) brings the total number of UMi  ACs in our study to 9+1. Main characteristics for these 10 ACs are summarized in Table~\ref{tab:acs_umi}.
Their $G$-band light curves are shown in the left and cetral panels of Figure \ref{fig:lc_acs}. 
\begin{figure}
    \centering
    \includegraphics[width=\hsize]{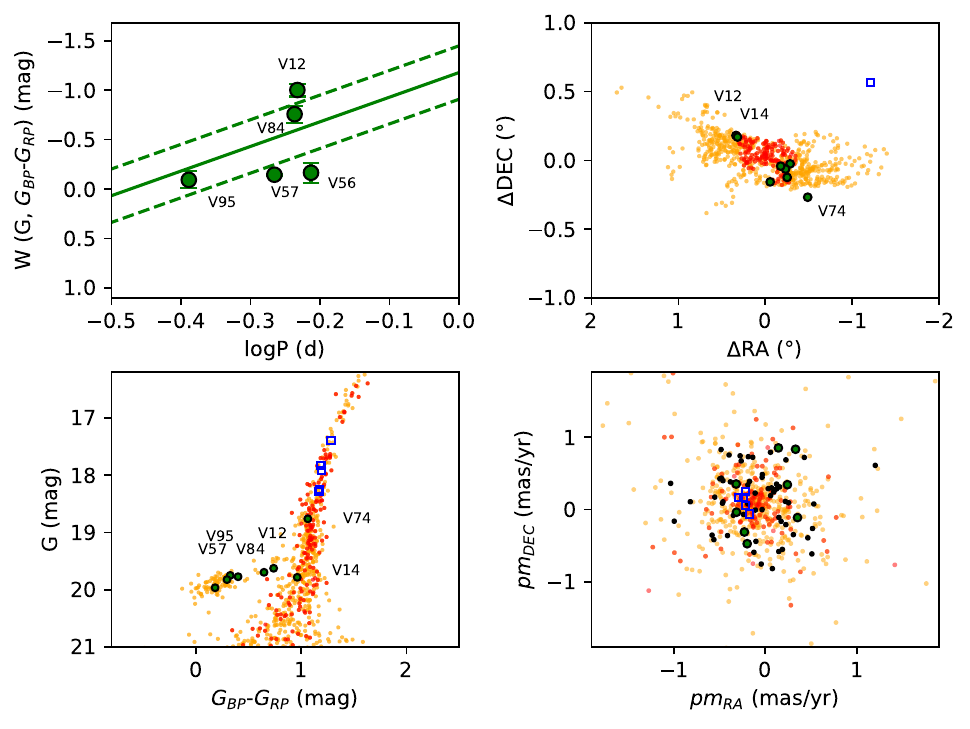}
\caption{{\it Top-left panel:} Position of V12, V57, V84 and V95 on the {\it Gaia} DR3 PW$_{G,G_{BP},G_{RP}}$ relation for RR Lyrae stars belonging to UMi's, after {\it fundamentalizing} these RRc stars by adding 0.127 to the logarithm of the first-overtone periods from N88. Also shown is V56 for which we suggest a classification as RRab star; {\it Top-right panel:} Spatial distribution of V12, V57, V84, V95, V14, V88, V94 and V74 with respect to UMi centre; {\it Bottom-left panel:} Position on the CMD; {\it Bottom-right panel:} Position on the proper motion diagram.} 
    \label{fig:extra_acs}
\end{figure}

\begin{table*}
 \caption{ACs belonging to UMi, confirmed or identified in this study.}
 \label{tab:acs_umi}
  \scriptsize
 \begin{tabular}{lcccccccccc}
  \hline
    Name$^*$ &  P&Type&source\_id& $<G_{BP}>$ &$<G>$  &$<G_{RP}>$& P& Type &P& Type  \\
   & N88 & N88 &{\it Gaia} DR3 & &&&  
   \texttt{vari\_rrlyrae} table & 
   \texttt{vari\_rrlyrae} table &this work& this work \\
  &  (days) &  &&(mag) &(mag)& (mag) & (days)&& (days)& \\
  \hline
  V1&0.470&AC&1645494025133820160&19.346&19.083&18.664&0.87841&RRab&&AC\\
  V6&0.725586&AC&1645486431630962176&18.167&18.058&17.800&0.7256&RRab&&AC\\
  V11&0.675&AC&1645485293464120320&19.122&18.950&18.593&0.67349&RRab&&AC\\
  V59&0.389981&AC&1645452720431944832&19.320&19.238&18.872&0.38998&RRab&&AC\\
  V62&0.729&AC&1645454816375975168&19.038&18.878&18.502&0.73183&RRab&&AC\\
  V105~(A)&-&-&1645394171437675520&18.627&18.460&18.105&0.98382&RRab&&AC\\
  V110~(H)&-&-&1645450285186128384&19.256&19.227&18.998&0.44199&RRab&&AC\\
  V143~(B)&-&-&1693632156025631872&18.437&18.364&17.993&0.63725&RRab&&AC\\
  V152~(D)&-&-&1694334086825050880&19.120&18.949&18.568&0.57969&RRab&&AC\\
  V178~(I)&-&-&1643960889310752128&19.041&18.801&18.673&0.72923&RRab&0.68690& candidate AC\\
  \hline
 \end{tabular}
 \tablefoot{$^{*}$Identifiers of the first five sources are from N88
 . We assigned identifiers to the other 5 variable stars in the table following the numbering scheme of N88.}
\end{table*}
\section{RR Lyrae stars in UMi: new members and known members with revised properties}\label{sec:new}
N88 did not confirm the variability of six stars  
classified as variable 
by \citet{Kholopov-1971} and \citet{vanAgt-1967}: V46, V65, V85, V87, V89 and V91. We do not have coordinates allowing us to identify these sources in the {\it Gaia} DR3 main catalogue. Furthermore, 
V86 V90 and V96, that are classified RRc stars by N88, are not present in the {\it Gaia} DR3 main catalogue. Thus, we cannot make further investigations of these 9 sources.
We list in Table~\ref{tab:rr-nosos} another 10  variable stars identified in UMi by N88 that are neither included in the {\it Gaia} DR3 \texttt{vari\_rrlyrae} table nor listed as RR Lyrae stars (`RR') in the \texttt{vari\_classifier\_result} table. 
Astrometry and mean magnitudes are available for these sources  
in the {\it Gaia} DR3 general catalogue but no 
 pulsation characteristics. 
Among them are: V56 and V80, for which we do not confirm
N88's classification as ACs  (see Sect.~\ref{sec:ac}); 
 4 RRc stars according to N88: V12, V57, V84 and V95; and, other 4 sources classified as variable stars by \citet{Kholopov-1971} and \citet{vanAgt-1967}, but not confirmed to vary (V14, V88 and V94), or to be RR Lyrae stars (V74)  by N88 \footnote{N88 find a clear peak at P$\sim$
0.218 days in the periodogram of V74, but notice that the star is too  red to be an RR Lyrae.}.
The spatial distribution in the top-right panel of Fig.~\ref{fig:extra_acs},  
and the distribution in proper motions (bottom-right panel) show that eight of these sources are unequivocally members of UMi. By adopting periods from N88  
we show that a classification as RRc stars belonging to UMi for V12, V57, V84 and V95 is plausible using the {\it Gaia} magnitudes, as they well-fit the RR Lyrae PW$_{G,G_{BP},G_{RP}}$ relation fixed at UMi's distance (top-left panel of Fig.~\ref{fig:extra_acs}) and lie on the HB of UMi CMD (bottom-left panel). 
V57 is classified eclipsing binary (ECL)  
in the \texttt{vari\_classifier\_result} table. We have analyzed its 
DR3 time series data confirming the classification as RRc star but deriving a shorter period (P=0.28932 days) than found by N88 (P=0.405 days, Table~\ref{tab:rr-nosos}). The star is plotted using our revised period in Fig.~\ref{fig:pw_rr_note}.
N88 derive two, equally probable periods for V95 (P=0.305/0.439 days; see Table~\ref{tab:rr-nosos}), both of which 
disagree with \citet{Kholopov-1971} period for this star (P=0.234 days).
V95 is plotted in the upper-left panel of 
Fig.~\ref{fig:lc_20di68}  
according to N88's shorter period, since it best places the star on the PW$_{G,G_{BP},G_{RP}}$ relationship. 
As discussed in Section~\ref{sec:ac}, classification as RRab star of V56, adopting the {\it Gaia} DR3 magnitudes and the period in N88 (0.612 days), seems to be more likely, since the star 
better fits the PW$_{G,G_{BP},G_{RP}}$ relation of UMi RR Lyrae stars (upper-left panel of Fig.~\ref{fig:extra_acs}) than the PW$_{G,G_{BP},G_{RP}}$ for UMi Cepheids (bottom panels of Fig.~\ref{fig:pw}). 
In addition, according to the position on the CMD (bottom-left panel of Fig.~\ref{fig:extra_acs}), V74 and V14 cannot be RR Lyrae stars, 
confirming N88 results.
Conversely, V88 and V94 being located on the HB could be RR Lyrae stars. But  
since we do not have period information from the literature or {\it Gaia} time series data for these stars we only classify them as 
potential RR Lyrae.\\


\begin{table*}
 \caption{
 Known 
 variable stars in UMi 
 that are neither included in the {\it Gaia} DR3 \texttt{vari\_rrlyrae} table nor classified as `RR' in \texttt{vari\_classifier\_result} table. {\it Gaia} DR3 time series data are available only for V80 and V57, which 
 are classified as variable `AGN' and `ECL' in the \texttt{vari\_classifier\_result} table.
 }
 \label{tab:rr-nosos}
 \begin{tabular}{lcccccccc}
  \hline
      Name &source\_id&ra  & dec &$G$& P&Type& P& Type 
    \\
          ~N88  &{\it Gaia} DR3 & &  && N88&N88& this work & this work \\
  &  & (deg) &(deg)& (mag) & (days)&& (days)&\\
  \hline
~V12 & 1645473512369330304 &227.6349 & 67.3956&19.625 $\pm$ 0.006&0.437&RRc\\
~V14  & 1645473404994625920&227.6175 & 67.3842& 19.782 $\pm$ 0.004&{-}&{-}\\
~V56${^*}$ &1645446741837484288&227.2629 & 67.1691&19.650 $\pm$ 0.019&0.612&AC&&
RRab\\
~V74 &1645343937500159232 &226.8097& 66.9467& 18.761$\pm$  0.002&{0.218}&{-}\\
~V84 & 1645453029669612672&227.0118& 67.1893&19.695 $\pm$ 0.006&0.433&RRc\\
~V88 &1645440350926081152&227.0460& 67.0908&19.961 $\pm$ 0.005&{-}&{-}\\
~V94 & 1645439491932606592& 227.2416& 67.0579&19.771 $\pm$  0.006&{-}&{-}\\
~V95 & 1645452308115077248 &227.0640& 67.1515&19.749 $\pm$ 0.005&0.305/0.439&RRc\\
    \hline
~V57 & 1645452827806790656& 227.1206 &  67.1727& 19.826 $\pm$ 0.006&0.405& RRc&0.28932& RRc\\ 
~V80${^*}$ & 1645461074144025088&227.4192&67.3246& 18.587 $\pm$ 0.003&0.498746&AC&$\geq$ 900& $-$\\
  \hline
 \end{tabular}
 \tablefoot{V14, V88 and V94 are not variable stars and V74 is not an RR Lyrae for N88. 
 ${^*}$V56 and V80 were discussed in the Sect.~\ref{sec:ac}, not confirming the classification as ACs by N88.}
\end{table*}
\begin{table*}
 \caption{Twenty RR Lyrae stars identified in UMi by N88 that are included in the {\it Gaia} DR3 \texttt{vari\_rrlyrae} table or are classified as `RR' in the DR3 \texttt{vari\_classifier\_result} table.}
 \label{tab:rr-psbag}
 \begin{tabular}{lccccccc}
  \hline
      Name &source\_id& P&Type &P&Type& P&Type\\
          ~N88  &{\it Gaia} DR3 & N88&N88 &
          \texttt{vari\_rrlyrae} table&\texttt{vari\_rrlyrae} table& this work&this work\\
           & & (days)& &(days) && (days)&\\         
  \hline
     ~V3 &1645474367067330048&0.383&c&0.27660&c&\\
     ~V7 &1645486289896542336&0.697&ab&0.69019&ab&\\
     ~V9$^*$ & 1645485430903072896& 0.455& c& $-$& $-$& 0.4555600&d?$^{**}$ \\
     ~V13 &1645462001856973312&0.652&ab&0.64615&ab&\\
     ~V16 &1645469625423395712&0.291&c&0.68903&ab&\\
     ~V21 &1645461349021933696&0.679&ab&0.67487&ab&\\
     ~V27 &1645460760610869376&0.395&c&0.66343&ab&\\
     ~V29 &1645448356750244096&0.292&c&0.40963&c&\\   
     ~ V47$^*$ & 1645446849212324992& 0.615219& ab& $-$& $-$&0.6152161& ab\\
     ~V61 &1645452067596906368&0.442&c&0.30569&c&\\
     ~V67 &1645437739590764416&0.714&ab&0.70698&ab&\\
     ~V71$^{***}$ &1645451071164429824&0.659&ab&0.65512&ab&\\
     ~V72 &1645434887727621248&0.2035&c&0.59755&ab&0.59746&ab\\
     ~V76 &1645354657738523008&$-$&$-$&0.63718&ab&\\
     ~V79 &1645461142863500928&0.458&c&0.31306&c&\\     
     ~V82 &1645442171992256640&0.547&ab&0.54217&ab&\\     
     ~V92 &1645360361455119744&0.518&ab&0.34110&c&\\
     ~V93 &1645340913846863232&0.676&ab&0.68137&ab&\\
     ~V97 &1645441038120910592&0.430093&c&0.75627&ab&\\     
     ~V98 &1645440213487130624&0.259527&c&0.35055&c&\\
  \hline
 \end{tabular}
\tablefoot{$^*$V9 and V47 are classified as `RR' in the \texttt{vari\_classifier\_result} table only. Pulsation periods and other characteristics (amplitudes, mean magnitude, etc.) for these stars are from our analysis of their {\it Gaia} DR3 time series data.\\
$^{**}$ According to the long first-overtone period and the rather noisy light curve, V9 could be a double mode RR Lyrae star.\\ 
$^{***}$The split light curve near the maximum light of V71 makes us suspect that it is a Blazhko RR Lyrae.}
\end{table*}
\subsection{Known RR Lyrae members of UMi with revised properties}\label{sec:know20}
\begin{figure*}
\center
\includegraphics[width=11.7cm]{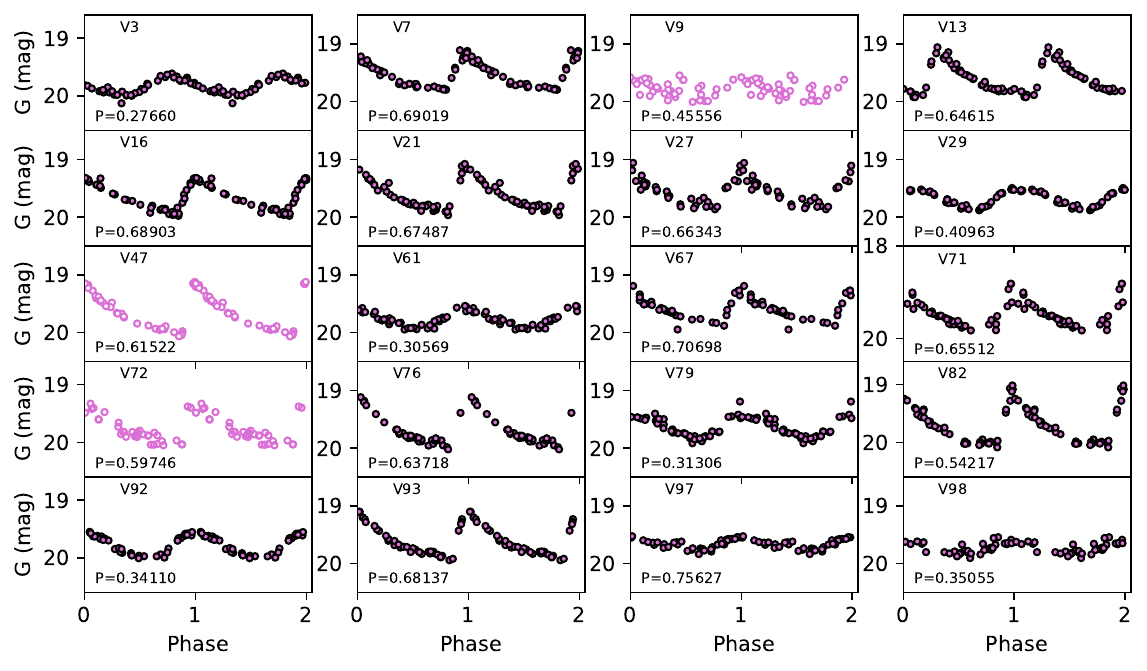}
    \caption{$G$-band light curves for twenty RR Lyrae stars identified in UMi by N88, that are not included in the Gold sample.  
   Sources are ordered according to  identifiers of N88 increasing from left to right and phased according to the periods listed in columns 5 (light curves with magenta-filled circles) and 7 (light curves with orchid open circles;  V9, V47 and V72) of Table~\ref{tab:rr-psbag}, respectively   
   (see discussion in Sect.~\ref{sec:new}).
 }
    \label{fig:lc_20di68}
\end{figure*}

\begin{figure}
 	\includegraphics[width=\hsize]{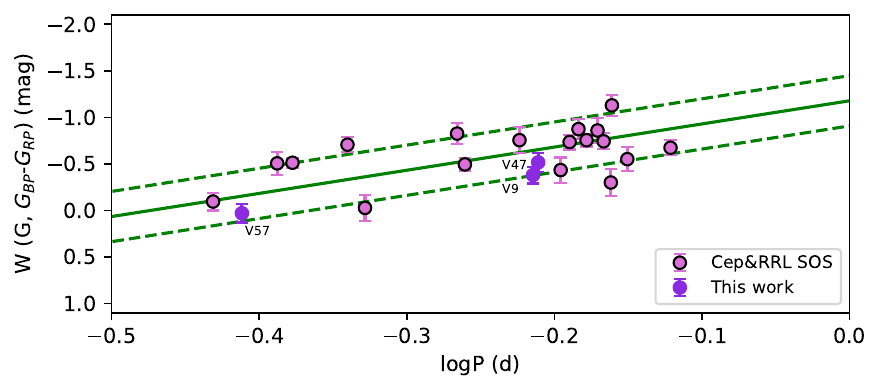}
    \caption{Position of the 20 sources in Table~\ref{tab:rr-psbag} on the PW$_{G,G_{BP},G_{RP}}$ for RR Lyrae stars of \citet{Garofalo-et-al-2022} (green solid line, along with its $\pm$ 3$\sigma$ deviation shown by dashed lines) scaled to the distance modulus of UMi [(m-M)$_{0}$=19.23 mag;  Sect.~\ref{sec:dist}]. Periods and intensity-averaged $G$, $G_{BP}$ and $G_{RP}$ magnitudes are those provided in the \texttt{vari\_rrlyrae} table (orchid-filled circles), except for V9 V47 and V57 (violet-filled circles; see Sects.~\ref{sec:new} and ~\ref{sec:know20}). RRc stars have been {\it fundamentalized}.  
    }
    \label{fig:pw_rr_note}
    \end{figure}

\begin{figure}
 	\includegraphics[width=\hsize]{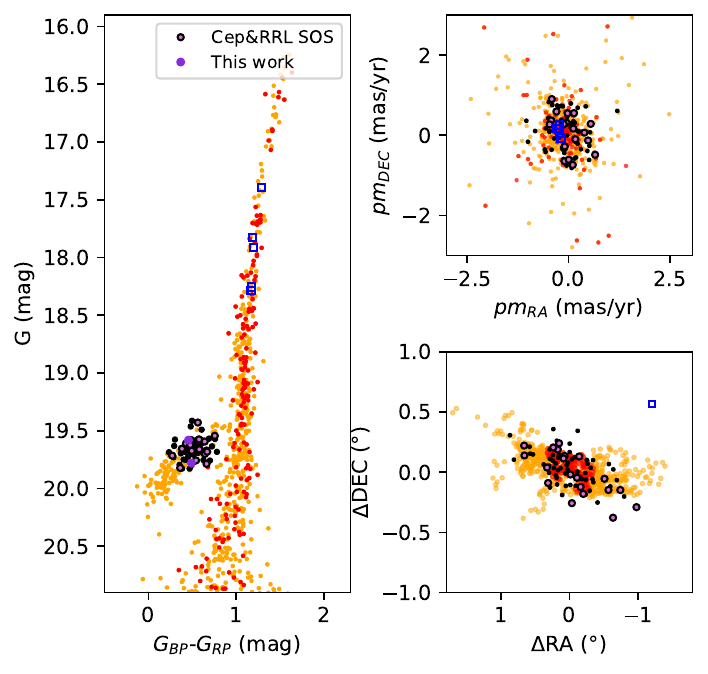}
    \caption{
    {\it Left panel:} Position of the 20 known RR Lyrae stars listed in Table~\ref{tab:rr-psbag} on the  $G$ {\it vs} $G_{BP}-G_{RP}$ CMD of 
     UMi 's RGB members (cyan, orange and red symbols)  and Gold sample RR Lyrae stars
     (black-filled circles).
    Mean magnitudes and colours of the 20 RR Lyrae stars are from the DR3 \texttt{vari\_rrlyrae} table (orchid-filled circles) and from our analysis of the {\it Gaia} time-series data for V9 and V47 (violet-filled circles);  
    {\it Right panels:} Spatial location (bottom) and proper motion  distribution (top)  of the 20 RR Lyrae stars (orchid-filled circles) using the {\it Gaia} DR3 astrometry.
    }
    \label{fig:cmdeco_note}
\end{figure}

As described in Section~\ref{sec:gs}, 82 
of the 168 variable sources identified within a radius of 3.5 degrees from UMi centre 
 (Fig.~\ref{fig:map1})
have a counterpart in the catalogue of N88. 
Excluding the 57 RR Lyrae in the Gold sample and the 43 bright sources discussed in Sect.~\ref{sec:ac} (that include 
5 ACs from N88 incorrectly classified RR Lyrae stars in the DR3 \texttt{vari\_rrlyrae} table), there are another  
20 
RR Lyrae stars identified in UMi by N88 to discuss, as well as 
48 variable sources  
classified as RR Lyrae stars 
in {\it Gaia} DR3 that are new potential members of UMi.
Starting from the 20 sources already classified as RR Lyrae stars of UMi by N88 we find that 18 of them are also classified 
as RR Lyrae in the DR3 \texttt{vari\_rrlyrae} table, but with slightly different pulsation periods for 17 of them:
$|P_{Gaia}-P_{N88}$|>0.003 days, while other 2 sources (V9 and V47) are classified `RR' in the \texttt{vari\_classifier\_result} table.
Identifier, period and type of these 20 RR Lyrae stars from both N88 and the \texttt{vari\_rrlyrae} table, or from our analysis of the DR3 light curves, are provided in Table~\ref{tab:rr-psbag}.
The $G$-band light curves of these sources folded according to the {\it Gaia} DR3 periods available in the \texttt{vari\_rrlyrae} table for 17 of them, and the periods derived from our analysis of the {\it Gaia} times series data with the GRATIS software for V9, V47 and V72 are presented in Fig.~\ref{fig:lc_20di68}. 
This figure shows that periods derived from the {\it Gaia} DR3 data 
are 
accurate and phase very well the observations, allowing us to confirm that these 
stars are RR Lyrae as they all have periods shorter than 1.0 day and the characteristic shape of the RR Lyrae light curves in the optical band. 
N88 did not publish a period for V76, but the period and type published in the \texttt{vari\_rrlyrae} table  (see columns 5 and 6 of Table~\ref{tab:rr-psbag}), 
as well as the star mean magnitudes confirm that V76 is an RR Lyrae of UMi.
For comparison, light curves folded with periods from N88
are presented in Fig.~\ref{fig:lc_app_4}. The {\it Gaia} DR3 data of 17 of the RR Lyrae stars listed in Table~\ref{tab:rr-psbag} are not well folded by N88 periods, and for five of them (V16, V27, V72, V92 and V97) we do not even confirm the classification in type of N88. Conversely, we confirm N88 period and classification of V47, and find for V9 a rather long first overtone period, P=0.455560 days, very much close to P=0.455 days from N88.  However, the light curves of V9 folded with the first overtone period are too noisy suggesting that the star could be a double mode RR Lyrae (RRd). 
The G$_{BP}$ and  G$_{RP}$ light curves of V47 and V9 folded with the periods computed in this work are shown in Fig.~\ref{fig:lc_app_2}.\\
Having verified that the periods derived from the {\it Gaia} DR3 data of  these 20 RR Lyrae stars are well constrained, we can use them along with the intensity-averaged $G$, $G_{BP}$ and $G_{RP}$ magnitudes published in the \texttt{vari\_rrlyrae} table 
to plot the sources on the PW$_{G,G_{BP},G_{RP}}$ relation for RR Lyrae stars of \citet{Garofalo-et-al-2022} scaled to the distance modulus of UMi [(m-M)$_{0}$=19.23 mag; see Sect.~\ref{sec:dist}], shown by the green solid 
line in Fig.~\ref{fig:pw_rr_note} along with its $\pm$3$\sigma$ deviation. 
We have {\it fundamentalized} the RRc stars by adding 0.127 to the logarithm of their periods.
V9 and V47 
are marked by violet-filled circles, and plotted adopting the periods and mean magnitudes derived in our analysis.  
N88 suggest V72 to be  
a candidate double-mode RR Lyrae (RRd) 
even though they found a rather short first overtone period (P=0.2035 days) for the star.
 V72 is classified as RRab star in the DR3 \texttt{vari\_rrlyrae} table with a much longer period (P=0.59755 days; see Table~\ref{tab:rr-psbag}). The SOS Cep\& RRL pipeline did not search V72 for a second period because 
 the star has only 39 phase points and the search for a second periodicity is activated 
only for stars with 40 or more epoch data \citep{Clementini-2023}.
We searched for a second periodicity the {\it Gaia} DR3 time-series data of V72, but only found a period 0.00009 days shorter (P=0.59746 days), widely confirming the period and classification as RRab star published in the DR3 \texttt{vari\_rrlyrae} table. However, 
the $G$-band light curve of V72 
is rather noisy (see Fig.~\ref{fig:lc_20di68}) with residuals from the best-fit model around 0.1 mag. 
The scatter could be caused by   Blazhko effect, but this hypothesis is difficult to confirm with the limited number of data available for the star.\\
The position of the 20 sources on the PW$_{G,G_{BP},G_{RP}}$ plane (Fig.~\ref{fig:pw_rr_note}) as well as their location on the HB in the CMD 
(left panel of Fig.~\ref{fig:cmdeco_note}) 
indicates that they all are RR Lyrae members of the UMi dSph.
This is further confirmed by the  location of these 20 RR Lyrae stars within the area enclosed by three times the galaxy half-light radius
(bottom-right panel of Fig.~\ref{fig:cmdeco_note}) 
and by their proper motion components
being consistent with those of the RGB members of UMi 
and the RR Lyrae stars in the Gold sample (top-right panel of Fig.~\ref{fig:cmdeco_note}).\\
\begin{figure*}
\center
\includegraphics[width=8.7cm]{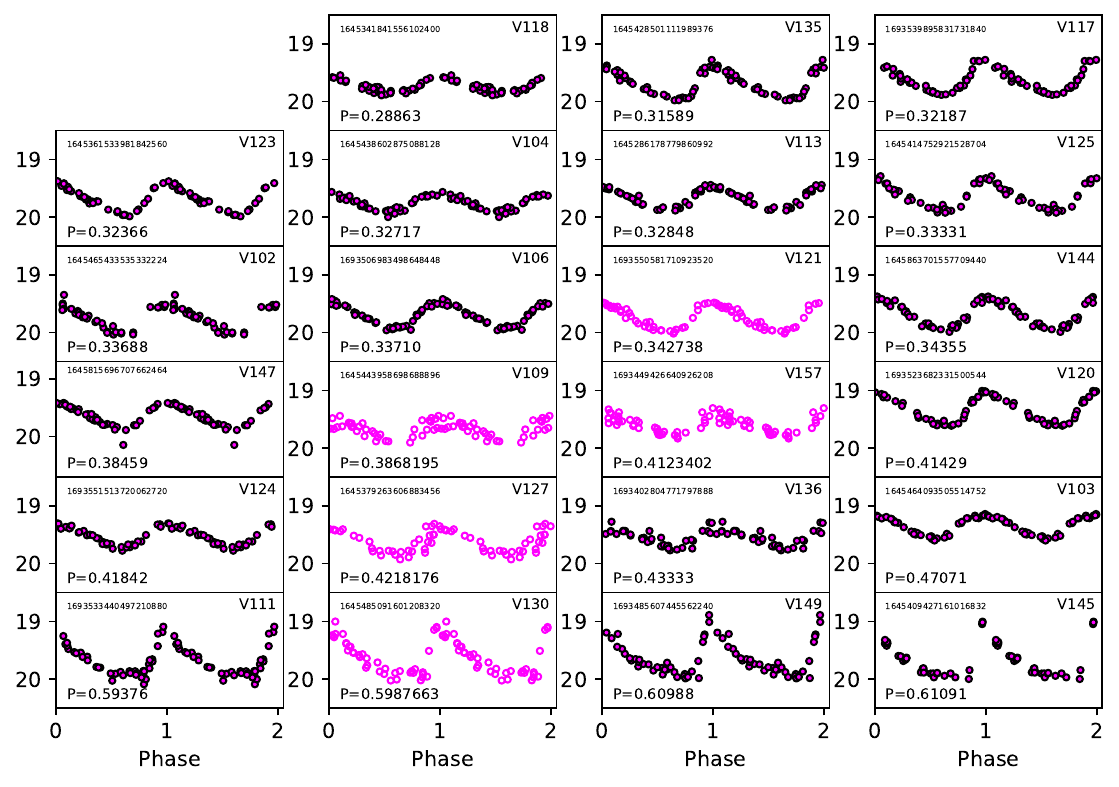}
~\includegraphics[width=8.9cm]{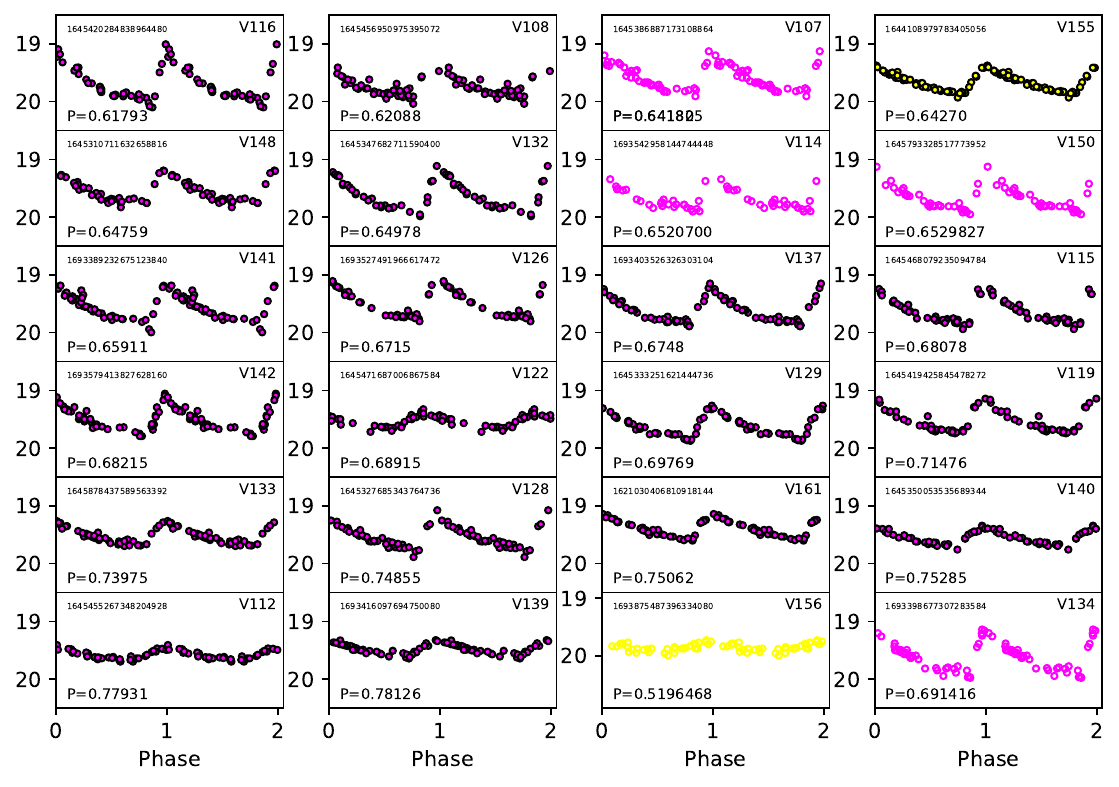}
    \caption{$G$-band light curves of the 47 
    RR Lyrae stars listed in Table~\ref{tab:rr_nuove}, all of them belong to UMi. 
    Magenta-filled circles are used to display light curves folded according to the periods provided in the {\it Gaia} DR3 \texttt{vari\_rrlyrae} table, magenta open circles are used instead for the light curves 
    of sources whose periods were re-determined from our analysis of the time series data.  V155 and V156 are located beyond 12 r$_{h}$, their light curves are displayed with yellow-filled and yellow open circles, respectively.
    }
    \label{fig:lc_48di68}
\end{figure*}
\subsection{New RR Lyrae members of UMi}\label{sec:new48}
In this section we discuss 48 variable sources classified as 
RR Lyrae stars in {\it Gaia} DR3 that are 
located      
 within a 3.5 degree radius from the UMi centre and are new potential members of the galaxy.
We start by 
rejecting 
V169 
as an RR Lyrae member of UMi, because (i) its proper motions (\texttt{pm\_ra}$=-9.793 \pm 1.375$, \texttt{pm\_dec}$=-2.619 \pm 1.365$ mas yr$^{-1}$) are not compatible with UMi dSph and (ii) its  
 mean $G$ magnitude in both the  \texttt{vari\_rrlyrae} table and the main catalogue of {\it Gaia} DR3 is $\sim$ 1.2 mag fainter compared to the typical mean magnitude of RR Lyrae stars belonging to this dSph and, together with the rather red colour ($G_{BP} - G_{RP}$), 
 places V169 in the faintest and reddest  part of the CMD (Fig.~
 \ref{fig:cmdall}). 
Identification and properties of these 47 stars are provided in Table~\ref{tab:rr_nuove}.  
V156 
and V134 
are not included in the \texttt{vari\_rrlyrae} table but are listed as 
`RR' in the \texttt{vari\_classifier} table. We analyzed their time series data with the GRATIS package obtaining the periods and intensity-averaged $G$ magnitudes that we provide in Table~\ref{tab:rr_nuove}, and the intensity-averaged G$_{BP}$, G$_{RP}$ magnitudes that we label 
in Fig.~\ref{fig:lc_app_5}.

\begin{figure}
    \includegraphics[width=\hsize]{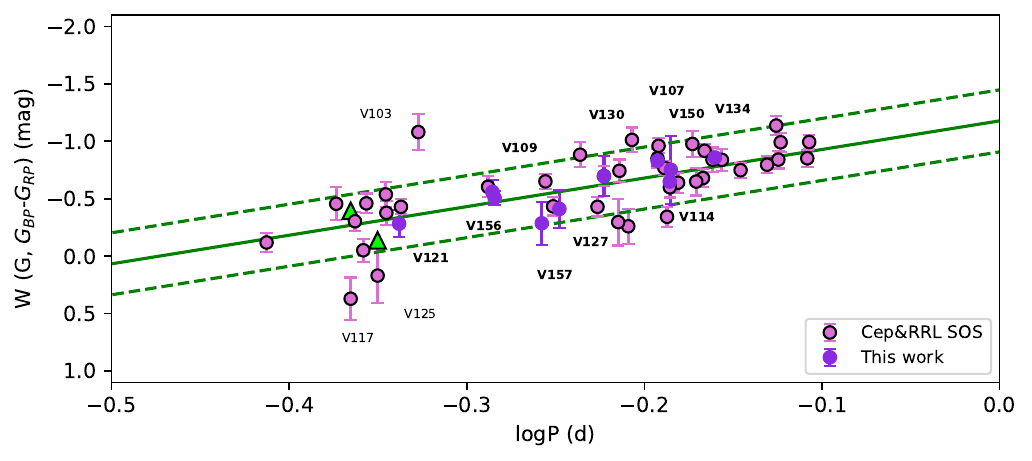}
    \caption{Position of the sources listed in 
    Table~\ref{tab:rr_nuove} on the PW$_{G,G_{BP},G_{RP}}$ relation for RR Lyrae stars of \citet{Garofalo-et-al-2022} (green solid line) 
    scaled to the distance modulus of UMi [(m-M)$_{0}$=19.23 mag;  Sect.~\ref{sec:dist}]. RRc and RRd stars have been {\it fundamentalized}.
    For the sources marked by orchid-filled circles (37 in total) periods and intensity-averaged $G$, $G_{BP}$ and $G_{RP}$ magnitudes adopted to compute the 
    Wesenheit magnitudes were taken from the \texttt{vari\_rrlyrae} table;  
    for 
    V107, V109, V121, V114, V127, V130, V134, V150, V156, V157 ( violet-filled circles) were  
    computed in this work (see text for details). Lime-filled triangles show V117 and V125 plotted adopting their $G_{BP}$ and $G_{RP}$ mean magnitudes in the {\it Gaia} DR3 main catalogue. 
    V103 lying well above the PWZ relation likely is an RR Lyrae of UMi evolved from the ZAHB.
}
    \label{fig:pw_new}
\end{figure}

\begin{figure*}
\center
\includegraphics[width=10cm]{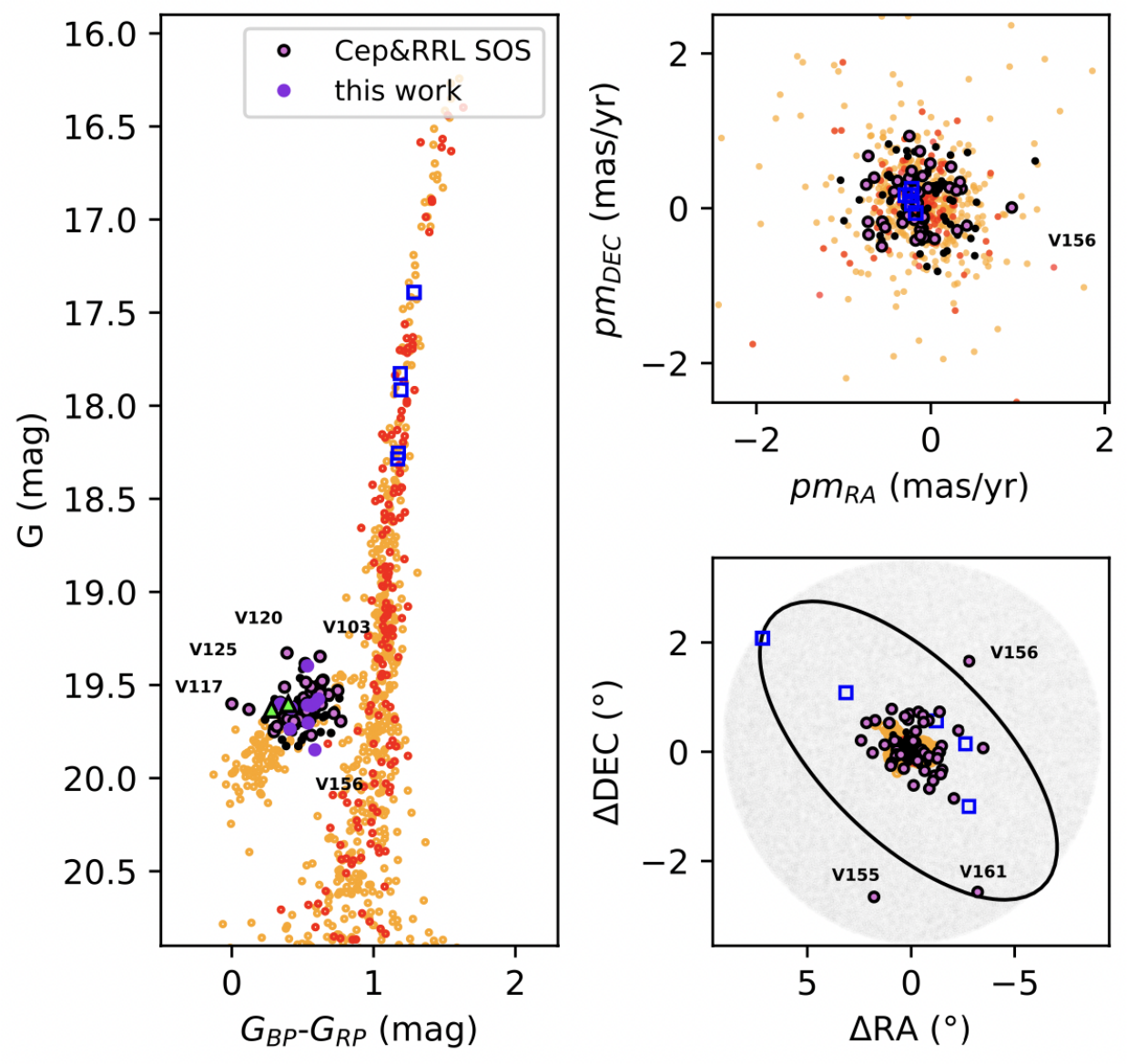}
    \caption{Same as in Fig.~\ref{fig:cmdeco_note} but for the 47 
    RR Lyrae stars that we confirm to be new members of UMi (see Sect.~\ref{sec:new48}). V103, V117, V120, V125 and V156 
    are discussed in the text. Lime-filled triangles show V117 and V125 plotted adopting their $G_{BP}$ and $G_{RP}$ magnitudes from the {\it Gaia} DR3 main catalogue. 
    }
    \label{fig:plot_48di68}
\end{figure*}

From our analysis, we find that V134 is an RRab star with pulsation period P=0.691416 days. 
We find for V156 a period P= 0.5196468 days, which also indicates that V156 can be an RRab star. However, the surprisingly small  amplitude in the $G$ band,  
Amp$G$=0.157 mag, makes this classification less certain. Hence, we classify the star as candidate RRab star.\\ 
For eight stars in Table~\ref{tab:rr_nuove}, (V107, V109, V114, V121, V127, V130 V150 and V157), 
we found the $G$ light curves to look noisy as if they were not perfectly folded with the periods in  the DR3 \texttt{vari\_rrlyrae} table. 
Also for them we analyzed the $G$ time series and confirmed the classification in type of V121 (RRc), V150 (RRab) and V107 (RRab) with  a very minor revision of  thier periods from the \texttt{vari\_rrlyrae} table. 
For the latter (V107) we also suggest it is 
blended with a companion causing the smaller amplitude than expected for the star period. 
In addition, we concluded that V130 and V114 could be affected by Blazhko effect,  while we reclassified 
V109, V127 and V157 as double-mode pulsators (RRd). They have less than 40 epochs in $G$, hence the 
SOS Cep\&RRL pipeline did not search the data of the 3 sources for a second periodicity, that we found instead in our analysis, and simply classifying them as RRc stars.  The periods in the \texttt{vari\_rrlyrae} table correspond indeed to the shorter periods as RRd stars. We list both primary and secondary period of the 3 stars in Table~\ref{tab:rr_nuove}. \\
Figure~\ref{fig:lc_48di68} shows the $G$-band light curves of the new variable stars plotted with filled circles when the data are phased with the periods published in the \texttt{vari\_rrlyrae} table (37 sources), and with empty circles if the data are folded with  periods derived from our analysis of the time series (10 sources).\\
In order to validate the classification as RR Lyrae of the new variable stars 
and test their membership to UMi, in Fig.~\ref{fig:pw_new} we plot them on the  PW$_{G,G_{BP},G_{RP}}$Z relation for RR Lyrae stars of \citet{Garofalo-et-al-2022}  scaled to the distance modulus of UMi [(m-M)$_{0}$=19.23 mag]. In the figure 
orchid-filled circles mark 37 sources with periods and  
$G$, $G_{BP}$ and $G_{RP}$ mean magnitudes from the \texttt{vari\_rrlyrae} table, whereas violet-filled circles mark 10 sources (labeled in the figure) all having periods and mean $G$ magnitudes from our analysis of the $G$-band time series data,  and $G_{BP}$, $G_{RP}$ mean magnitudes from our analysis for V134 and V156; and  from the \texttt{vari\_rrlyrae} table for the other 8 stars.
All sources in  Fig.~\ref{fig:pw_new} are compatible with the PW$_{G,G_{BP},G_{RP}}$Z relation of RR Lyrae stars
(green solid line with its $\pm$3$\sigma$ deviation) thus confirming that they are RR Lyrae members of UMi, except V117, V125 and V103.
We reach the same conclusion from the CMD 
in the left panel of Fig.~\ref{fig:plot_48di68} showing that they all RR Lyrae in this sample 
lie on UMi HB,  
and from the top-right panel of Fig.~\ref{fig:plot_48di68} showing that they all also have proper motions well 
consistent with those of UMi spectroscopically confirmed members. 
The spatial distribution of this sample (bottom-right panel) 
further supports 
the hypothesis that these 
RR Lyrae stars belong to UMi since most of them are located within the area enclosing the most peripheral boundary of the dSph determined so far (12 r$_{h}$; black solid line). 
Only V155 and V156 lie outside this region. However, it is hard to believe that they are MW field stars which are on the HB of UMi and have proper motions that clearly indicate their belonging to the dSph. Furthermore,
it is extremely difficult to find MW RR Lyrae field stars at such great distances from the centre of our Galaxy.  Since their population decreases with the Galactocentric radius, they are infact quite rare at distances greater than $\sim$50 kpc, as shown by a number of studies in the literature \citep{Vivas-Zinn-2006,Medina-et-al-2018,Garofalo-et-al-2021}.\\ 
On the other hand, 
 V117, V125 and V103 are outside $\pm3\sigma$ from
 the PWZ relation of RR Lyrae stars, 
 V125 significantly below and V103 being significantly above the green dashed lines in Fig.~\ref{fig:pw_new}. V117 and V125 also have  bluer colours ($G_{BP}-G_{RP}\sim$ 0) than other RR Lyrae stars on the CMD in the left panel of Fig.~\ref{fig:plot_48di68}. 
We found that, like V43, V60 and V63 in the Gold sample, V117 and V125 have very poor $G_{BP}$ light curves, resulting in poorly data-constrained best-fit models that produce too bright intensity-averaged $G_{BP}$ magnitudes for the stars. In fact the intensity-averaged $G_{BP}$ magnitudes of V117 and V125 in the \texttt{vari\_rrlyrae} table
are much brighter than $<G_{BP}>$  values for the stars in the {\it Gaia} DR3 main catalogue\footnote{Mean magnitudes (\texttt{phot\_bp\_mean\_mag} and \texttt{phot\_rp\_mean\_mag}) in the {\it Gaia} DR3 main catalogue are computed from the BP$-$band and RP$-$band mean flux applying the magnitude zero$-$point in the Vega scale \citep{Riello-et-al-2021}.
}.
V117 and V125 nicely fall near the PWZ and well within the bulk of the RR Lyrae stars on the CMD when we adopt the   $<G_{BP}>$ values in the {\it Gaia} DR3 main catalogue 
(lime triangles in  Figs.~\ref{fig:pw_new}  and ~\ref{fig:plot_48di68}), thus confirming that they are RR Lyrae belonging to UMi. 
The position of V103 and V120 on the HB of UMi (left panel of Figure~\ref{fig:plot_48di68}) makes them the brightest  RR Lyrae stars in the sample being about 0.3 mag more luminous than the mean $G$ magnitude of the Gold RR Lyrae dataset. 
Their $G$ mean magnitudes are based on well-sampled light curves as shown in Fig.~\ref{fig:lc_48di68}. We thus suggest that V103 and V120 are likely RR Lyrae stars evolved from the ZAHB, as star V73, discussed in Section~\ref{sec:ac}.\\
To summarize, we have found that 47 of the sources discussed in this section are RR Lyrae stars (46 confirmed and 1 candidate -- V156) belonging to the UMi dSph. 
This sample contains: 29 RRab stars (28 confirmed and one candidate), 15 RRc and 3 RRd stars. 
Among them 37 are known RR Lyrae first identified by the ZTF survey, 6 were discovered by {\it Gaia} DR2 (V112, V118, V135, V140, V155 and V157) and 3 were first identified by {\it Gaia} DR3 (V103, V122 and V139). 
%

\begin{table*}
 \caption{Identification and properties of the 47 RR Lyrae stars discussed in Sect.~\ref{sec:new48}. 
 The sources are ordered by increasing ID number. Column (1):  {\it Gaia} DR3 sourceid, column (2): ID number following N88 numbering scheme,  
 columns (3) and (4): {\it Gaia} DR3 coordinates (ra and dec).  Pulsation period (column 5), type/pulsation mode (column 6), intensity-averaged $G$ magnitude (column 7), and peak-to-peak $G$ amplitude (column 8), are from the {\it Gaia} DR3 \texttt{vari\_rrlyrae} table, unless differently stated in the footnotes.  Photometric [Fe/H] values (column 9) are from \cite[see Section~\ref{sec:final}]{Muraveva-et-al-2024}. }
 \label{tab:rr_nuove}
 \begin{tabular}{clllllccc}
  \hline
{\it Gaia} DR3 source\_id & name & ~~~~~ra & ~~~dec & ~~~~~~P&Type& $G$  &AmpG&$\rm[Fe/H]$\\
  &  & ~~~(deg) &~~(deg)& ~~(days) &  &(mag)& (mag)&(dex)\\
  \hline
1645465433535332224&V102&227.1974&67.4126&0.33688&RRc&19.769&0.534&$-$\\
1645464093505514752&V103&227.0982&67.3858&0.47071&RRab&19.346&0.4183&$-$\\
1645438602875088128&V104&226.9891&67.0522&0.32717&RRc&19.750&0.314&$-$\\
1693506983498648448&V106&227.0785&67.5903&0.33710&RRc&19.694&0.450&$-$\\
    1645386887173108864 & V107$^{*}$ & 227.6513 & 66.9065 & 0.641825  & RRab & 19.575 & 0.655 
    &$-$1.47$^{***}$\\    
 1645456950975395072 & V108 & 226.8012 & 67.2482 & 0.62088 &RRab & 19.695 & 0.464 &$-$\\
 1645443958698688896 & V109$^{*}$ & 227.7974 & 67.1624 &  0.3868195  & RRd & 19.704& 0.375 &$-$\\
   &  & &  & 0.5187564 &  &  & \\
   1693533440497210880 & V111 & 227.4483 & 67.8019 & 0.59376 & RRab & 19.642 & 0.801 & $-$2.27\\
  1645455267348204928 & V112 & 226.6747 & 67.2005 & 0.77931 & RRab & 19.583 & 0.208 & $-$\\
  1645286178779860992 & V113 & 227.1829 & 66.5994 & 0.32848 & RRc & 19.682  & 0.410 &$-$\\
1693542958144744448 & V114$^{*}$ & 227.3919 & 67.9061 & 0.6520700 & RRab & 19.608 & 0.826 &$-$\\
 1645468079235094784 & V115 & 227.9968 & 67.2423 & 0.68077 & RRab & 19.605& 0.671 &$-$2.25\\
 1645420284838964480 & V116 & 228.0029 & 67.1646 & 0.61793 & RRab & 19.675&  0.996 &$-$1.67\\
   1693539895831731840 & V117$^{**}$ & 227.5977 & 67.8612 &   0.32187 & RRc & 19.601 & 0.642 &$-$\\
  1645341841556102400 & V118 & 226.6486 & 66.8655&  0.28863 & RRc & 19.722 & 0.273 &$-$\\  
  1645419425845478272 & V119 & 228.0396 & 67.1304 & 0.71477 & RRab & 19.485& 0.567 &$-$\\
 1693523682331500544 & V120 & 226.7027 & 67.7560 &   0.41428&RRc & 19.328 & 0.584 &$-$\\
1693550581710923520 & V121$^{*}$ & 226.9498 & 67.9432 &  0.342738    & RRc & 19.736& 0.523 &$-$\\
  1645471687006867584 & V122 & 228.0974 & 67.4197 & 0.68915  & RRab & 19.539 & 0.281 &$-$\\
 1645361533981842560 & V123 & 226.4399 & 67.1284& 0.32366 & RRc & 19.683 & 0.581 &$-$\\
  1693551513720062720 & V124 & 226.7320 & 67.8815 &  0.41842 & RRc & 19.511 & 0.417 &$-$\\
  1645414752921528704 & V125$^{**}$ & 228.2284 & 67.0135  & 0.33330 & RRc & 19.631 & 0.585 &$-$\\
  1693527491966617472 & V126 & 226.4644 & 67.7885 & 0.67149 & RRab & 19.479 & 0.677 &$-$2.24\\
  1645379263606883456 & V127$^{*}$ & 228.2800 & 66.9611 &  0.4218176 & RRd & 19.623  & 0.460 &$-$\\
 &  & &  &  0.5656201 &      &  & \\
  1645327685343764736 & V128 & 226.4265 & 66.5394 & 0.74855& RRab & 19.530 & 0.557 &$-$2.13\\
  1645333251621444736 & V129 & 226.2438 & 66.6845 & 0.69769 & RRab & 19.605 & 0.570 &$-$2.01\\
1645485091601208320& V130$^{*}$ & 228.3545 & 67.7414 & 0.5987663  & RRab & 19.611 & 0.901 
    &$-$1.52$^{***}$\\
  1645347682711590400 & V132 & 226.1694 & 66.7759 & 0.64977 &  RRab & 19.617& 0.870 &$-$1.96\\  
  1645878437589563392 & V133 & 228.2361 & 67.9971 & 0.73975& RRab & 19.509 &  0.407 & $-$2.38\\
1693398677307283584 & V134$^{*}$ & 226.0354 & 67.0934 & 0.691416    & RRab & 19.609 &  0.827  & $-$\\ 
  1645428501111989376 & V135 & 228.5738 & 67.3433 & 0.31589 & RRc & 19.692 & 0.605 &$-$2.10\\
 1693402804771797888 & V136 & 225.9916 & 67.2046 &   0.43333 & RRc & 19.543 & 0.352 &$-$\\
 1693403526326303104 & V137 & 225.9244 & 67.2113 & 0.67480& RRab & 19.597& 0.585 &$-$2.47\\
  1693416097694750080 & V139 & 225.8337 & 67.3204 & 0.78126  & RRab & 19.474 & 0.272&$-$\\
 1645350053535689344 & V140 & 225.8740& 66.8071 & 0.75285& RRab & 19.551 & 0.314 & $-$\\
  1693389232675123840 & V141 & 225.8143 & 66.8830 & 0.65911& RRab & 19.568& 0.732 &$-$2.14\\
 1693579413827628160 & V142 & 225.9330 & 67.9459 & 0.68215  & RRab & 19.505 & 0.699 &$-$2.25\\
 1645863701557709440 & V144 & 229.0312 & 67.7893 & 0.34354 & RRc & 19.668 & 0.529 &$-$1.81\\
 1645409427161016832 & V145 & 229.1561 & 67.1967 & 0.61090 & RRab & 19.621 & 1.142 & $-$2.01\\
  1645815696707662464 & V147 & 229.4490 & 67.7438 & 0.38459 & RRc & 19.639 & 0.476 &$-$\\
  1645310711632658816 & V148 & 225.2429 & 66.3610 & 0.64759   & RRab & 19.538 & 0.594 & $-$2.07\\
  1693485607445562240 & V149 & 225.0232 & 67.6033 & 0.60988 & RRab & 19.593 & 1.115 &$-$1.84\\
   1645793328517773952 & V150$^{*}$ & 229.7093 & 67.4215 & 0.6529827  & RRab & 19.600 & 0.847 &$-$2.47\\
 1644108979783405056 & V155 & 229.0879 & 64.5577 & 0.64270  & RRab & 19.652  & 0.477 &$-$1.52\\
1693875487396334080 & V156$^{*}$ & 224.4934 & 68.8726 & 0.5196556   &  RRab?& 19.848  & 0.179 &$-$\\
    1693449426640926208 & V157$^{*}$ & 223.8107 & 67.2827 & 0.4123402 & RRd & 19.602 & 0.360 &$-$\\
      &  & &  & 0.5524454 &  &    & \\
  1621030406810918144 & V161 & 224.0765 & 64.6480 & 0.75062  & RRab & 19.384 & 0.469 &$-$1.93\\
  \hline
 \end{tabular}
 \tablefoot{
 $^{*}$P, type, $G$ mean magnitude and $G$ amplitude derived in this work. First and second periodicity are provided for RRd stars. The $G$ amplitude for RRd stars corresponds to the amplitude of the light curve folded with the first periodicity, before the pre-whitening. $^{**}$$G_{BP}$  and $G_{RP}$ values provided by the {\it Gaia} main catalogue, see text for details.
 $^{***}$Metallicity estimates for V107 and V130 may be unreliable (see Sect.~\ref{sec:final}).}
\end{table*}

\section{Variables in UMi: final sample and properties}\label{sec:final}
\begin{figure}
\center
\includegraphics[width=\hsize]{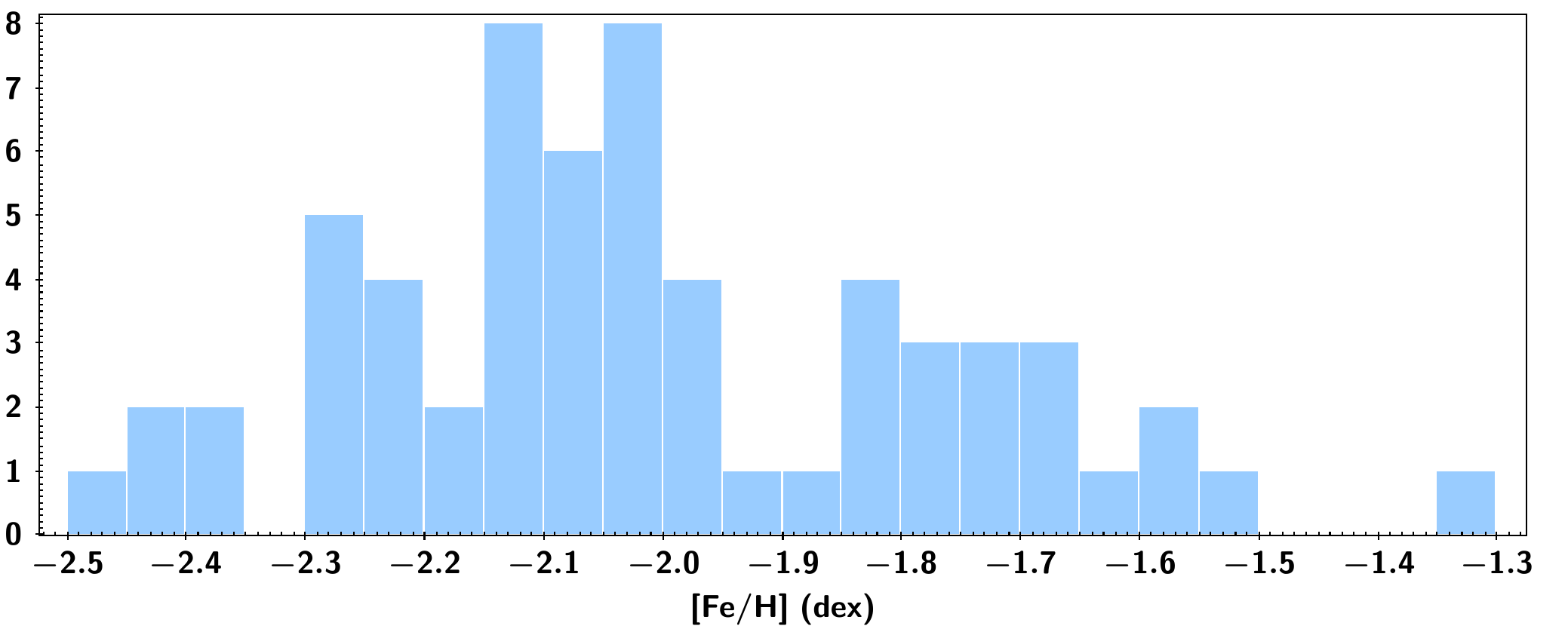}
    \caption{Distribution of the photometric metallicities from   \citet{Muraveva-et-al-2024} for 62 RR Lyrae stars that we confirm to belong to UMi. 
    }
    \label{fig:isot_met}
\end{figure}

\begin{figure*}
\center
\includegraphics[width=13.5cm]{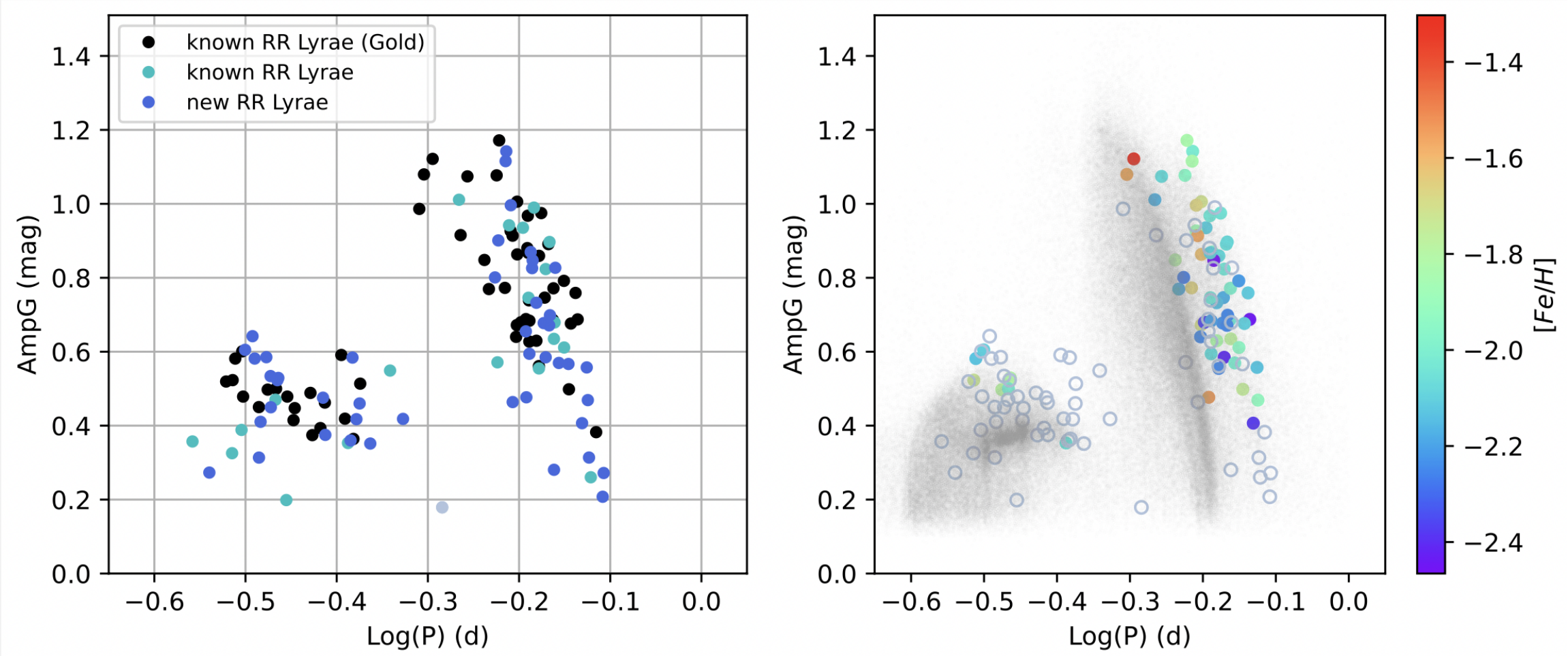}
\caption{Bailey diagrams of the RR Lyrae stars belonging to UMi. {\it Left panel}: Black filled-circles mark the Gold sample RR Lyrae, known and new RR Lyrae members of UMi confirmed in this work 
are plotted with teal- and blue-filled circles, respectively. A total of 125 sources are shown in the figure.  {\it Right panel}: Same as in the left panel, but with 62 UMi RR Lyrae stars with photometric metallicities from \citet{Muraveva-et-al-2024} marked by filled circles colour-coded according to their metallicity, and UMi RR Lyrae members lacking a metallicity value shown by empty circles. Grey points in the background show the all sky RR Lyrae sample published in the {\it Gaia} DR3 \texttt{vari\_rrlyrae} table ($\sim$ 270\,000 sources).}
    \label{fig:bailey_new}
\end{figure*}
\begin{figure*}
\center
\includegraphics[width=8.9cm]{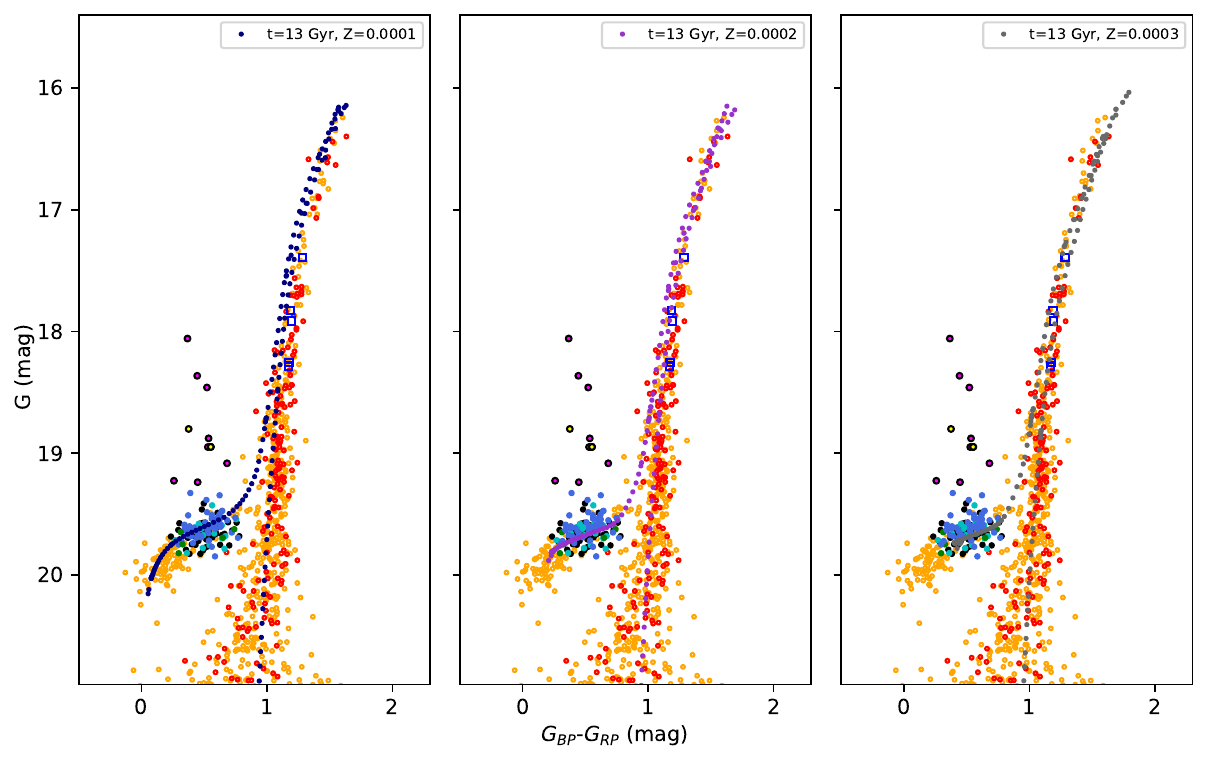}
~\includegraphics[width=8.9cm]{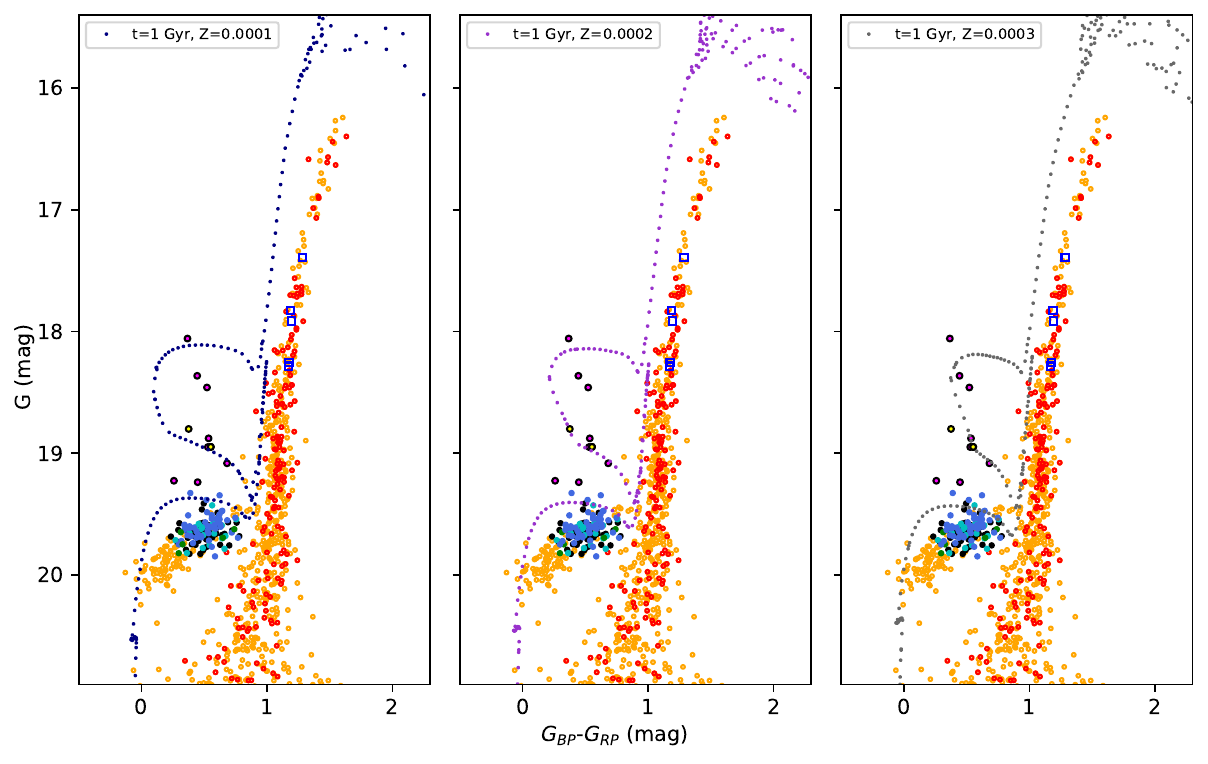}
    \caption{CMD of UMi's member stars from \citet{Sestito-et-al-2023,Pace-et-al-2020,Kirby-et-al-2010} 
    (blue, orange and red symbols, respectively),     
    overlaid by Padua stellar isochrones \citep{Bressan-et-al-2012} for different combinations of age and metallicity. 
    {\it Left panels:} Stellar isochrones with metal abundance: Z=0.0001, Z =0.0002 and Z = 0.0003, and age t= 13 Gyr. 
    {\it Right panels:}  Stellar isochrones with metal abundance: Z=0.0001, Z =0.0002 and Z = 0.0003, and age t= 1 Gyr.  
    Filled symbols in different colours distinguish the various RR Lyrae subsamples: Gold (black), known (cyan), new members (navy), and known, but not classified by {\it Gaia} DR3 as RR Lyrae stars (green). Magenta- and yellow-filled circles are known and new ACs belonging to UMi located within and beyond 12 r$_{h}$.
    }
    \label{fig:isocrone_cmd_rr}
\end{figure*}
With our analysis, we bring the total number of 
confirmed and candidate variable stars hosted by the UMi dSph galaxy to  139 (this number includes V12, V56 V57, V84, and V95 from Table~\ref{tab:rr-nosos})
In particular, we found 129 RR Lyrae stars, of which 47 (1 candidate and 46 confirmed RR Lyrae) are new members of UMi, and 10 ACs, of which 5 (1 candidate and 4 confirmed ACs) are also new members of the galaxy. 
This has allowed us to have a more complete picture of the pulsation properties of the variable star population in UMi. \\
We adopt for our updated list of RR Lyrae stars belonging to UMi, the photometric metallicities  
recently published by \citet{Muraveva-et-al-2024}. 
These Authors published individual [Fe/H] values for $\sim$ 134,000 RR Lyrae stars based on new $P$-$\phi_{31}$-[Fe/H] (for RRab) and P-$\phi_{31}$-A${_2}$-[Fe/H] (for RRc) relationships that they determined directly from the parameters (period, $\phi_{31}$ and  A${_2}$\footnote{$\phi_{31}$ and  A${_2}$ are parameters of the Fourier decomposition of the star light curve folded with period.}) of the $G$-band light curves published in the {\it Gaia} DR3 \texttt{vari\_rrlyrae} table, for RR Lyrae stars with accurate metal abundances from spectroscopy. 
Metallicity measurements from \citet{Muraveva-et-al-2024} are available for 64 RR Lyrae stars belonging to UMi, corresponding to about 
50\% of the sample. They are reported in column 10 of Table~\ref{tab:rrnemec} and column 9 of 
Table~\ref{tab:rr_nuove}. The mean uncertainty of the individual metallicity estimates is around 0.48 dex, obtained by combining the dispersion of \citet{Muraveva-et-al-2024} relations (0.28  and 0.21 dex for RRab and RRc, respectively) and the uncertainty in the period and Fourier parameter ($\phi_{31}$ and A${_2}$) of each source, 
a main contribution, due to the reduced reliability of the pulsation parameters for stars at the magnitude level of UMi RR Lyrae ($G \sim$ 19.5-20 mag).\\
The metallicity distribution 
is shown in Fig. ~\ref{fig:isot_met} and is based  on 62 RR Lyrae stars. We dropped V107 and V130, whose  $\phi_{31}$ and  A${_2}$ parameters, hence metallicity estimates, may be unreliable, since V107 is suspected to be blended with a companion, and V130 is suspected to be affected by Blazhko.\\ 
Metallicities 
range from [Fe/H]= $-$1.30 dex (for V45) to $-$2.47 dex for (V150) and  peak  at [Fe/H] $\sim-2.0/-$2.1 dex,  in excellent agreement with the mean metallicity of UMi determined spectroscopically by \citet{Kirby-et-al-2011,Pace-et-al-2020} using different datasets and instruments. Although the metallicity information is only available for 50\% of the RR Lyrae sample, this subsample seems to be rather representative of the entire RR Lyrae population in UMi. \citet{Kirby-et-al-2011} and \citet{Pace-et-al-2020} find a metallicity spread 
around 0.30 dex in UMi. The standard deviation associated with the mean metallicity of our sub-sample, 0.25 dex, is fairly consistent with that value.\\
The left panel of Fig.~\ref{fig:bailey_new} shows the Bailey diagram for our final sample of RR Lyrae stars belonging to UMi. A total of 125 
RR Lyrae stars are plotted in this figure, since we do not have and could not calculate the $G$ amplitude for all stars listed in 
Table~\ref{tab:rr-nosos}, except V57.
Gold sample, known, and new RR Lyrae members of UMi are plotted with black-, royal blue-, and cyan-filled circles, respectively. The dubious RR Lyrae star V156 
is plotted in lighter blue. In the right panel of Fig.~\ref{fig:bailey_new}, we compare the $G$ amplitudes and periods of UMi RR Lyrae stars with the Bailey diagram of the approximately 270 thousand RR Lyrae stars published in the 
{\it Gaia} DR3 \texttt{vari\_rrlyrae} table \citep{Clementini-2023}.  
Almost all UMi's RRab stars are placed on the Oosterhoff~II (Oo~II) line, the locus of RR Lyrae stars with periods $\sim$0.65 days and a low metal content ([Fe/H]$\sim -$2.0 dex). UMi RR Lyrae stars with photometric metallicity estimates by \citet{Muraveva-et-al-2024} are marked with filled circles colour-coded according to their metallicity 
and confirm the predominantly metal-poor content of the RR Lyrae population in this MW satellite. 
A few more metal-rich RR Lyrae stars can be seen near the Oosterhoff~I (Oo~I) locus. With future {\it Gaia} data releases, 
a more complete picture of the metallicity spread in this dSph galaxy will certainly be possible.\\
In Sec.~\ref{sec:gs}, we derived $<{\rm P_{ab}}>$=0.637 $\pm 0.062$ days (average on 38 stars) and 
${\rm<P_{c+d}>=0.3560 \pm 0.038}$ days (average on 19 stars), and a classification as Oo~II system for UMi, 
from the 57 RR Lyrae in the Gold sample. We can now revisit these numbers 
with all data at our disposal. We have periods for 125 (124 confirmed and 1 candidate) RR Lyrae members of UMi, 80 RRab and 45 RRc+d. Excluding the RRab candidate (V156) we obtain: $<{\rm P_{ab}}>$=0.652 $\pm$ 0.063 days (average on 79 stars) and ${\rm<P_{c+d}>}$=0.369 $\pm$0.042d  (average on 45 stars; for V95, we adopted the period that fits the PWZ relation, P=0.305 days, see Tab.~\ref{tab:rr-nosos}); excluding the RRc stars in Tab.~\ref{tab:rr-nosos}, the average does not change:  ${\rm<P_{c+d}>}$=0.368 $\pm$ 0.043. These numbers confirm that UMi is an Oo~II system. 
The fraction of RRc+d stars: 35\%  (or 34\%, if we exclude 2 suspected RRd stars in the Gold sample, V49 and V81, see Sec.~\ref{sec:gs}) is also closer to the typical fraction of RRc stars in Oo~II systems.\\ 
Finally, Figure~\ref{fig:isocrone_cmd_rr} shows UMi's CMD 
in the {\it Gaia} DR3 bands,  
built with the spectroscopically-confirmed RGB and HB member stars 
from \citet{Sestito-et-al-2023,Pace-et-al-2020,Kirby-et-al-2010} and over-plotted our final catalogue of ACs and RR Lyrae stars belonging to the galaxy.
%
We overlaid the CMD by stellar isochrones of different age and chemical compositions from the PARSEC database \citep{Bressan-et-al-2012}, to fit UMi's stellar populations. Isochrones were corrected by adopting the distance modulus of UMi we derived from the RR Lyrae stars [$(m-M)_{0}$=19.23 mag, Sect.~\ref{sec:dist}] and the reddening E($G_{BP}-G_{RP}$)=0.043 mag, from the $A_{V}$ value converted into A$_{G_{BP}}$ and A$_{G_{RP}}$, using relations from \citet{Bono-at-al-2019}. We selected isochrones with old age t=13 Gyr and a combination of low metal abundances [Z=0.0001, Z=0.0002, and Z=0.0003 ([Fe/H]=$\sim-$1.7 dex)] to fit the predominantly old stellar population and the large number of UMi RR Lyrae stars.  We found the isochrones that best reproduce both UMi extended HB and the RGB stars 
to be those with age t=13 Gyr and metallicity Z=0.0001 ([Fe/H]=$\sim-$2.2) and Z=0.0002 ([Fe/H]=$\sim-$2.0) (see left panels of Fig.~\ref{fig:isocrone_cmd_rr}).\\
Our final catalogue of UMi variable stars lists 10 ACs (9 confirmed and 1 candidate). There are at least two accepted scenarios for the formation of ACs. A first one considers ACs intermediate age, metal poor single stars produced by a 1-2 Gyr old star formation burst and,  a second scenario in which ACs are supposed to form via mass transfer in binary systems as old as the other stars in the host galaxy. We over-plotted the galaxy CMD with isochrones of intermediate age (t=1 Gyr)
and low metal abundance (Z=0.0001, Z=0.0002, and Z=0.0003) and found that 1 Gyr old isochrones can reproduce the location of UMi ACs reasonably well (see right panels of Fig.~\ref{fig:isocrone_cmd_rr}). 
However, so far no indisputable evidence of an intermediate-age population or presence of gas has been found in the UMi dSph. \cite{Carrera-et-al-2002} studied UMi star formation history based on a CMD reaching 1.5 magnitudes below the main sequence (MS) turn-off (TO). They show  
that UMi hosts a predominantly old stellar population formed more than 10 Gyr ago, and conclude 
 the blue plume extending about 2 magnitudes above the TO (see their fig.~4) to be most likely formed by blue stragglers originating in close binary stars rather than an intermediate-age population. 
 UMi is the only MW dSph satellite with a pure old stellar population, so far. In this framework, it seems more plausible that UMi ACs formed via mass transfer in close binaries.

\subsection{MU\~NOZ~1 with  {\it Gaia} DR3 Data: is it there or not?}
\begin{figure*}
\center
\includegraphics[width=8.7cm]{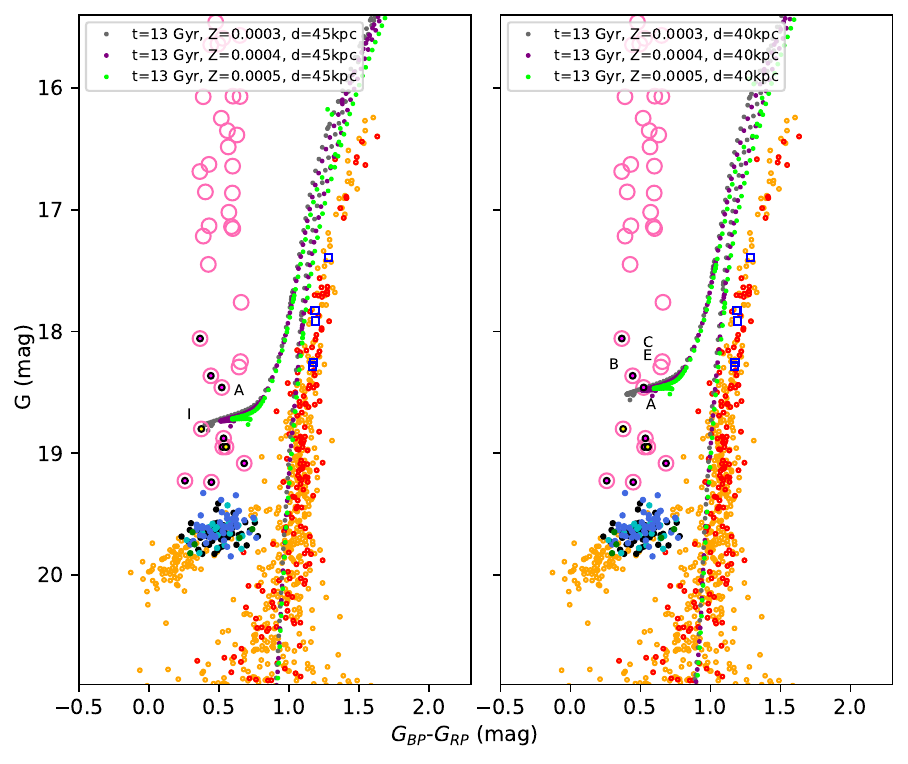}
    \caption{CMD of UMi members (RGB, HB and RR Lyrae stars) using the same symbols and colour-coding as in Fig.~\ref{fig:isocrone_cmd_rr}. Overlaid on the CMD are Padua 
    stellar isochrones \citep{Bressan-et-al-2012} for a fixed age, t=13 Gyr, and three different metal abundances (Z=0.0003 grey, Z =0.0004 purple and Z = 0.0005 lime). Isochrones are fixed at distance d=45 kpc on the left panel and at d=40 kpc on the right panel. 
   Pink circles highlight 43 variable stars brighter than $G=$19.25 mag discussed in Sect.~\ref{sec:ac}, among which are 10 ACs of UMi.}
    \label{fig:mu1_cmd}
\end{figure*}

Based on deep photometry obtained with MegaCam at CFHT, 
\citet{Munoz-et-al-2012} detected the second "ultra-faint" star cluster ever identified in the vicinity of a MW dSph, Mu\~noz~1.
They found  Mu\~noz~1  to be located at a distance of 
 45 $\pm$ 5 kpc and at a projected distance of 
45$\arcmin$ from the centre of UMi, to which it dose not seem to be 
associated 
being separated by 
$\sim$ 30 kpc in distance. The best-fit isochrone, from the Dartmouth database \citep{Dotter-et-al-2008} overplotted on the CMD 
suggested that Mu\~noz~1 hosts an ancient stellar population (t=12.5 Gyr) with [Fe/H]= $-$1.5 dex. 
More recently,  using  {\it Gaia} EDR3 astrometry \citet{Vasiliev-Baumgardt-2021} published proper motions for 170 clusters including Mu\~noz~1 for which they give: 
$<pm_{ra}>=-$0.100 $\pm$ 0.203 $<pm_{dec}>=-$0.020 $\pm$ 0.207 based on 5 member stars. 
These values are almost identical to those of the UMi dSph, as shown by the right panel of  Fig.~\ref{fig:pm_members}.
We tried to find out if any RR Lyrae star in our final catalogue, which covers up to  UMi's extended halo redefined by \citet{Sestito-et-al-2023}, could belong to Mu\~noz~1.\\
Since \citet{Munoz-et-al-2012} did not provide coordinates of Mu\~noz~1 member stars we were unable to counter-identify them in the DR3 catalogue. However, adopting for the cluster the   
distance  d=45 or 40 kpc (corresponding to distance moduli in the {\it Gaia} $G$ band of 18.266 or 18.010 mag, respectively) and the metallicity [Fe/H]=$-1.5$ dex from \citet{Munoz-et-al-2012}, we estimate the mean magnitude of any potential RR Lyrae stars on the cluster HB to be $G\sim$18.90 (or $G\sim$18.64 mag). 
%
There are 9 RR Lyrae stars in our final catalogue within an area  of 0.4 $\deg$ in radius 
 from Mu\~noz~1 centre (lime cross in Figure~\ref{fig:map1}): 
 V140, V141, V132, V129, V134, V136, V137, V76 and V128. Their $<G>$ mean magnitudes are $\sim$0.7 (or $\sim$1) mag fainter than Mu\~noz~1's HB, while being perfectly consistent with the mean magnitude of UMi's HB. Hence, there are no RR Lyrae stars in our catalogue 
 that should belong to the cluster. We then checked whether any of the ACs or bright variable stars with G$<$19.25 discussed in Sect.~\ref{sec:ac}) could be RR Lyrae stars belonging to Mu\~noz~1. In Fig.~\ref{fig:mu1_cmd} we show the CMD of UMi member stars (RGB, HB and RR Lyrae stars) using the same symbols and colour-coding as in Fig.~\ref{fig:isocrone_cmd_rr} and highlighting with pink circles 43 bright variable sources with G$<$19.25 mag. \\
%
We over-plotted on the CMD, stellar isochrones from the PARSEC dataset with fixed age t=13 Gyr and three different metal abundances, Z=0.0005 ([Fe/H]=$\sim-1.5$ dex), Z=0.0004 and Z=0.0003, to trace the ancient population of Mu\~noz~1. Isochrones were set to d=45 kpc in the left panel, and at d=40kpc in the right panel. 
In both panels of Fig.~\ref{fig:mu1_cmd} none of the bright stars marked by pink empty circles  is found to lie on the HB of the isochrones we overlaid on the CMD. Only star I (V178), a candidate AC, reaches near the blue end of the Z=0.0003 isochrone, the most metal poor one. However, I (V178) is too blue to be an RR Lyrae of a cluster with a metallicity of Z=0.0003 and definitely much bluer than the isochrone that best corresponds to the metallicity of Mu\~noz~1, the one with Z =0.0005. But more importantly, the star's proper motions are not compatible with those of UMi, and therefore with those of Mu\~noz~1.
%
If Mu\~noz~1 were at the shorter distance adopted in the right panel of Fig.~\ref{fig:mu1_cmd} (d=40 kpc), and  were the cluster metallicity lower than  estimated in the discovery paper ([Fe/H]=$-1$.5 dex), star A (V105) 
could perhaps be an  RR Lyrae beloging to Mu\~noz~1. However, assuming for the cluster the metallicity and distance values provided by \citet{Munoz-et-al-2012}, in our catalogue of 168 variable stars identified within a radius of 3.5 degrees from UMi centre we do not find RR Lyrae stars that can be reliably attributed to the Mu\~noz~1.

\begin{figure*}
\center
\includegraphics[width=8.3cm]{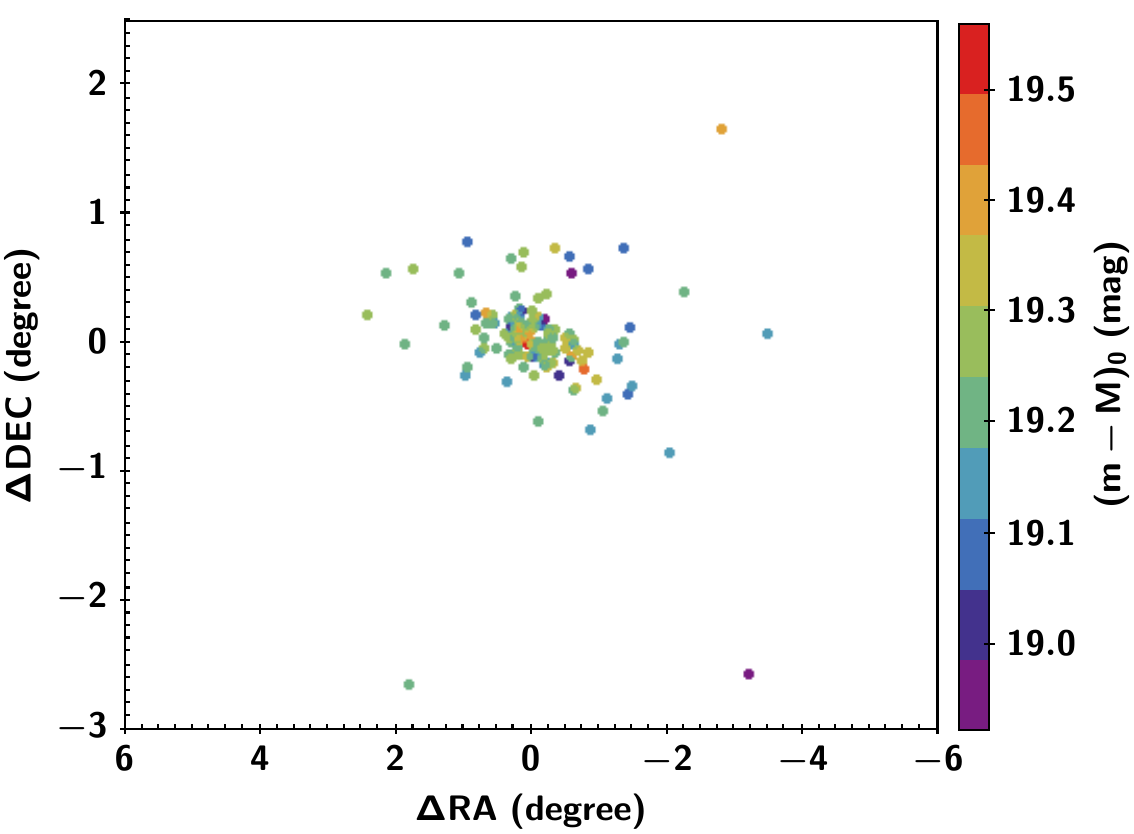}~\includegraphics[width=8.3cm]{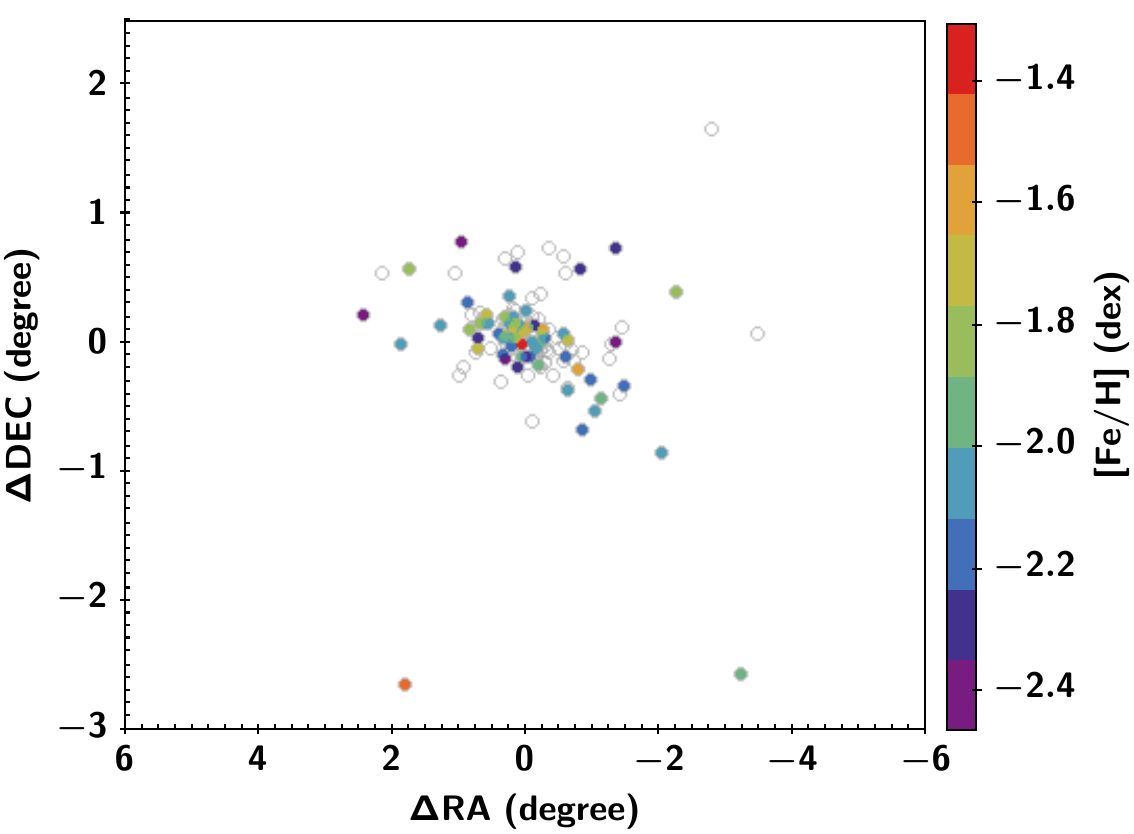}
    \caption{Spatial distribution of 129 RR Lyrae in UMi (128 bona-fide and 1 candidate) with sources colour-coded according to their individual distance moduli derived from the M$_{G}-$[Fe/H] relation for fixed metallicity: [Fe/H]=$-$2.13 dex (left panel) and, according to their individual photometric metallicity values from \citet{Muraveva-et-al-2024} (right panel).}
    \label{fig:mod-met}
\end{figure*}

\section{Discussion and Conclusions}\label{sec:concl}
We have analyzed 168 variable stars classified as RR Lyrae stars in the {\it Gaia} DR3 \texttt{vari\_rrlyrae} and \texttt{vari\_classifier\_result} tables. The sources are located within an area of 3.5 degree in radius from the centre of the UMi dSph galaxy. We have reclassified and recharacterized
known variable stars and identified new members of this MW satellite.
We adopted different tools in our analaysis, based on {\it Gaia} DR3 astrometry and photometry: (i) proper motions, (ii) Period-Wesenheit-Metallicity relations of RR Lyrae stars and Anomalous Cepheids (ACs), (iii) spatial distributions, and (iv) colour-magnitude diagrams with overlaid stellar isochrones to discriminate between RR Lyrae and ACs, and between members and non-members.
We built a reference sample, that we named Gold sample, which included 57 known RR Lyrae members of UMi studied by N88 having a counterpart in the {\it Gaia} DR3 catalogue of RR Lyrae (\texttt{vari\_rrlyrae}) and pulsation periods in agreement with the literature.
From the Gold sample RR Lyrae and the M$_{G}$-[Fe/H] relation published in \citet{Garofalo-et-al-2022} we derived  UMi's distance modulus: (m-M)$_{0}$=19.23$\pm$0.09 mag, which is in excellent agreement with the values published in different studies. 
We confirmed 129 sources to be RR Lyrae stars, 128 bona-fide (among them 46 are new members of the galaxy) and 1 candidate (V156). We have periods for 125 of them: 80 RRab, 42 RRc and 3 RRd stars.
The average period of RRab stars is: $<{\rm P_{ab}}>$=0.652 $\pm$ 0.063 days (average on 79 stars) and the avergage period of RRc+RRd stars is: ${\rm<P_{c+d}>}$=0.369 $\pm$0.042d  (average on 45 stars).
These numbers confirm that UMi is an Oosterhoff II system.
We classify 10 sources as Anomalous
Cepheids (among them 4 new members) and 1 candidate (V178). 
The variable stars are distributed everywhere around the centre of UMi, and also beyond the elliptical area delimited by 12 r$_{h}$, that represents the maximum distance at which  spectroscopically confirmed members has been found so far \citep{Sestito-et-al-2023}.
Using our final sample of 128 bona-fide RR Lyrae belonging to UMi 
we obtain a distance modulus for the galaxy of (m-M)$_{0}$=19.23$\pm$0.11 mag, in excellent agreement with the distance modulus obtained 
from the  Gold sample RR Lyrae,  that are all placed within 5 r$_{h}$ from UMi's centre.
In Fig.~\ref{fig:mod-met} we compare the spatial distribution of our final sample of RR Lyrae stars in UMi with the sources colour-coded according to their individual distance moduli in left panel of the figure,    
and according to their individual photometric metallicities in the right panel. 
Individual distances were obtained from the M$_{G}-$[Fe/H] relation computed at mean metallicity value of [Fe/H]=$-$2.13 dex from \citet{Kirby-et-al-2011,Pace-et-al-2020}. As individual photometric metallicities of the RR Lyrae stars 
we adopted those derived by \citet{Muraveva-et-al-2024}, which are available for about half the RR Lyrae stars in our catalogue. 
We do not see any effect due to distance, no gradients, and no trends, in the plot on the left pannel,  while in the distribution on the right, we notice a weak trend that is expected from literature \citep{Pace-et-al-2020}: the metal-poor RR Lyrae stars [Fe/H]<$-$2.2 dex (purplish symbols) are placed at a greater distance from the centre of the dSph and have a distribution with smaller ellipticity than the RR Lyrae less metal-poor (greenish symbols) which seem to be located at a smaller distance with a larger value of ellipticity. The sample of RR Lyrae stars in UMi that we updated could reflect the two populations that \citet{Pace-et-al-2020} have distinguished in terms of metallicity but, since we only have individual metallicities for 50 \% of the sample we cannot draw a solid conclusion.   
Among the 128 bona-fide RR Lyrae stars, there are 4 sources which are known RR Lyrae of UMi, that do not have a counterpart in the {\it Gaia} DR3 \texttt{vari\_rrlyrae} catalogue. We confirmed them by adopting the periods inferred by N88 and checking the average magnitudes provided by the {\it Gaia} DR3 main catalogue (they are not computed as an intensity-average over the full  pulsation cycle like the mean magnitudes in the \texttt{vari\_rrlyrae} table) on the PWZ relation, on the CMD and the proper motions diagram. In N88 catalogue there are 2 variable sources without a counterpart in {\it Gaia} DR3, V88 and V94, we could not derive their periods as {\it Gaia} DR3 time-series are not available for them. For this reason, we have not included them among the confirmed RR Lyrae even though, their proper motions and {\it Gaia} mean magnitudes suggest that they may be RR Lyrae of UMi (see Section~\ref{sec:new} and Figure~\ref{fig:extra_acs}). 
Finally, we explore the possibility that some RR Lyrae or ACs we found in the outskirt of UMi could be instead RR Lyrae members of Mu\~noz~1, a faint cluster located at 40-45 kpc from us,  with mean proper motions that are indistinguishable from those of UMi. Adopting distances and pulsation properties of our dataset of variables, together with CMDs and {\it ad hoc} stellar isochrones to mimic the Mu\~noz~1 stellar population, we conclude that the metal content suggested for the cluster in the discovery paper \citep{Munoz-et-al-2012}, is too high to produce RR Lyrae stars that could be cluster members.
Our updated catalogue of variable stars in UMi reaches the most peripheral regions of the galaxy at the same time complementing the sample of known variables in the galaxy inner regions, with this increasing 
by more than 1/3 the number of RR Lyrae stars known to be member of this dSph. The catalogue 
represents a powerful tool to study the stellar populations in the outskirt of UMi and reconstruct their origin trying to understand if they hide the signature of stellar streams and tidal stripping events from past interactions.
\begin{acknowledgements}
 This work made use of data from the European Space Agency
(ESA) mission Gaia (\url{https://www.cosmos.esa.int/gaia}), processed
by the Gaia Data Processing and Analysis Consortium (DPAC; \url{https://www.cosmos.esa.int/web/gaia/dpac/consortium}). Funding for
the DPAC has been provided by national institutions, in particular,
the institutions participating in the Gaia Multilateral Agreement.
 Support to this study has been provided by INAF Mini-Grant (PI: Tatiana Muraveva), by the Agenzia Spaziale Italiana (ASI) through contract and ASI 2018-24-HH.0, and by Premiale 2015, MIning The Cosmos - Big Data and Innovative Italian Technology for Frontiers Astrophysics and Cosmology (MITiC; P.I.B.Garilli). This research was also supported by the International Space Science Institute (ISSI)
in Bern, through ISSI International Team project 490, ‘SH0T: The
Stellar Path to the H0 Tension in the Gaia, TESS, LSST, and JWST
Era’ (PI: G. Clementini).
 This research also made use of TOPCAT \citep{Taylor-2005} an interactive graphical viewer and editor for tabular data. 
 This research has made use of 
 numpy \citep{vanderWalt-2011} and the Astropy library \citep{AstropyCollaboration-2013,Astropy-Collaboration2018}. 
 This research has used the VizieR tool. The original description of theVizieR service was published in \citet{Ochsenbein-2000}. 
 Most of the figures in this paper were produced with Matplotlib \citep{Hunter-2007}.
\end{acknowledgements}

%
  \bibliographystyle{aa} 
   \bibliography{pumi} 

\begin{thebibliography}{70}
\expandafter\ifx\csname natexlab\endcsname\relax\def\natexlab#1{#1}\fi

\bibitem[{{Astropy Collaboration} {et~al.}(2018){Astropy Collaboration}, {Price-Whelan}, {Sip{\H{o}}cz}, {G{\"u}nther}, {Lim}, {Crawford}, {Conseil}, {Shupe}, {Craig}, {Dencheva}, {Ginsburg}, {VanderPlas}, {Bradley}, {P{\'e}rez-Su{\'a}rez}, {de Val-Borro}, {Aldcroft}, {Cruz}, {Robitaille}, {Tollerud}, {Ardelean}, {Babej}, {Bach}, {Bachetti}, {Bakanov}, {Bamford}, {Barentsen}, {Barmby}, {Baumbach}, {Berry}, {Biscani}, {Boquien}, {Bostroem}, {Bouma}, {Brammer}, {Bray}, {Breytenbach}, {Buddelmeijer}, {Burke}, {Calderone}, {Cano Rodr{\'\i}guez}, {Cara}, {Cardoso}, {Cheedella}, {Copin}, {Corrales}, {Crichton}, {D'Avella}, {Deil}, {Depagne}, {Dietrich}, {Donath}, {Droettboom}, {Earl}, {Erben}, {Fabbro}, {Ferreira}, {Finethy}, {Fox}, {Garrison}, {Gibbons}, {Goldstein}, {Gommers}, {Greco}, {Greenfield}, {Groener}, {Grollier}, {Hagen}, {Hirst}, {Homeier}, {Horton}, {Hosseinzadeh}, {Hu}, {Hunkeler}, {Ivezi{\'c}}, {Jain}, {Jenness}, {Kanarek}, {Kendrew}, {Kern}, {Kerzendorf}, {Khvalko}, {King}, {Kirkby}, {Kulkarni},
  {Kumar}, {Lee}, {Lenz}, {Littlefair}, {Ma}, {Macleod}, {Mastropietro}, {McCully}, {Montagnac}, {Morris}, {Mueller}, {Mumford}, {Muna}, {Murphy}, {Nelson}, {Nguyen}, {Ninan}, {N{\"o}the}, {Ogaz}, {Oh}, {Parejko}, {Parley}, {Pascual}, {Patil}, {Patil}, {Plunkett}, {Prochaska}, {Rastogi}, {Reddy Janga}, {Sabater}, {Sakurikar}, {Seifert}, {Sherbert}, {Sherwood-Taylor}, {Shih}, {Sick}, {Silbiger}, {Singanamalla}, {Singer}, {Sladen}, {Sooley}, {Sornarajah}, {Streicher}, {Teuben}, {Thomas}, {Tremblay}, {Turner}, {Terr{\'o}n}, {van Kerkwijk}, {de la Vega}, {Watkins}, {Weaver}, {Whitmore}, {Woillez}, {Zabalza}, \& {Astropy Contributors}}]{Astropy-Collaboration2018}
{Astropy Collaboration}, {Price-Whelan}, A.~M., {Sip{\H{o}}cz}, B.~M., {et~al.} 2018, \aj, 156, 123

\bibitem[{{Astropy Collaboration} {et~al.}(2013){Astropy Collaboration}, {Robitaille}, {Tollerud}, {Greenfield}, {Droettboom}, {Bray}, {Aldcroft}, {Davis}, {Ginsburg}, {Price-Whelan}, {Kerzendorf}, {Conley}, {Crighton}, {Barbary}, {Muna}, {Ferguson}, {Grollier}, {Parikh}, {Nair}, {Unther}, {Deil}, {Woillez}, {Conseil}, {Kramer}, {Turner}, {Singer}, {Fox}, {Weaver}, {Zabalza}, {Edwards}, {Azalee Bostroem}, {Burke}, {Casey}, {Crawford}, {Dencheva}, {Ely}, {Jenness}, {Labrie}, {Lim}, {Pierfederici}, {Pontzen}, {Ptak}, {Refsdal}, {Servillat}, \& {Streicher}}]{AstropyCollaboration-2013}
{Astropy Collaboration}, {Robitaille}, T.~P., {Tollerud}, E.~J., {et~al.} 2013, \aap, 558, A33

\bibitem[{{Battaglia} {et~al.}(2022){Battaglia}, {Taibi}, {Thomas}, \& {Fritz}}]{Battaglia-et-al-2022}
{Battaglia}, G., {Taibi}, S., {Thomas}, G.~F., \& {Fritz}, T.~K. 2022, \aap, 657, A54

\bibitem[{{Bellazzini} {et~al.}(2002){Bellazzini}, {Ferraro}, {Origlia}, {Pancino}, {Monaco}, \& {Oliva}}]{Bellazzini-et-al-2002}
{Bellazzini}, M., {Ferraro}, F.~R., {Origlia}, L., {et~al.} 2002, \aj, 124, 3222

\bibitem[{{Bellm} {et~al.}(2019){Bellm}, {Kulkarni}, {Graham}, {Dekany}, {Smith}, {Riddle}, {Masci}, {Helou}, {Prince}, {Adams}, {Barbarino}, {Barlow}, {Bauer}, {Beck}, {Belicki}, {Biswas}, {Blagorodnova}, {Bodewits}, {Bolin}, {Brinnel}, {Brooke}, {Bue}, {Bulla}, {Burruss}, {Cenko}, {Chang}, {Connolly}, {Coughlin}, {Cromer}, {Cunningham}, {De}, {Delacroix}, {Desai}, {Duev}, {Eadie}, {Farnham}, {Feeney}, {Feindt}, {Flynn}, {Franckowiak}, {Frederick}, {Fremling}, {Gal-Yam}, {Gezari}, {Giomi}, {Goldstein}, {Golkhou}, {Goobar}, {Groom}, {Hacopians}, {Hale}, {Henning}, {Ho}, {Hover}, {Howell}, {Hung}, {Huppenkothen}, {Imel}, {Ip}, {Ivezi{\'c}}, {Jackson}, {Jones}, {Juric}, {Kasliwal}, {Kaspi}, {Kaye}, {Kelley}, {Kowalski}, {Kramer}, {Kupfer}, {Landry}, {Laher}, {Lee}, {Lin}, {Lin}, {Lunnan}, {Giomi}, {Mahabal}, {Mao}, {Miller}, {Monkewitz}, {Murphy}, {Ngeow}, {Nordin}, {Nugent}, {Ofek}, {Patterson}, {Penprase}, {Porter}, {Rauch}, {Rebbapragada}, {Reiley}, {Rigault}, {Rodriguez}, {van Roestel}, {Rusholme}, {van
  Santen}, {Schulze}, {Shupe}, {Singer}, {Soumagnac}, {Stein}, {Surace}, {Sollerman}, {Szkody}, {Taddia}, {Terek}, {Van Sistine}, {van Velzen}, {Vestrand}, {Walters}, {Ward}, {Ye}, {Yu}, {Yan}, \& {Zolkower}}]{Bellm-et-al-2019}
{Bellm}, E.~C., {Kulkarni}, S.~R., {Graham}, M.~J., {et~al.} 2019, \pasp, 131, 018002

\bibitem[{{Bono} {et~al.}(2019){Bono}, {Iannicola}, {Braga}, {Ferraro}, {Stetson}, {Magurno}, {Matsunaga}, {Beaton}, {Buonanno}, {Chaboyer}, {Dall'Ora}, {Fabrizio}, {Fiorentino}, {Freedman}, {Gilligan}, {Madore}, {Marconi}, {Marengo}, {Marinoni}, {Marrese}, {Martinez-Vazquez}, {Mateo}, {Monelli}, {Neeley}, {Nonino}, {Sneden}, {Thevenin}, {Valenti}, \& {Walker}}]{Bono-at-al-2019}
{Bono}, G., {Iannicola}, G., {Braga}, V.~F., {et~al.} 2019, \apj, 870, 115

\bibitem[{{Bressan} {et~al.}(2012){Bressan}, {Marigo}, {Girardi}, {Salasnich}, {Dal Cero}, {Rubele}, \& {Nanni}}]{Bressan-et-al-2012}
{Bressan}, A., {Marigo}, P., {Girardi}, L., {et~al.} 2012, \mnras, 427, 127

\bibitem[{{Carrera} {et~al.}(2002){Carrera}, {Aparicio}, {Mart{\'\i}nez-Delgado}, \& {Alonso-Garc{\'\i}a}}]{Carrera-et-al-2002}
{Carrera}, R., {Aparicio}, A., {Mart{\'\i}nez-Delgado}, D., \& {Alonso-Garc{\'\i}a}, J. 2002, \aj, 123, 3199

\bibitem[{{Catelan} \& {Smith}(2015)}]{Catelan-2015}
{Catelan}, M. \& {Smith}, H.~A. 2015, {Pulsating Stars}

\bibitem[{{Chiti} {et~al.}(2021){Chiti}, {Frebel}, {Simon}, {Erkal}, {Chang}, {Necib}, {Ji}, {Jerjen}, {Kim}, \& {Norris}}]{Chiti-2021}
{Chiti}, A., {Frebel}, A., {Simon}, J.~D., {et~al.} 2021, Nature Astronomy, 5, 392

\bibitem[{{Clementini} {et~al.}(2000){Clementini}, {Di Tomaso}, {Di Fabrizio}, {Bragaglia}, {Merighi}, {Tosi}, {Carretta}, {Gratton}, {Ivans}, {Kinard}, {Marconi}, {Smith}, {Wilhelm}, {Woodruff}, \& {Sneden}}]{Clementini-et-al-2000}
{Clementini}, G., {Di Tomaso}, S., {Di Fabrizio}, L., {et~al.} 2000, \aj, 120, 2054

\bibitem[{{Clementini} {et~al.}(2023){Clementini}, {Ripepi}, {Garofalo}, {Molinaro}, {Muraveva}, {Leccia}, {Rimoldini}, {Holl}, {Jevardat de Fombelle}, {Sartoretti}, {Marchal}, {Audard}, {Nienartowicz}, {Andrae}, {Marconi}, {Szabados}, {Evans}, {Lecoeur-Taibi}, {Mowlavi}, {Musella}, \& {Eyer}}]{Clementini-2023}
{Clementini}, G., {Ripepi}, V., {Garofalo}, A., {et~al.} 2023, \aap, 674, A18

\bibitem[{{Clementini} {et~al.}(2019){Clementini}, {Ripepi}, {Molinaro}, {Garofalo}, {Muraveva}, {Rimoldini}, {Guy}, {Jevardat de Fombelle}, {Nienartowicz}, {Marchal}, {Audard}, {Holl}, {Leccia}, {Marconi}, {Musella}, {Mowlavi}, {Lecoeur-Taibi}, {Eyer}, {De Ridder}, {Regibo}, {Sarro}, {Szabados}, {Evans}, \& {Riello}}]{Clementini-et-2019}
{Clementini}, G., {Ripepi}, V., {Molinaro}, R., {et~al.} 2019, \aap, 622, A60

\bibitem[{{Cudworth} {et~al.}(1986){Cudworth}, {Olszewski}, \& {Schommer}}]{Cudworth-et-al1986}
{Cudworth}, K.~M., {Olszewski}, E.~W., \& {Schommer}, R.~A. 1986, \aj, 92, 766

\bibitem[{{Dolphin}(2002)}]{Dolphin-2002}
{Dolphin}, A.~E. 2002, \mnras, 332, 91

\bibitem[{{Dotter} {et~al.}(2008){Dotter}, {Chaboyer}, {Jevremovi{\'c}}, {Kostov}, {Baron}, \& {Ferguson}}]{Dotter-et-al-2008}
{Dotter}, A., {Chaboyer}, B., {Jevremovi{\'c}}, D., {et~al.} 2008, \apjs, 178, 89

\bibitem[{{Drake} {et~al.}(2014){Drake}, {Graham}, {Djorgovski}, {Catelan}, {Mahabal}, {Torrealba}, {Garc{\'\i}a-{\'A}lvarez}, {Donalek}, {Prieto}, {Williams}, {Larson}, {Christen sen}, {Belokurov}, {Koposov}, {Beshore}, {Boattini}, {Gibbs}, {Hill}, {Kowalski}, {Johnson}, \& {Shelly}}]{Drake-et-al-2014}
{Drake}, A.~J., {Graham}, M.~J., {Djorgovski}, S.~G., {et~al.} 2014, \apjs, 213, 9

\bibitem[{{Frayn} \& {Gilmore}(2003)}]{Frayn-2003}
{Frayn}, C.~M. \& {Gilmore}, G.~F. 2003, \mnras, 339, 887

\bibitem[{{Gaia Collaboration} {et~al.}(2023){Gaia Collaboration}, {Vallenari}, {Brown}, {Prusti}, {de Bruijne}, {Arenou}, {Babusiaux}, {Biermann}, {Creevey}, {Ducourant}, {Evans}, {Eyer}, {Guerra}, {Hutton}, {Jordi}, {Klioner}, {Lammers}, {Lindegren}, {Luri}, {Mignard}, {Panem}, {Pourbaix}, {Randich}, {Sartoretti}, {Soubiran}, {Tanga}, {Walton}, {Bailer-Jones}, {Bastian}, {Drimmel}, {Jansen}, {Katz}, {Lattanzi}, {van Leeuwen}, {Bakker}, {Cacciari}, {Casta{\~n}eda}, {De Angeli}, {Fabricius}, {Fouesneau}, {Fr{\'e}mat}, {Galluccio}, {Guerrier}, {Heiter}, {Masana}, {Messineo}, {Mowlavi}, {Nicolas}, {Nienartowicz}, {Pailler}, {Panuzzo}, {Riclet}, {Roux}, {Seabroke}, {Sordo}, {Th{\'e}venin}, {Gracia-Abril}, {Portell}, {Teyssier}, {Altmann}, {Andrae}, {Audard}, {Bellas-Velidis}, {Benson}, {Berthier}, {Blomme}, {Burgess}, {Busonero}, {Busso}, {C{\'a}novas}, {Carry}, {Cellino}, {Cheek}, {Clementini}, {Damerdji}, {Davidson}, {de Teodoro}, {Nu{\~n}ez Campos}, {Delchambre}, {Dell'Oro}, {Esquej},
  {Fern{\'a}ndez-Hern{\'a}ndez}, {Fraile}, {Garabato}, {Garc{\'\i}a-Lario}, {Gosset}, {Haigron}, {Halbwachs}, {Hambly}, {Harrison}, {Hern{\'a}ndez}, {Hestroffer}, {Hodgkin}, {Holl}, {Jan{\ss}en}, {Jevardat de Fombelle}, {Jordan}, {Krone-Martins}, {Lanzafame}, {L{\"o}ffler}, {Marchal}, {Marrese}, {Moitinho}, {Muinonen}, {Osborne}, {Pancino}, {Pauwels}, {Recio-Blanco}, {Reyl{\'e}}, {Riello}, {Rimoldini}, {Roegiers}, {Rybizki}, {Sarro}, {Siopis}, {Smith}, {Sozzetti}, {Utrilla}, {van Leeuwen}, {Abbas}, {{\'A}brah{\'a}m}, {Abreu Aramburu}, {Aerts}, {Aguado}, {Ajaj}, {Aldea-Montero}, {Altavilla}, {{\'A}lvarez}, {Alves}, {Anders}, {Anderson}, {Anglada Varela}, {Antoja}, {Baines}, {Baker}, {Balaguer-N{\'u}{\~n}ez}, {Balbinot}, {Balog}, {Barache}, {Barbato}, {Barros}, {Barstow}, {Bartolom{\'e}}, {Bassilana}, {Bauchet}, {Becciani}, {Bellazzini}, {Berihuete}, {Bernet}, {Bertone}, {Bianchi}, {Binnenfeld}, {Blanco-Cuaresma}, {Blazere}, {Boch}, {Bombrun}, {Bossini}, {Bouquillon}, {Bragaglia}, {Bramante}, {Breedt},
  {Bressan}, {Brouillet}, {Brugaletta}, {Bucciarelli}, {Burlacu}, {Butkevich}, {Buzzi}, {Caffau}, {Cancelliere}, {Cantat-Gaudin}, {Carballo}, {Carlucci}, {Carnerero}, {Carrasco}, {Casamiquela}, {Castellani}, {Castro-Ginard}, {Chaoul}, {Charlot}, {Chemin}, {Chiaramida}, {Chiavassa}, {Chornay}, {Comoretto}, {Contursi}, {Cooper}, {Cornez}, {Cowell}, {Crifo}, {Cropper}, {Crosta}, {Crowley}, {Dafonte}, {Dapergolas}, {David}, {David}, {de Laverny}, {De Luise}, {De March}, {De Ridder}, {de Souza}, {de Torres}, {del Peloso}, {del Pozo}, {Delbo}, {Delgado}, {Delisle}, {Demouchy}, {Dharmawardena}, {Di Matteo}, {Diakite}, {Diener}, {Distefano}, {Dolding}, {Edvardsson}, {Enke}, {Fabre}, {Fabrizio}, {Faigler}, {Fedorets}, {Fernique}, {Fienga}, {Figueras}, {Fournier}, {Fouron}, {Fragkoudi}, {Gai}, {Garcia-Gutierrez}, {Garcia-Reinaldos}, {Garc{\'\i}a-Torres}, {Garofalo}, {Gavel}, {Gavras}, {Gerlach}, {Geyer}, {Giacobbe}, {Gilmore}, {Girona}, {Giuffrida}, {Gomel}, {Gomez}, {Gonz{\'a}lez-N{\'u}{\~n}ez},
  {Gonz{\'a}lez-Santamar{\'\i}a}, {Gonz{\'a}lez-Vidal}, {Granvik}, {Guillout}, {Guiraud}, {Guti{\'e}rrez-S{\'a}nchez}, {Guy}, {Hatzidimitriou}, {Hauser}, {Haywood}, {Helmer}, {Helmi}, {Sarmiento}, {Hidalgo}, {Hilger}, {H{\l}adczuk}, {Hobbs}, {Holland}, {Huckle}, {Jardine}, {Jasniewicz}, {Jean-Antoine Piccolo}, {Jim{\'e}nez-Arranz}, {Jorissen}, {Juaristi Campillo}, {Julbe}, {Karbevska}, {Kervella}, {Khanna}, {Kontizas}, {Kordopatis}, {Korn}, {K{\'o}sp{\'a}l}, {Kostrzewa-Rutkowska}, {Kruszy{\'n}ska}, {Kun}, {Laizeau}, {Lambert}, {Lanza}, {Lasne}, {Le Campion}, {Lebreton}, {Lebzelter}, {Leccia}, {Leclerc}, {Lecoeur-Taibi}, {Liao}, {Licata}, {Lindstr{\o}m}, {Lister}, {Livanou}, {Lobel}, {Lorca}, {Loup}, {Madrero Pardo}, {Magdaleno Romeo}, {Managau}, {Mann}, {Manteiga}, {Marchant}, {Marconi}, {Marcos}, {Marcos Santos}, {Mar{\'\i}n Pina}, {Marinoni}, {Marocco}, {Marshall}, {Martin Polo}, {Mart{\'\i}n-Fleitas}, {Marton}, {Mary}, {Masip}, {Massari}, {Mastrobuono-Battisti}, {Mazeh}, {McMillan}, {Messina}, {Michalik},
  {Millar}, {Mints}, {Molina}, {Molinaro}, {Moln{\'a}r}, {Monari}, {Mongui{\'o}}, {Montegriffo}, {Montero}, {Mor}, {Mora}, {Morbidelli}, {Morel}, {Morris}, {Muraveva}, {Murphy}, {Musella}, {Nagy}, {Noval}, {Oca{\~n}a}, {Ogden}, {Ordenovic}, {Osinde}, {Pagani}, {Pagano}, {Palaversa}, {Palicio}, {Pallas-Quintela}, {Panahi}, {Payne-Wardenaar}, {Pe{\~n}alosa Esteller}, {Penttil{\"a}}, {Pichon}, {Piersimoni}, {Pineau}, {Plachy}, {Plum}, {Poggio}, {Pr{\v{s}}a}, {Pulone}, {Racero}, {Ragaini}, {Rainer}, {Raiteri}, {Rambaux}, {Ramos}, {Ramos-Lerate}, {Re Fiorentin}, {Regibo}, {Richards}, {Rios Diaz}, {Ripepi}, {Riva}, {Rix}, {Rixon}, {Robichon}, {Robin}, {Robin}, {Roelens}, {Rogues}, {Rohrbasser}, {Romero-G{\'o}mez}, {Rowell}, {Royer}, {Ruz Mieres}, {Rybicki}, {Sadowski}, {S{\'a}ez N{\'u}{\~n}ez}, {Sagrist{\`a} Sell{\'e}s}, {Sahlmann}, {Salguero}, {Samaras}, {Sanchez Gimenez}, {Sanna}, {Santove{\~n}a}, {Sarasso}, {Schultheis}, {Sciacca}, {Segol}, {Segovia}, {S{\'e}gransan}, {Semeux}, {Shahaf}, {Siddiqui}, {Siebert},
  {Siltala}, {Silvelo}, {Slezak}, {Slezak}, {Smart}, {Snaith}, {Solano}, {Solitro}, {Souami}, {Souchay}, {Spagna}, {Spina}, {Spoto}, {Steele}, {Steidelm{\"u}ller}, {Stephenson}, {S{\"u}veges}, {Surdej}, {Szabados}, {Szegedi-Elek}, {Taris}, {Taylor}, {Teixeira}, {Tolomei}, {Tonello}, {Torra}, {Torra}, {Torralba Elipe}, {Trabucchi}, {Tsounis}, {Turon}, {Ulla}, {Unger}, {Vaillant}, {van Dillen}, {van Reeven}, {Vanel}, {Vecchiato}, {Viala}, {Vicente}, {Voutsinas}, {Weiler}, {Wevers}, {Wyrzykowski}, {Yoldas}, {Yvard}, {Zhao}, {Zorec}, {Zucker}, \& {Zwitter}}]{Vallenari-2023}
{Gaia Collaboration}, {Vallenari}, A., {Brown}, A.~G.~A., {et~al.} 2023, \aap, 674, A1

\bibitem[{{Garofalo} {et~al.}(2022){Garofalo}, {Delgado}, {Sarro}, {Clementini}, {Muraveva}, {Marconi}, \& {Ripepi}}]{Garofalo-et-al-2022}
{Garofalo}, A., {Delgado}, H.~E., {Sarro}, L.~M., {et~al.} 2022, \mnras, 513, 788

\bibitem[{{Garofalo} {et~al.}(2021){Garofalo}, {Tantalo}, {Cusano}, {Clementini}, {Calura}, {Muraveva}, {Paris}, \& {Speziali}}]{Garofalo-et-al-2021}
{Garofalo}, A., {Tantalo}, M., {Cusano}, F., {et~al.} 2021, \apj, 916, 10

\bibitem[{{Hunter}(2007)}]{Hunter-2007}
{Hunter}, J.~D. 2007, Computing in Science and Engineering, 9, 90

\bibitem[{{Iben}(1974)}]{Iben-1974}
{Iben}, I., J. 1974, \araa, 12, 215

\bibitem[{{Iorio} \& {Belokurov}(2021)}]{Iorio-et-al-2021}
{Iorio}, G. \& {Belokurov}, V. 2021, \mnras, 502, 5686

\bibitem[{{Irwin} \& {Hatzidimitriou}(1995)}]{Irwin-1995}
{Irwin}, M. \& {Hatzidimitriou}, D. 1995, \mnras, 277, 1354

\bibitem[{{Jensen} {et~al.}(2024){Jensen}, {Hayes}, {Sestito}, {McConnachie}, {Waller}, {Smith}, {Navarro}, \& {Venn}}]{Jensen-et-al-2024}
{Jensen}, J., {Hayes}, C.~R., {Sestito}, F., {et~al.} 2024, \mnras, 527, 4209

\bibitem[{{Kalirai} {et~al.}(2013){Kalirai}, {Anderson}, {Dotter}, {Richer}, {Fahlman}, {Hansen}, {Hurley}, {Reid}, {Rich}, \& {Shara}}]{Kalirai-et-al-2013}
{Kalirai}, J.~S., {Anderson}, J., {Dotter}, A., {et~al.} 2013, \apj, 763, 110

\bibitem[{{Kholopov}(1971)}]{Kholopov-1971}
{Kholopov}, P.~N. 1971, Peremennye Zvezdy, 18, 117

\bibitem[{{Kirby} {et~al.}(2010){Kirby}, {Guhathakurta}, {Simon}, {Geha}, {Rockosi}, {Sneden}, {Cohen}, {Sohn}, {Majewski}, \& {Siegel}}]{Kirby-et-al-2010}
{Kirby}, E.~N., {Guhathakurta}, P., {Simon}, J.~D., {et~al.} 2010, \apjs, 191, 352

\bibitem[{{Kirby} {et~al.}(2011){Kirby}, {Lanfranchi}, {Simon}, {Cohen}, \& {Guhathakurta}}]{Kirby-et-al-2011}
{Kirby}, E.~N., {Lanfranchi}, G.~A., {Simon}, J.~D., {Cohen}, J.~G., \& {Guhathakurta}, P. 2011, \apj, 727, 78

\bibitem[{{Lee} {et~al.}(1990){Lee}, {Demarque}, \& {Zinn}}]{Lee-1990}
{Lee}, Y.-W., {Demarque}, P., \& {Zinn}, R. 1990, \apj, 350, 155

\bibitem[{{Li} {et~al.}(2023){Li}, {Huang}, {Liu}, {Beers}, \& {Zhang}}]{Li-et-al-2023}
{Li}, X.-Y., {Huang}, Y., {Liu}, G.-C., {Beers}, T.~C., \& {Zhang}, H.-W. 2023, \apj, 944, 88

\bibitem[{{Mateo}(1998)}]{Mateo-1998}
{Mateo}, M.~L. 1998, \araa, 36, 435

\bibitem[{{McConnachie}(2012)}]{McConnachie-2012}
{McConnachie}, A.~W. 2012, \aj, 144, 4

\bibitem[{{McConnachie} \& {Venn}(2020)}]{McConnachie&Venn-2020}
{McConnachie}, A.~W. \& {Venn}, K.~A. 2020, \aj, 160, 124

\bibitem[{{Medina} {et~al.}(2018){Medina}, {Mu{\~n}oz}, {Vivas}, {Carlin}, {F{\"o}rster}, {Mart{\'\i}nez}, {Galbany}, {Gonz{\'a}lez-Gait{\'a}n}, {Hamuy}, {de Jaeger}, {Maureira}, \& {San Mart{\'\i}n}}]{Medina-et-al-2018}
{Medina}, G.~E., {Mu{\~n}oz}, R.~R., {Vivas}, A.~K., {et~al.} 2018, \apj, 855, 43

\bibitem[{{Mighell} \& {Burke}(1999)}]{Mighell-1999}
{Mighell}, K.~J. \& {Burke}, C.~J. 1999, \aj, 118, 366

\bibitem[{{Mu{\~n}oz} {et~al.}(2018){Mu{\~n}oz}, {C{\^o}t{\'e}}, {Santana}, {Geha}, {Simon}, {Oyarz{\'u}n}, {Stetson}, \& {Djorgovski}}]{Munoz-et-al-2018}
{Mu{\~n}oz}, R.~R., {C{\^o}t{\'e}}, P., {Santana}, F.~A., {et~al.} 2018, \apj, 860, 66

\bibitem[{{Mu{\~n}oz} {et~al.}(2012){Mu{\~n}oz}, {Geha}, {C{\^o}t{\'e}}, {Vargas}, {Santana}, {Stetson}, {Simon}, \& {Djorgovski}}]{Munoz-et-al-2012}
{Mu{\~n}oz}, R.~R., {Geha}, M., {C{\^o}t{\'e}}, P., {et~al.} 2012, \apjl, 753, L15

\bibitem[{Muraveva {et~al.}(2024)Muraveva, Giannetti, Clementini, Garofalo, \& Monti}]{Muraveva-et-al-2024}
Muraveva, T., Giannetti, A., Clementini, G., Garofalo, A., \& Monti, L. 2024, Metallicity of RR Lyrae stars from the Gaia Data Release 3 catalogue computed with Machine Learning algorithms

\bibitem[{{Nemec} {et~al.}(1988){Nemec}, {Wehlau}, \& {Mendes de Oliveira}}]{nemec88}
{Nemec}, J.~M., {Wehlau}, A., \& {Mendes de Oliveira}, C. 1988, \aj, 96, 528

\bibitem[{{Ochsenbein} {et~al.}(2000){Ochsenbein}, {Bauer}, \& {Marcout}}]{Ochsenbein-2000}
{Ochsenbein}, F., {Bauer}, P., \& {Marcout}, J. 2000, \aaps, 143, 23

\bibitem[{{Olszewski} \& {Aaronson}(1985)}]{Olszewski-1985}
{Olszewski}, E.~W. \& {Aaronson}, M. 1985, \aj, 90, 2221

\bibitem[{{Pace} {et~al.}(2020){Pace}, {Kaplinghat}, {Kirby}, {Simon}, {Tollerud}, {Mu{\~n}oz}, {C{\^o}t{\'e}}, {Djorgovski}, \& {Geha}}]{Pace-et-al-2020}
{Pace}, A.~B., {Kaplinghat}, M., {Kirby}, E., {et~al.} 2020, \mnras, 495, 3022

\bibitem[{{Prudil} {et~al.}(2024){Prudil}, {Kunder}, {D{\'e}k{\'a}ny}, \& {Koch-Hansen}}]{Prudil-et-al-2024}
{Prudil}, Z., {Kunder}, A., {D{\'e}k{\'a}ny}, I., \& {Koch-Hansen}, A.~J. 2024, \aap, 684, A176

\bibitem[{{Riello} {et~al.}(2021){Riello}, {De Angeli}, {Evans}, {Montegriffo}, {Carrasco}, {Busso}, {Palaversa}, {Burgess}, {Diener}, {Davidson}, {Rowell}, {Fabricius}, {Jordi}, {Bellazzini}, {Pancino}, {Harrison}, {Cacciari}, {van Leeuwen}, {Hambly}, {Hodgkin}, {Osborne}, {Altavilla}, {Barstow}, {Brown}, {Castellani}, {Cowell}, {De Luise}, {Gilmore}, {Giuffrida}, {Hidalgo}, {Holland}, {Marinoni}, {Pagani}, {Piersimoni}, {Pulone}, {Ragaini}, {Rainer}, {Richards}, {Sanna}, {Walton}, {Weiler}, \& {Yoldas}}]{Riello-et-al-2021}
{Riello}, M., {De Angeli}, F., {Evans}, D.~W., {et~al.} 2021, \aap, 649, A3

\bibitem[{{Rimoldini} {et~al.}(2023){Rimoldini}, {Holl}, {Gavras}, {Audard}, {De Ridder}, {Mowlavi}, {Nienartowicz}, {Jevardat de Fombelle}, {Lecoeur-Ta{\"\i}bi}, {Karbevska}, {Evans}, {{\'A}brah{\'a}m}, {Carnerero}, {Clementini}, {Distefano}, {Garofalo}, {Garc{\'\i}a-Lario}, {Gomel}, {Klioner}, {Kruszy{\'n}ska}, {Lanzafame}, {Lebzelter}, {Marton}, {Mazeh}, {Molinaro}, {Panahi}, {Raiteri}, {Ripepi}, {Szabados}, {Teyssier}, {Trabucchi}, {Wyrzykowski}, {Zucker}, \& {Eyer}}]{Rimoldini-et-al-2023}
{Rimoldini}, L., {Holl}, B., {Gavras}, P., {et~al.} 2023, \aap, 674, A14

\bibitem[{{Ripepi} {et~al.}(2023){Ripepi}, {Clementini}, {Molinaro}, {Leccia}, {Plachy}, {Moln{\'a}r}, {Rimoldini}, {Musella}, {Marconi}, {Garofalo}, {Audard}, {Holl}, {Evans}, {Jevardat de Fombelle}, {Lecoeur-Taibi}, {Marchal}, {Mowlavi}, {Muraveva}, {Nienartowicz}, {Sartoretti}, {Szabados}, \& {Eyer}}]{Ripepi-2023}
{Ripepi}, V., {Clementini}, G., {Molinaro}, R., {et~al.} 2023, \aap, 674, A17

\bibitem[{{Ripepi} {et~al.}(2019){Ripepi}, {Molinaro}, {Musella}, {Marconi}, {Leccia}, \& {Eyer}}]{Ripepi-at-al-2019}
{Ripepi}, V., {Molinaro}, R., {Musella}, I., {et~al.} 2019, A\&A, 625, A14

\bibitem[{{Ruhland} {et~al.}(2011){Ruhland}, {Bell}, {Rix}, \& {Xue}}]{Ruhland-et-al-2011}
{Ruhland}, C., {Bell}, E.~F., {Rix}, H.-W., \& {Xue}, X.-X. 2011, \apj, 731, 119

\bibitem[{{Samus'} {et~al.}(2017){Samus'}, {Kazarovets}, {Durlevich}, {Kireeva}, \& {Pastukhova}}]{Samus-2017}
{Samus'}, N.~N., {Kazarovets}, E.~V., {Durlevich}, O.~V., {Kireeva}, N.~N., \& {Pastukhova}, E.~N. 2017, Astronomy Reports, 61, 80

\bibitem[{{Sarajedini}(2011)}]{Sarajedini-2011}
{Sarajedini}, A. 2011, in RR Lyrae Stars, Metal-Poor Stars, and the Galaxy, ed. A.~{McWilliam}, Vol.~5, 181

\bibitem[{{Schlafly} \& {Finkbeiner}(2011)}]{Schlafly-and-Finkbeiner-2011}
{Schlafly}, E.~F. \& {Finkbeiner}, D.~P. 2011, \apj, 737, 103

\bibitem[{{Schlegel} {et~al.}(1998){Schlegel}, {Finkbeiner}, \& {Davis}}]{Schlegel-98}
{Schlegel}, D.~J., {Finkbeiner}, D.~P., \& {Davis}, M. 1998, \apj, 500, 525

\bibitem[{{Searle} \& {Zinn}(1978)}]{SearleZinn-1978}
{Searle}, L. \& {Zinn}, R. 1978, \apj, 225, 357

\bibitem[{{Sesar} {et~al.}(2017){Sesar}, {Hernitschek}, {Mitrovi{\'c}}, {Ivezi{\'c}}, {Rix}, {Cohen}, {Bernard}, {Grebel}, {Martin}, {Schlafly}, {Burgett}, {Draper}, {Flewelling}, {Kaiser}, {Kudritzki}, {Magnier}, {Metcalfe}, {Tonry}, \& {Waters}}]{Sesar-et-al-2017}
{Sesar}, B., {Hernitschek}, N., {Mitrovi{\'c}}, S., {et~al.} 2017, \aj, 153, 204

\bibitem[{{Sestito} {et~al.}(2023){Sestito}, {Zaremba}, {Venn}, {D'Aoust}, {Hayes}, {Jensen}, {Navarro}, {Jablonka}, {Fern{\'a}ndez-Alvar}, {Glover}, {McConnachie}, \& {Chen{\'e}}}]{Sestito-et-al-2023}
{Sestito}, F., {Zaremba}, D., {Venn}, K.~A., {et~al.} 2023, arXiv e-prints, arXiv:2301.13214

\bibitem[{{Smith}(1995)}]{Smith-1995}
{Smith}, H.~A. 1995, Cambridge Astrophysics Series, 27

\bibitem[{{Tammann} {et~al.}(2008){Tammann}, {Sandage}, \& {Reindl}}]{Tammann2008}
{Tammann}, G.~A., {Sandage}, A., \& {Reindl}, B. 2008, \apj, 679, 52

\bibitem[{{Tau} {et~al.}(2024){Tau}, {Vivas}, \& {Mart{\'\i}nez-V{\'a}zquez}}]{Tau-et-al-2024}
{Tau}, E.~A., {Vivas}, A.~K., \& {Mart{\'\i}nez-V{\'a}zquez}, C.~E. 2024, \aj, 167, 57

\bibitem[{{Taylor}(2005)}]{Taylor-2005}
{Taylor}, M.~B. 2005, in Astronomical Society of the Pacific Conference Series, Vol. 347, Astronomical Data Analysis Software and Systems XIV, ed. P.~{Shopbell}, M.~{Britton}, \& R.~{Ebert}, 29

\bibitem[{{Tully} {et~al.}(2013){Tully}, {Courtois}, {Dolphin}, {Fisher}, {H{\'e}raudeau}, {Jacobs}, {Karachentsev}, {Makarov}, {Makarova}, {Mitronova}, {Rizzi}, {Shaya}, {Sorce}, \& {Wu}}]{Tully-et-al-2013}
{Tully}, R.~B., {Courtois}, H.~M., {Dolphin}, A.~E., {et~al.} 2013, \aj, 146, 86

\bibitem[{{van Agt}(1967)}]{vanAgt-1967}
{van Agt}, S.~L. T.~J. 1967, \bain, 19, 275

\bibitem[{{van der Walt} {et~al.}(2011){van der Walt}, {Colbert}, \& {Varoquaux}}]{vanderWalt-2011}
{van der Walt}, S., {Colbert}, S.~C., \& {Varoquaux}, G. 2011, Computing in Science and Engineering, 13, 22

\bibitem[{{Vasiliev} \& {Baumgardt}(2021)}]{Vasiliev-Baumgardt-2021}
{Vasiliev}, E. \& {Baumgardt}, H. 2021, \mnras, 505, 5978

\bibitem[{{Vivas} \& {Zinn}(2006)}]{Vivas-Zinn-2006}
{Vivas}, A.~K. \& {Zinn}, R. 2006, \aj, 132, 714

\bibitem[{{Webbink}(1985)}]{Webbink1985}
{Webbink}, R.~F. 1985, in Dynamics of Star Clusters, ed. J.~{Goodman} \& P.~{Hut}, Vol. 113, 541--577

\bibitem[{{Weisz} {et~al.}(2014){Weisz}, {Dolphin}, {Skillman}, {Holtzman}, {Gilbert}, {Dalcanton}, \& {Williams}}]{Weisz-et-al-2014}
{Weisz}, D.~R., {Dolphin}, A.~E., {Skillman}, E.~D., {et~al.} 2014, \apj, 789, 147

\bibitem[{{Wilson}(1955)}]{Wilson-1955}
{Wilson}, A.~G. 1955, \pasp, 67, 27

\bibitem[{{Zgirski} {et~al.}(2023){Zgirski}, {Pietrzy{\'n}ski}, {G{\'o}rski}, {Gieren}, {Wielg{\'o}rski}, {Karczmarek}, {Hajdu}, {Lewis}, {Chini}, {Graczyk}, {Ka{\l}uszy{\'n}ski}, {Narloch}, {Pilecki}, {Garc{\'\i}a}, {Suchomska}, \& {Taormina}}]{Zgirski-et-al-2023}
{Zgirski}, B., {Pietrzy{\'n}ski}, G., {G{\'o}rski}, M., {et~al.} 2023, \apj, 951, 114

\end{thebibliography}
%

\begin{appendix} 
\section{$G$, G$_{BP}$ and G$_{RP}$ light Curves}\label{sec:appendix}
\begin{figure*}
    \centering
    \includegraphics[width=12cm]{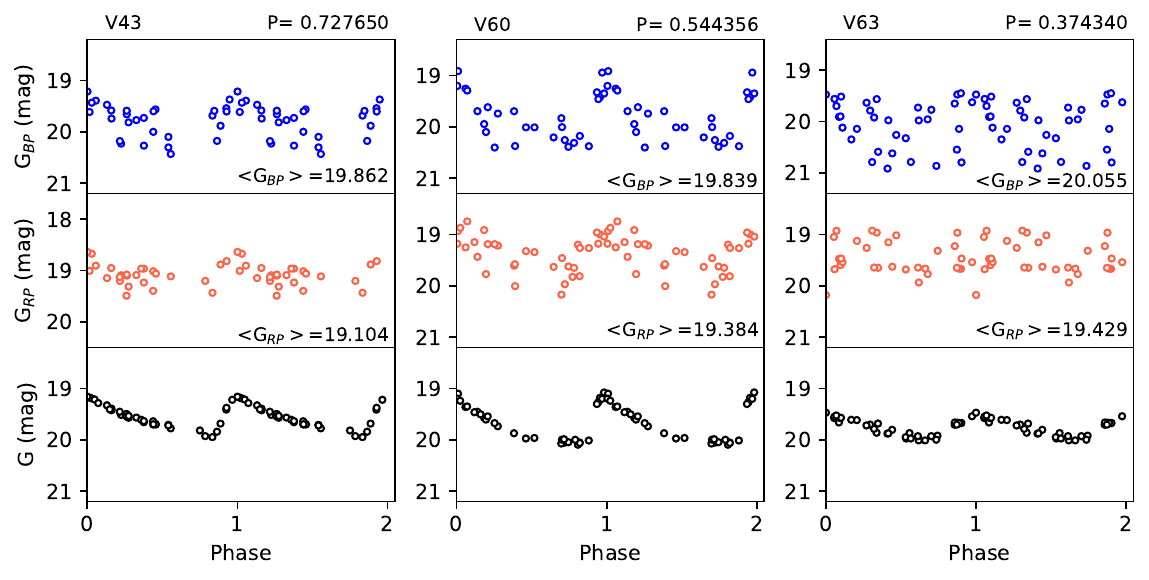}
    \caption{From top to bottom: {\it Gaia} DR3 $G_{BP}$, $G_{RP}$ and $G$ light curves for the Gold sample RR Lyrae stars V43, V60 and V63. 
    All curves are folded with the periods obtained from our analysis. They are perfectly consistent with the periods published for these sources in the {\it Gaia} DR3 \texttt{vari\_rrlyrae} table and 
    in N88.}
    \label{fig:lc_app_1}
\end{figure*}

\begin{figure*}
    \centering
    \includegraphics[width=12cm]{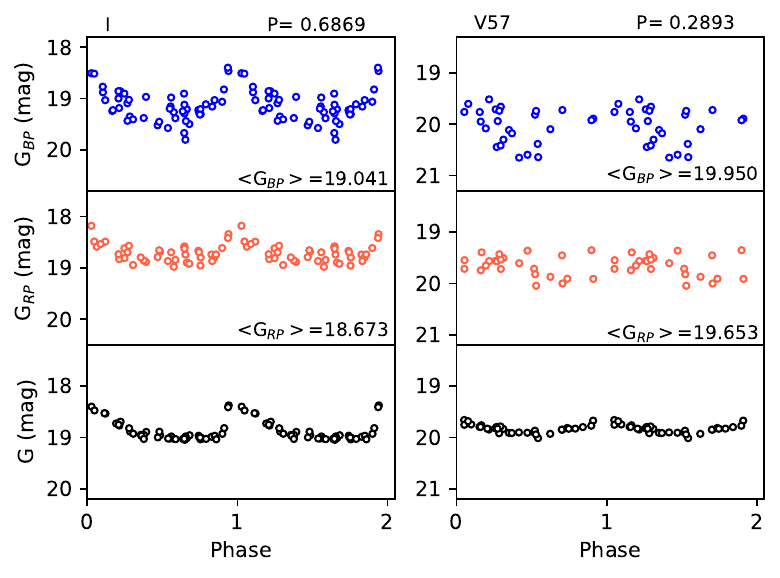}
    \caption{{\it Gaia} DR3 $G_{BP}$ and $G_{RP}$ light curves of the candidate AC V178 (I) and the RRc star V57. The periods adopted to fold the light curves are those redetermined in this work using  {\it Gaia} DR3 time-series (see Sec.~\ref{sec:ac}).}
    \label{fig:lc_app_3}
\end{figure*}

Here are collected the light curves folded with the periods that we derived in this work.\\ Figure~\ref{fig:lc_app_1} shows the {\it Gaia} DR3 light curves of the RR Lyrae stars V43, V60, and V63 discussed in Section~\ref{sec:dist}.
  The light curves of the candidate AC V178 (I) and the RRc star V57, discussed in Section~\ref{sec:ac}, are shown in Figure~\ref{fig:lc_app_3}.
  The $G_{BP}$ and $G_{RP}$ light curves of the RR Lyrae stars V9 and V47, discussed in Section~\ref{sec:new}, are shown in Figure~\ref{fig:lc_app_2} while those of RR Lyrae stars V134 and candidate V156 discussed in Section~\ref{sec:new48} are shown in Figure~\ref{fig:lc_app_5} .
\\

\begin{figure*}
    \centering
    \includegraphics[width=11.7cm]{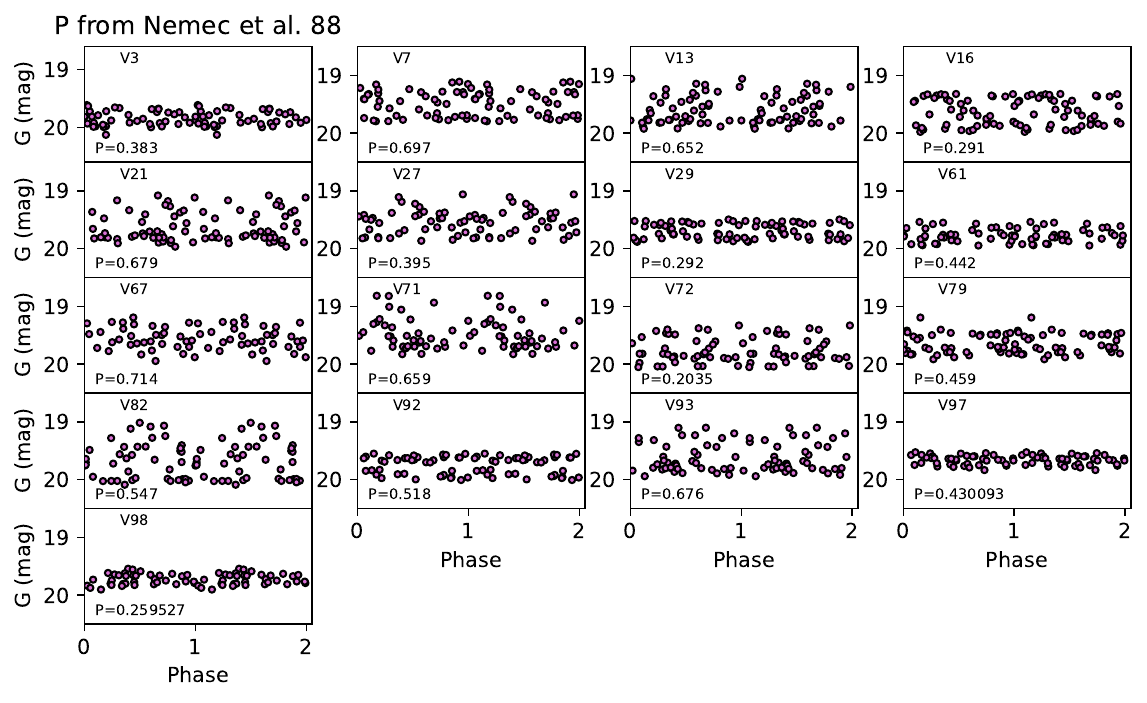}
    \caption{{\it Gaia} DR3 $G$-band  light curves for 17 of the 20 RR Lyrae stars discussed in Sec.~\ref{sec:new}. The periods adopted to fold the light curves are those from N88 (see 
 column 3 of Table~\ref{tab:rr-psbag}).}
    \label{fig:lc_app_4}
\end{figure*}
\begin{figure*}
    \centering
    \includegraphics[width=12cm]{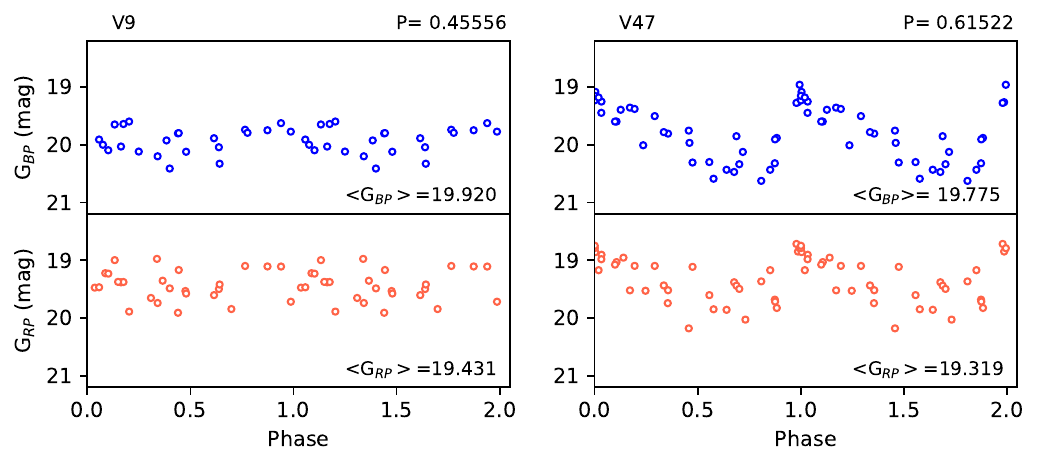}
    \caption{{\it Gaia} DR3 $G_{BP}$ and $G_{RP}$ light curves of the RR Lyrae stars V9 and V47. The periods adopted to fold the light curves are those determined in this work using the {\it Gaia} DR3 $G$-band time-series data (see Sec.~\ref{sec:new}).}
    \label{fig:lc_app_2}
\end{figure*}
\begin{figure*}
    \centering
    \includegraphics[width=12cm]{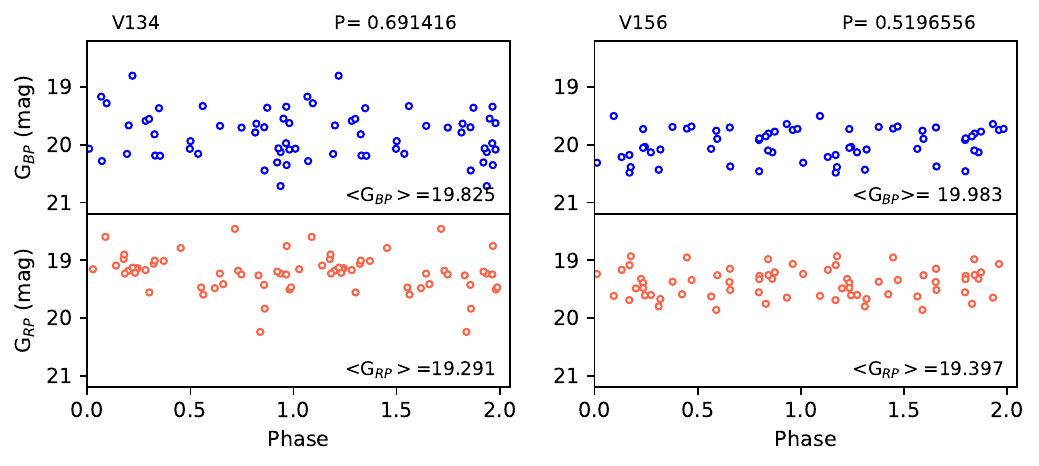}
    \caption{Same as in Fig.~\ref{fig:lc_app_2} for the confirmed RR Lyrae star V134 and the candidate RR Lyrae V156. The periods adopted to fold the light curves are those determined in this work using the {\it Gaia} DR3 $G$-band time-series data (see Sec.~\ref{sec:new48}).}
    \label{fig:lc_app_5}
\end{figure*}


\end{appendix}

\end{document}